%%%%%%%%%%%%%%%%%%  tex macros for preprints, cm version %%%%%%%%%%%%%%
%         (P. Ginsparg <ginsparg@lanl.gov>, last updated 7/94)
%         hypertex extensions (still provisional), 7/26/94
%	  Some modifications by C.R.Mafra, 2012

%comment out this line to restore non-hyper functionality
%\input hyperbasics

\input amssym.tex % for blackboard bold

\def\unredoffs{}
\tolerance=1000\hfuzz=2pt
\catcode`\@=11 % This allows us to modify PLAIN macros.
\ifx\hyperdef\UNd@FiNeD\def\hyperdef#1#2#3#4{#4}\def\hyperref#1#2#3#4{#4}\def\href#1#2{#2}\fi
\magnification=1200\unredoffs\baselineskip=16pt plus 2pt minus 1pt
\def\Date#1{\vfill\leftline{#1}\tenpoint\supereject%
\footline={\hss\tenrm\hyperdef\hypernoname{page}\folio\folio\hss}}%
% (restores pagenumbers)

%%%%%% Hour:Minute %%%%%%%%%%%%%%%%%
{\count255=\time\divide\count255 by 60 \xdef\hourmin{\number\count255}
 \multiply\count255 by-60\advance\count255 by\time
 \xdef\hourmin{\hourmin:\ifnum\count255<10 0\fi\the\count255}
}
\def\date{\number\day.\number\month.\number\year\ at \hourmin}

%%%%%%%%%%%% Draft mode %%%%%%%%%%%%%
% puts date/time on each page in big mode, writes labels in margins

% use \nolabels to get rid of eqn, ref, and fig labels in draft mode
\def\nolabels{\def\wrlabeL##1{}\def\eqlabeL##1{}\def\reflabeL##1{}}
\def\writelabels{\def\wrlabeL##1{\leavevmode\vadjust{\rlap{\smash%
{\line{{\escapechar=` \hfill\rlap{\sevenrm\hskip.03in\string##1}}}}}}}%
\def\eqlabeL##1{{\escapechar-1\rlap{\sevenrm\hskip.05in\string##1}}}%
\def\reflabeL##1{\noexpand\llap{\noexpand\sevenrm\string\string\string##1}}}
\nolabels

% tagged sec numbers
\global\newcount\secno \global\secno=0
\global\newcount\meqno \global\meqno=1
\def\s@csym{}

%%%%%%%%% Section %%%%%%%%%%%%%
\def\newsec#1\par{\global\advance\secno by1%
{\toks0{#1}\message{(\the\secno. \the\toks0)}}%
\global\subsecno=0\eqnres@t\let\s@csym\secsym\xdef\secn@m{\the\secno}\noindent
{\bf\hyperdef\hypernoname{section}{\the\secno}{\the\secno.} #1}%
\writetoca{{\string\hyperref{}{section}{\the\secno}{\bf \the\secno\quad}} {\bf #1}}\par%
\nobreak\medskip\nobreak\noindent\ignorespaces}
\def\eqnres@t{\xdef\secsym{\the\secno.}\global\meqno=1\bigbreak\bigskip}
\def\sequentialequations{\def\eqnres@t{\bigbreak}}\xdef\secsym{}

%%%%%%%% Subsection %%%%%%%%%%%
\global\newcount\subsecno \global\subsecno=0
\def\subsec#1\par{\global\advance\subsecno by1%
{\toks0{#1}\message{(\s@csym\the\subsecno. \the\toks0)}}%
\global\subsubsecno=0%
\ifnum\lastpenalty>9000\else\bigbreak\fi
\noindent{\it\hyperdef\hypernoname{subsection}{\secn@m.\the\subsecno}%
{\secn@m.\the\subsecno.} #1}\writetoca{\string\hskip1.45cm
{\string\hyperref{}{subsection}{\secn@m.\the\subsecno}{\secn@m.\the\subsecno.}}
{#1}}\par\nobreak\medskip\nobreak\noindent\ignorespaces}

%%%%%%%%%%%%%%% Subsubsection %%%%%%%%%%%%%%%%%%%%%%%%%%%%%%%%%%%%
\global\newcount\subsubsecno \global\subsubsecno=0
\def\subsubsec#1\par{\global\advance\subsubsecno by1%
{\toks0{#1}\message{(\secn@m.\the\subsecno.\the\subsubsecno. \the\toks0)}}%
\global\subsubsubsecno=0%
\ifnum\lastpenalty>9000\else\bigbreak\fi
\noindent{\it\hyperdef\hypernoname{subsubsection}{\secn@m.\the\subsecno\the\subsubsecno}%
{\secn@m.\the\subsecno.\the\subsubsecno.} #1}
%%% Add Subsubsections to Index
%\writetoca{\string\quad{\string\hyperref{}{subsubsection}{\the\secno\the\subsecno\the
%\subsubsecno}{\baselineskip=9pt\it\the\secno.\the\subsecno.\the\subsubsecno.}}
% {\baselineskip=9pt\it\ #1}}
\par\nobreak\medskip\nobreak\noindent\ignorespaces}

%%%%%%%%%%%%%%% Subsubsubsection %%%%%%%%%%%%%%%%%%%%%%%%%%%%%%%%%%%%
\global\newcount\subsubsubsecno \global\subsubsubsecno=0
\def\subsubsubsec#1\par{\global\advance\subsubsubsecno by1%
{\toks0{#1}\message{(\secn@m.\the\subsecno.\the\subsubsecno.\the\subsubsubsecno \the\toks0)}}%
\ifnum\lastpenalty>9000\else\bigbreak\fi
\noindent{\it\hyperdef\hypernoname{subsubsection}{\secn@m.\the\subsecno\the\subsubsecno\the\subsubsubsecno}%
{\secn@m.\the\subsecno.\the\subsubsecno.\the\subsubsubsecno.} #1}%
\par\nobreak\medskip\nobreak\noindent\ignorespaces}

%%%%%%%%% sections with automatic labels %%%%%%%%%%%%%%%%%%%%%%%%%%

%%%%%%%%% section with label %%%%%%%%%%%%%%%%%%%%%%%%%%%%%%%%
\def\newnewsec#1#2\par{\global\advance\secno by1%
{\toks0{#2}\message{(\secn@m. \the\toks0)}}%
\global\subsecno=0\global\subsubsecno=0\eqnres@t\let\s@csym\secsym\xdef\secn@m{\the\secno}\noindent
\ifnum\lastpenalty>9000\else\bigbreak\fi
\noindent{\bf\hyperdef\hypernoname{section}{\secn@m}{\secn@m.} #2}%
\writetoca{{\string\hyperref{}{section}{\the\secno}{\bf \the\secno\quad}} {\bf #2}}
%define section label
\DefWarn#1%
\xdef#1{\noexpand\hyperref{}{section}{\the\secno}%
{\the\secno}}\writedef{#1\leftbracket#1}\wrlabeL{#1=#1}%
\par\nobreak\medskip\nobreak\noindent\ignorespaces}

%%%%%%%%% subsection with label %%%%%%%%%%%%%%%%%%%%%%%%%%%%%%%%
\def\newsubsec#1#2\par{\global\advance\subsecno by1%
{\toks0{#2}\message{(\secn@m.\the\subsecno. \the\toks0)}}%
\global\subsubsecno=0%
\ifnum\lastpenalty>9000\else\bigbreak\fi
\noindent{\it\hyperdef\hypernoname{subsection}{\secn@m.\the\subsecno}%
{\secn@m.\the\subsecno.} #2}
%define section label
\DefWarn#1%
\xdef#1{\noexpand\hyperref{}{subsection}{\secn@m.\the\subsecno}%
{\secn@m.\the\subsecno}}\writedef{#1\leftbracket#1}\wrlabeL{#1=#1}%
\writetoca{\string\hskip1.45cm
{\string\hyperref{}{subsection}{\secn@m.\the\subsecno}{\secn@m.\the\subsecno.}}
{#2}}%
\par\nobreak\medskip\nobreak\noindent\ignorespaces}

%%%%%%%%% subsubsection with label %%%%%%%%%%%%%%%%%%%%%%%%%%%%%%%%
\def\newsubsubsec#1#2\par{\global\advance\subsubsecno by1%
{\toks0{#2}\message{(\secn@m.\the\subsecno.\the\subsubsecno. \the\toks0)}}%
\global\subsubsubsecno=0%
\ifnum\lastpenalty>9000\else\bigbreak\fi
\noindent{\it\hyperdef\hypernoname{subsubsection}{\secn@m.\the\subsecno\the\subsubsecno}%
{\secn@m.\the\subsecno.\the\subsubsecno.} #2}
%define section label
\DefWarn#1%
\xdef#1{\noexpand\hyperref{}{subsubsection}{\secn@m.\the\subsecno.\the\subsubsecno}%
{\secn@m.\the\subsecno.\the\subsubsecno}}\writedef{#1\leftbracket#1}\wrlabeL{#1=#1}%
\par\nobreak\medskip\nobreak\noindent\ignorespaces}

%%%%%%%%% subsubsubsection with label %%%%%%%%%%%%%%%%%%%%%%%%%%%%%%%%
\def\newsubsubsubsec#1#2\par{\global\advance\subsubsubsecno by1%
{\toks0{#2}\message{(\secn@m.\the\subsecno.\the\subsubsecno.\the\subsubsubsecno \the\toks0)}}%
\ifnum\lastpenalty>9000\else\bigbreak\fi
\noindent{\it\hyperdef\hypernoname{subsubsection}{\secn@m.\the\subsecno\the\subsubsecno\the\subsubsubsecno}%
{\secn@m.\the\subsecno.\the\subsubsecno.\the\subsubsubsecno.} #2}
%define section label
\DefWarn#1%
\xdef#1{\noexpand\hyperref{}{subsubsubsection}{\secn@m.\the\subsecno.\the\subsubsecno.\the\subsubsubsecno}%
{\secn@m.\the\subsecno.\the\subsubsecno.\the\subsubsubsecno}}\writedef{#1\leftbracket#1}\wrlabeL{#1=#1}%
\par\nobreak\medskip\nobreak\noindent\ignorespaces}

%%%%%%% Appendix %%%%%%%%%%%%%%
\def\appendix#1#2{\global\meqno=1\global\subsecno=0\global\subsubsecno=0\xdef\secsym{\hbox{#1.}}%
\bigbreak\bigskip\noindent{\bf Appendix \hyperdef\hypernoname{appendix}{#1}%
{#1.} #2}{\toks0{(#1. #2)}\message{\the\toks0}}%
\xdef\s@csym{#1.}\xdef\secn@m{#1}%
\writetoca{{\string\hyperref{}{appendix}{#1}{\bf {#1}\quad}} {\bf #2}}%
\par\nobreak\medskip\nobreak}

% \eqn\label{a+b=c}   gives displayed equation, numbered consecutively within sections.
% \eqnn, \eqna        define labels in advance, use \eqna\label before an eqalign and
%                     later \label a, \label b etc inside eqalign to get (2.3a), (2.3b) etc
%
\def\checkm@de#1#2{\ifmmode{\def\f@rst##1{##1}\hyperdef\hypernoname{equation}%
{#1}{#2}}\else\hyperref{}{equation}{#1}{#2}\fi}
\def\eqnn#1{\DefWarn#1\xdef #1{(\noexpand\relax\noexpand\checkm@de%
{\s@csym\the\meqno}{\secsym\the\meqno})}%
\wrlabeL#1\writedef{#1\leftbracket#1}\global\advance\meqno by1}
\def\f@rst#1{\c@t#1a\em@ark}\def\c@t#1#2\em@ark{#1}
\def\eqna#1{\DefWarn#1\wrlabeL{#1$\{\}$}%
\xdef #1##1{(\noexpand\relax\noexpand\checkm@de%
{\s@csym\the\meqno\noexpand\f@rst{##1}1}{\hbox{$\secsym\the\meqno##1$}})}
\writedef{#1\numbersign1\leftbracket#1{\numbersign1}}\global\advance\meqno by1}
\def\eqn#1#2{\DefWarn#1%
\xdef #1{(\noexpand\hyperref{}{equation}{\s@csym\the\meqno}%
{\secsym\the\meqno})}$$#2\eqno(\hyperdef\hypernoname{equation}%
{\s@csym\the\meqno}{\secsym\the\meqno})\eqlabeL#1$$%
\writedef{#1\leftbracket#1}\global\advance\meqno by1}
\def\xeqn{\expandafter\xe@n}\def\xe@n(#1){#1}
\def\xeqna#1{\expandafter\xe@n#1}
\def\eqns#1{(\e@ns #1{\hbox{}})}
\def\e@ns#1{\ifx\UNd@FiNeD#1\message{eqnlabel \string#1 is undefined.}%
\xdef#1{(?.?)}\fi{\let\hyperref=\relax\xdef\next{#1}}%
\ifx\next\em@rk\def\next{}\else%
\ifx\next#1\xeqn#1\else\def\n@xt{#1}\ifx\n@xt\next#1\else\xeqna#1\fi
\fi\let\next=\e@ns\fi\next}
\def\DefWarn#1{}
%
% footnotes
\newskip\footskip\footskip14pt plus 1pt minus 1pt %sets footnote baselineskip
\def\footnotefont{\ninepoint}\def\f@t#1{\footnotefont #1\@foot}
\def\f@@t{\baselineskip\footskip\bgroup\footnotefont\aftergroup\@foot\let\next}
\setbox\strutbox=\hbox{\vrule height9.5pt depth4.5pt width0pt}
\global\newcount\ftno \global\ftno=0
\def\foot{\global\advance\ftno by1\def\foot@rg{\hyperref{}{footnote}%
{\the\ftno}{\the\ftno}\xdef\foot@rg{\noexpand\hyperdef\noexpand\hypernoname%
{footnote}{\the\ftno}{\the\ftno}}}\footnote{$^{\foot@rg}$}}
%
%
%     \ref\label{text}
% generates a number, assigns it to \label, generates an entry.
% To list the refs on a separate page,  \listrefs
%
\global\newcount\refno \global\refno=1
\newwrite\rfile
\def\ref{[\hyperref{}{reference}{\the\refno}{\the\refno}]\nref}
\def\nref#1{\DefWarn#1%
\xdef#1{[\noexpand\hyperref{}{reference}{\the\refno}{\the\refno}]}%
\writedef{#1\leftbracket#1}%
\ifnum\refno=1\immediate\openout\rfile=\jobname.refs\fi
\chardef\wfile=\rfile\immediate\write\rfile{\noexpand\item{[\noexpand\hyperdef%
\noexpand\hypernoname{reference}{\the\refno}{\the\refno}]\ }%
\reflabeL{#1\hskip.31in}\pctsign}\global\advance\refno by1\findarg}
%	horrible hack to sidestep tex \write limitation
\def\findarg#1#{\begingroup\obeylines\newlinechar=`\^^M\pass@rg}
{\obeylines\gdef\pass@rg#1{\writ@line\relax #1^^M\hbox{}^^M}%
\gdef\writ@line#1^^M{\expandafter\toks0\expandafter{\striprel@x #1}%
\edef\next{\the\toks0}\ifx\next\em@rk\let\next=\endgroup\else\ifx\next\empty%
\else\immediate\write\wfile{\the\toks0}\fi\let\next=\writ@line\fi\next\relax}}
\def\striprel@x#1{} \def\em@rk{\hbox{}}
\def\lref{\begingroup\obeylines\lr@f}
\def\lr@f#1#2{\DefWarn#1\gdef#1{\let#1=\UNd@FiNeD\ref#1{#2}}\endgroup\unskip}
\def\semi{;\hfil\break}
\def\addref#1{\immediate\write\rfile{\noexpand\item{}#1}} %now unnecessary
\def\listrefs{\vfill\supereject\immediate\closeout\rfile\writestoppt
\baselineskip=\footskip\centerline{{\bf References}}\bigskip{\parindent=20pt%
\frenchspacing\escapechar=` \input \jobname.refs\vfill\eject}\nonfrenchspacing}
\def\startrefs#1{\immediate\openout\rfile=\jobname.refs\refno=#1}
\def\xref{\expandafter\xr@f}\def\xr@f[#1]{#1}
\def\refs#1{\count255=1[\r@fs #1{\hbox{}}]}
\def\r@fs#1{\ifx\UNd@FiNeD#1\message{reflabel \string#1 is undefined.}%
\nref#1{need to supply reference \string#1.}\fi%
\vphantom{\hphantom{#1}}{\let\hyperref=\relax\xdef\next{#1}}%
\ifx\next\em@rk\def\next{}%
\else\ifx\next#1\ifodd\count255\relax\xref#1\count255=0\fi%
\else#1\count255=1\fi\let\next=\r@fs\fi\next}
%

%
% this is ugly, but moore insists
\newwrite\ffile\global\newcount\figno \global\figno=1
\def\fig{fig.~\hyperref{}{figure}{\the\figno}{\the\figno}\nfig}
\def\nfig#1{\DefWarn#1%
\xdef#1{fig.~\noexpand\hyperref{}{figure}{\the\figno}{\the\figno}}%
\writedef{#1\leftbracket fig.\noexpand~\xfig#1}%
\ifnum\figno=1\immediate\openout\ffile=\jobname.figs\fi\chardef\wfile=\ffile%
{\let\hyperref=\relax
\immediate\write\ffile{\noexpand\medskip\noexpand\item{Fig.\ %
\noexpand\hyperdef\noexpand\hypernoname{figure}{\the\figno}{\the\figno}. }
\reflabeL{#1\hskip.55in}\pctsign}}\global\advance\figno by1\findarg}
\def\xfig{\expandafter\xf@g}\def\xf@g fig.\penalty\@M\ {}
\def\figs#1{figs.~\f@gs #1{\hbox{}}}
\def\f@gs#1{{\let\hyperref=\relax\xdef\next{#1}}\ifx\next\em@rk\def\next{}\else
\ifx\next#1\xfig #1\else#1\fi\let\next=\f@gs\fi\next}
%
%% because TeXlive 2011 is buggy wrt to tikz pictures with plain TeX..
\def\figin{\epsfcheck\figin}\def\figins{\epsfcheck\figins}
\def\epsfcheck{\ifx\epsfbox\UnDeFiNeD
\message{(NO epsf.tex, FIGURES WILL BE IGNORED)}
\gdef\figin##1{\vskip2in}\gdef\figins##1{\hskip.5in}% blank space instead
\else\message{(FIGURES WILL BE INCLUDED)}%
\gdef\figin##1{##1}\gdef\figins##1{##1}\fi}
\def\figinsert{\goodbreak\topinsert}
\def\ifig#1#2#3{\DefWarn#1\xdef#1{fig.~\the\figno}
\writedef{#1\leftbracket fig.\noexpand~\the\figno}%
\figinsert\figin{\centerline{#3}}
\smallskip
\leftskip=0pt \rightskip=0pt
\baselineskip12pt\noindent
{{\bf Fig.~\the\figno}\ \ninepoint #2}
\medskip
\global\advance\figno by1\par\endinsert}
%%%%%%%%%%%%%%%%%%%%%%%%%%%%%%%%%%%%%%%%%%%%%%%%%%%%%%%%%
\newwrite\lfile
{\escapechar-1\xdef\pctsign{\string\%}\xdef\leftbracket{\string\{}
\xdef\rightbracket{\string\}}\xdef\numbersign{\string\#}}
\def\writedefs{\immediate\openout\lfile=label.defs \def\writedef##1{%
{\let\hyperref=\relax\let\hyperdef=\relax\let\hypernoname=\relax
 \immediate\write\lfile{\string\checkdef\string##1\rightbracket}}}}%
\def\writestop{\def\writestoppt{\immediate\write\lfile{\string\pageno
 \the\pageno\string\startrefs\leftbracket\the\refno\rightbracket
 \string\def\string\secsym\leftbracket\secsym\rightbracket
 \string\secno\the\secno\string\meqno\the\meqno}\immediate\closeout\lfile}}
\def\writestoppt{}\def\writedef#1{}

% Section, subsection and appendix labels %
% Note that there must be a blanck line after \newsec,\subsec and before \seclab,\subseclab!
\def\seclab#1\par{\DefWarn#1%
\xdef #1{\noexpand\hyperref{}{section}{\the\secno}{\the\secno}}%
\writedef{#1\leftbracket#1}\wrlabeL{#1=#1}\par%
\nobreak\medskip\nobreak\noindent\ignorespaces}
\def\subseclab#1\par{\DefWarn#1%
\xdef #1{\noexpand\hyperref{}{subsection}{\the\secno.\the\subsecno}%
{\the\secno.\the\subsecno}}\writedef{#1\leftbracket#1}\wrlabeL{#1=#1}\par%
\nobreak\medskip\nobreak\noindent\ignorespaces}
\def\subsubseclab#1\par{\DefWarn#1%
\xdef#1{\noexpand\hyperref{}{subsubsection}{\the\secno.\the\subsecno.\the\subsubsecno}%
{\the\secno.\the\subsecno.\the\subsubsecno}}\writedef{#1\leftbracket#1}\wrlabeL{#1=#1}\par%
\nobreak\medskip\nobreak\noindent\ignorespaces}
\def\applab#1\par{\DefWarn#1%
\xdef#1{\noexpand\hyperref{}{appendix}{\secn@m}{\secn@m}}%
\writedef{#1\leftbracket#1}\wrlabeL{#1=#1}%
\par\nobreak\medskip\nobreak\noindent\ignorespaces}
\def\appsublab#1{\DefWarn#1%
\xdef #1{\noexpand\hyperref{}{appendix}{\secn@m.\the\subsecno}{\secn@m.\the\subsecno}}%
\writedef{#1\leftbracket#1}\wrlabeL{#1=#1}}
\newwrite\tfile \def\writetoca#1{}
\def\leaderfill{\leaders\hbox to 1em{\hss.\hss}\hfill}
% use this to write file with table of contents
\def\writetoc{\immediate\openout\tfile=\jobname.toc
   \def\writetoca##1{{\edef\next{\write\tfile{\noindent ##1
   \string\leaderfill{
% comment this line if you don't want hyperlinked page numbers on TOC
   \string\hyperref{}{page}{\noexpand\number\pageno}%
   {\noexpand\number\pageno}} \par}}\next}}
}
% and this lists table of contents on second pass
\newread\ch@ckfile
\def\listtoc{\immediate\closeout\tfile\immediate\openin\ch@ckfile=\jobname.toc
\ifeof\ch@ckfile\message{no file \jobname.toc, no table of contents this pass}%
\else\closein\ch@ckfile\centerline{\bf Contents}\nobreak\medskip%
{\baselineskip=15.5pt\footnotefont\parskip=0pt\catcode`\@=11\input\jobname.toc
\catcode`\@=12\bigbreak\bigskip}\fi}
\catcode`\@=12 % at signs are no longer letters
\def\tenpoint{\def\rm{\fam0\tenrm}% switch back to 10-point type
\textfont0=\tenrm \scriptfont0=\sevenrm \scriptscriptfont0=\fiverm
\textfont1=\teni  \scriptfont1=\seveni  \scriptscriptfont1=\fivei
\textfont2=\tensy \scriptfont2=\sevensy \scriptscriptfont2=\fivesy
\textfont\itfam=\tenit \def\it{\fam\itfam\tenit}\def\footnotefont{\ninepoint}%
\textfont\bffam=\tenbf \def\bf{\fam\bffam\tenbf}\def\sl{\fam\slfam\tensl}\rm}
\font\ninerm=cmr9 \font\sixrm=cmr6 \font\ninei=cmmi9 \font\sixi=cmmi6
\font\ninesy=cmsy9 \font\sixsy=cmsy6 \font\ninebf=cmbx9
\font\nineit=cmti9 \font\ninesl=cmsl9 \skewchar\ninei='177
\skewchar\sixi='177 \skewchar\ninesy='60 \skewchar\sixsy='60
\def\ninepoint{\def\rm{\fam0\ninerm}% switch to footnote font
\textfont0=\ninerm \scriptfont0=\sixrm \scriptscriptfont0=\fiverm
\textfont1=\ninei \scriptfont1=\sixi \scriptscriptfont1=\fivei
\textfont2=\ninesy \scriptfont2=\sixsy \scriptscriptfont2=\fivesy
\textfont\itfam=\ninei \def\it{\fam\itfam\nineit}\def\sl{\fam\slfam\ninesl}%
\textfont\bffam=\ninebf \def\bf{\fam\bffam\ninebf}\rm}
%
%---------------------------------------------------------------------
\hyphenation{anom-aly anom-alies coun-ter-term coun-ter-terms}

% Caption for inline tikzpictures
%\def\DefWarn#1{}
\def\tikzcaption#1#2{\DefWarn#1\xdef#1{Fig.~\the\figno}
\writedef{#1\leftbracket Fig.\noexpand~\the\figno}%
{
\smallskip
\leftskip=20pt \rightskip=20pt \baselineskip12pt\noindent
{{\bf Fig.~\the\figno}\ \ninepoint #2}
\bigskip
\global\advance\figno by1 \par}}

% convert numbers [1-9] to upper case letters [A-I]
\def\ntoalpha#1{%
\ifcase#1%
@%
\or A\or B\or C\or D\or E\or F\or G\or H\or I\or J\or K\or L\or M%
\fi
}

% Appendix label
\global\newcount\appno \global\appno=1
\def\applab#1{\xdef #1{\ntoalpha{\appno}}\writedef{#1\leftbracket#1}\wrlabeL{#1=#1}
\global\advance\appno by1}

% Clean up the title page definitions
\def\preprint#1 #2\par{\rightline{\vbox{\baselineskip12pt\hbox{#1}\hbox{#2}}}\vskip2cm}
% title with more than one line (note the blanck line in between)
%\title some line
%
%\tile another line
\def\title#1\par{\centerline{\bf #1}\nopagenumbers\pageno=0}
\def\author#1\par{\bigskip\bigskip\centerline{#1}}

\newcount\addressno

\def\email#1#2{%\unskip$^#1$
\footnote{\null}{\kern-\parindent \llap{$^#1$\hskip1pt}email: #2}}

% centermode for address lines
\def\startcenter{%
  \par
  \begingroup
  \leftskip=0pt plus 1fil
  \rightskip=\leftskip
  \parindent=0pt
  \parfillskip=0pt
}
\def\stopcenter{\endgroup}

\def\address{\bigskip%
  \ifnum\the\addressno=0\else\stopcenter\endgroup\fi
  \advance\addressno by 1%
  \begingroup
  \startcenter
  \it
  \obeylines
  \addressAux
}
\def\addressAux#1{#1}

% need to stop center mode and obeylines from address
\def\abstract{\stopcenter\endgroup\bigskip\bigskip\noindent}

% some sample definitions
\def\Dsl{\,\raise.15ex\hbox{/}\mkern-13.5mu D} %this one can be subscripted
\def\dsl{\raise.15ex\hbox{/}\kern-.57em\partial}
 
\def\boxeqn#1{\vcenter{\vbox{\hrule\hbox{\vrule\kern3pt\vbox{\kern3pt
	\hbox{${\displaystyle #1}$}\kern3pt}\kern3pt\vrule}\hrule}}}

 %pound sterling

\def\ap{{\alpha^{\prime}}}

\def\g{{\gamma}}

\def\l{\lambda}

\def\t{{\theta}}

\def\half{{1\over 2}}

\def\({\left(}
\def\){\right)}

\def\cK{{\cal K}}

\def\cZ{{\cal Z}}

% blackboard bold

% primed summation symbol

% length of words, |P|
\def\len#1{{%
\def\Dlen{\left|\mkern-1mu #1\mkern -0.5mu\right|}%
\def\Sslen{\left|\mkern-1.3mu #1\mkern -1.3mu\right|}%
\def\SSlen{\left|\mkern-2.8mu #1\mkern-1.3mu\right|}%
\mathchoice{\Dlen}{\Dlen}{\Sslen}{\SSlen}}}

\def\perm#1{{\rm perm}#1}

\def\Im{\mathop{{\rm Im}}} %redefine plain TeX \Im..
% small inlined fractions, from the TeXbook
\def\sfrac#1/#2{\kern.1em\raise.5ex\hbox{\the\scriptfont0 #1}%
\kern-.1em/\kern-.15em\lower.25ex\hbox{\the\scriptfont0 #2}}

%shuffle product

%\owedge

% From Knuth's \pfbox macro
\def\qed{\hbox{\hskip 3pt
%\lower2pt
\vbox{\hrule\hbox to 7pt{\vrule height 7pt\hfill\vrule}
\hrule}}\hskip3pt}

% do not display overfull marks
\overfullrule=0pt\relax

\frenchspacing

% DefWarn-like behavior for labels defined in advance
\def\checkdef#1#2{
\ifx\UndeFined#1%
	\def#1{#2}
%\immediate\write16{*** define label \string#1 by #2 ***}
\else
	\immediate\write16{*** BUG ***: the label \string#1 is already defined ***}
\fi
}
% define labels in advance
\newread\instream

\openin\instream= label.defs
\ifeof\instream\message{No labels in advance yet. Wait till next pass.}
\else\closein\instream \input label.defs
\fi
\writedefs

%%% References with hyperlinks to arxiv.org; both styles accepted
% Change arXiv to \arXiv ie
% [arXiv:hep-th/1234567].     --> [\arXiv:hep-th/1234567].
% [arXiv:1234.5678 [hep-th]]. --> [\arXiv:1234.5678 [hep-th]].
% Need to strip trailing [hep-th] (if present) to define valid URL
\def\arXiv:#1].{\hepthStrip#1 \nil}
\def\hepthStrip#1 #2\nil{\href{http://arxiv.org/abs/#1}{arXiv:#1 #2\unskip}].}

%%% Fraktur fonts for Berends-Giele components

\input epsf.tex

\def\frac#1#2{{#1\over #2}}

\def\cN{{\cal N}}
\def\sgn{\mathord{\rm sgn}}
\def\dist{\mathord{\rm dist}}
\def\Ord{\mathord{\rm Ord}}
\def\textbf#1{{\bf #1}}

\title Local BCJ numerators for ten-dimensional SYM at one loop

\author
Elliot Bridges\email{\star}{e.n.bridges@soton.ac.uk}$^\star$, and
Carlos R. Mafra\email{\ap}{c.r.mafra@soton.ac.uk}$^\ap$

\address
Mathematical Sciences and STAG Research Centre, University of Southampton,
Highfield, Southampton, SO17 1BJ, United Kingdom

\abstract
We obtain local numerators satisfying the BCJ color-kinematics duality at one
loop for super-Yang--Mills theory in ten dimensions. This is done explicitly
for six points via the field-theory limit of the genus-one open
superstring correlators for different color orderings, in an analogous manner to an earlier
derivation of local BCJ-satisfying numerators at tree level from disk correlators.
These results solve
an outstanding puzzle from a previous analysis where the six-point numerators
did not satisfy the color-kinematics duality.

\lref\bcjreview{
	Z.~Bern, J.~J.~Carrasco, M.~Chiodaroli, H.~Johansson and R.~Roiban,
	``The Duality Between Color and Kinematics and its Applications,''
	[arXiv:1909.01358 [hep-th]].
}
\lref\genJac{
	Z.~Bern, J.~J.~Carrasco, W.~M.~Chen, H.~Johansson and R.~Roiban,
	``Gravity Amplitudes as Generalized Double Copies of Gauge-Theory Amplitudes,''
	Phys. Rev. Lett. \textbf{118}, no.18, 181602 (2017)
	%doi:10.1103/PhysRevLett.118.181602
	[arXiv:1701.02519 [hep-th]].
}
\lref\fivebcj{
	Z.~Bern, J.~J.~M.~Carrasco, W.~M.~Chen, H.~Johansson, R.~Roiban and M.~Zeng,
	``Five-loop four-point integrand of $N=8$ supergravity as a generalized double copy,''
	Phys. Rev. D \textbf{96}, no.12, 126012 (2017)
	%doi:10.1103/PhysRevD.96.126012
	[arXiv:1708.06807 [hep-th]].
}

\lref\monodVanhove{
	N.E.J.~Bjerrum-Bohr, P.H.~Damgaard, P.~Vanhove,
  	``Minimal Basis for Gauge Theory Amplitudes,''
	Phys.\ Rev.\ Lett.\  {\bf 103}, 161602 (2009).
	[arXiv:0907.1425 [hep-th]].
	%%CITATION = arXiv:0907.1425%%
}
\lref\monodStie{
	S.~Stieberger,
	``Open \& Closed vs. Pure Open String Disk Amplitudes,''
	[arXiv:0907.2211 [hep-th]].
	%%CITATION = arXiv:0907.2211%%
}

\lref\BCJ{
	Z.~Bern, J.J.M.~Carrasco and H.~Johansson,
	``New Relations for Gauge-Theory Amplitudes,''
	Phys.\ Rev.\ D {\bf 78}, 085011 (2008).
	[arXiv:0805.3993 [hep-ph]].
	%%CITATION = arXiv:0805.3993%%
}
\lref\psf{
 	N.~Berkovits,
	``Super-Poincare covariant quantization of the superstring,''
	JHEP {\bf 0004}, 018 (2000)
	[arXiv:hep-th/0001035].
	%%CITATION = JHEPA,0004,018;%%
	\semi
	N.~Berkovits,
  	``ICTP lectures on covariant quantization of the superstring,''
	ICTP Lect.\ Notes Ser.\  {\bf 13}, 57 (2003).
	[hep-th/0209059].
	%%CITATION = hep-th/0209059%%
}

\lref\FORM{
        J.A.M.~Vermaseren,
        ``New features of FORM,''
        arXiv:math-ph/0010025.
        %%CITATION = MATH-PH/0010025;%%
\semi
        M.~Tentyukov and J.A.M.~Vermaseren,
        ``The multithreaded version of FORM,''
        arXiv:hep-ph/0702279.
        %%CITATION = HEP-PH/0702279;%%
}

\lref\towardsOne{
	C.R.~Mafra and O.~Schlotterer,
  	``Towards one-loop SYM amplitudes from the pure spinor BRST cohomology,''
	Fortsch.\ Phys.\  {\bf 63}, no. 2, 105 (2015).
	[arXiv:1410.0668 [hep-th]].
	%%CITATION = AEI-2014-053%%
}

\lref\worldline{
	Z.~Bern and D.~A.~Kosower,
  	``Efficient calculation of one loop QCD amplitudes,''
	Phys.\ Rev.\ Lett.\  {\bf 66}, 1669 (1991)..
	%%CITATION = FERMILAB-PUB-90-225-T%%
\semi
  	Z.~Bern and D.~A.~Kosower,
  	``The Computation of loop amplitudes in gauge theories,''
	Nucl.\ Phys.\ B {\bf 379}, 451 (1992)..
	%%CITATION = FERMILAB-PUB-91-111-T%%
\semi
	M.~J.~Strassler,
 	``Field theory without Feynman diagrams: One loop effective actions,''
	Nucl.\ Phys.\ B {\bf 385}, 145 (1992).
	[hep-ph/9205205].
	%%CITATION = hep-ph/9205205%%
\semi
 	Z.~Bern, D.~C.~Dunbar and T.~Shimada,
  	``String based methods in perturbative gravity,''
	Phys.\ Lett.\ B {\bf 312}, 277 (1993).
	[hep-th/9307001].
	%%CITATION = hep-th/9307001%%
\semi
	D.~C.~Dunbar and P.~S.~Norridge,
  	``Calculation of graviton scattering amplitudes using string based methods,''
	Nucl.\ Phys.\ B {\bf 433}, 181 (1995).
	[hep-th/9408014].
	%%CITATION = hep-th/9408014%%
\semi
	C.~Schubert,
  	``Perturbative quantum field theory in the string inspired formalism,''
	Phys.\ Rept.\  {\bf 355}, 73 (2001).
	[hep-th/0101036].
	%%CITATION = hep-th/0101036%%
\semi
 	N.~E.~J.~Bjerrum-Bohr and P.~Vanhove,
  	``Absence of Triangles in Maximal Supergravity Amplitudes,''
	JHEP {\bf 0810}, 006 (2008).
	[arXiv:0805.3682 [hep-th]].
	%%CITATION = arXiv:0805.3682%%
}
\lref\oneloopI{
	C.R.~Mafra and O.~Schlotterer,
  	``Towards the n-point one-loop superstring amplitude. Part I. Pure spinors and superfield kinematics,''
	JHEP {\bf 0819}, 090 (2019).
	[arXiv:1812.10969 [hep-th]].
	%%CITATION = arXiv:1812.10969%%
	}
\lref\oneloopII{
	C.R.~Mafra and O.~Schlotterer,
  	``Towards the n-point one-loop superstring amplitude. Part II. Worldsheet functions and their duality to kinematics,''
	JHEP {\bf 0819}, 091 (2019).
	[arXiv:1812.10970 [hep-th]].
	%%CITATION = arXiv:1812.10970%%
	}
\lref\oneloopIII{
	C.R.~Mafra and O.~Schlotterer,
  	``Towards the n-point one-loop superstring amplitude. Part III. One-loop correlators and their double-copy structure,''
	JHEP {\bf 0819}, 092 (2019).
	[arXiv:1812.10971 [hep-th]].
	%%CITATION = arXiv:1812.10971%%
	}
\lref\BGDoubleCurrent{
	C.R.~Mafra,
  	``Berends-Giele recursion for double-color-ordered amplitudes,''
	JHEP {\bf 0716}, 080 (2016).
	[arXiv:1603.09731 [hep-th]].
	%%CITATION = arXiv:1603.09731%%
	}
\lref\OneLoopDiffEq{
	C.R.~Mafra and O.~Schlotterer,
  	``One-loop open-string integrals from differential equations: all-order $\alpha^\prime$-expansions at $n$ points,''
	JHEP {\bf 0320}, 007 (2020).
	[arXiv:1908.10830 [hep-th]].
	%%CITATION = arXiv:1908.10830%%
}
\lref\EOMbbs{
	C.R.~Mafra and O.~Schlotterer,
  	``Multiparticle SYM equations of motion and pure spinor BRST blocks,''
	JHEP {\bf 1407}, 153 (2014).
	[arXiv:1404.4986 [hep-th]].
	%%CITATION = AEI-2014-011%%
}
\lref\MPS{
	N.~Berkovits,
	``Multiloop amplitudes and vanishing theorems using the pure spinor formalism for the superstring,''
	JHEP {\bf 0409}, 047 (2004).
	[hep-th/0406055].
	%%CITATION = hep-th/0406055%%
}
\lref\anomalypaper{
	N.~Berkovits and C.R.~Mafra,
	``Some Superstring Amplitude Computations with the Non-Minimal Pure Spinor Formalism,''
	JHEP {\bf 0611}, 079 (2006).
	[hep-th/0607187].
	%%CITATION = hep-th/0607187%%
}
\lref\anomalysix{
	C.R.~Mafra and O.~Schlotterer,
	``One-loop superstring six-point amplitudes and anomalies in pure spinor superspace,''
	JHEP {\bf 1604}, 148 (2016).
	[arXiv:1603.04790 [hep-th]].
	%%CITATION = arXiv:1603.04790%%
}
\lref\BCJtree{
	C.R.~Mafra, O.~Schlotterer and S.~Stieberger,
  	``Explicit BCJ Numerators from Pure Spinors,''
	JHEP {\bf 1107}, 092 (2011).
	[arXiv:1104.5224 [hep-th]].
	%%CITATION = arXiv:1104.5224%%
}
\lref\yutin{
	M.~Bianchi, A.~L.~Guerrieri, Y.~t.~Huang, C.~J.~Lee and C.~Wen,
	``Exploring soft constraints on effective actions,''
	JHEP \textbf{10}, 036 (2016)
	%doi:10.1007/JHEP10(2016)036
	[arXiv:1605.08697 [hep-th]].
}
\lref\mizera{
	E.~Casali, S.~Mizera and P.~Tourkine,
	``Loop amplitudes monodromy relations and color-kinematics duality,''
	[arXiv:2005.05329 [hep-th]].
}
\lref\nptMethod{
	C.R.~Mafra, O.~Schlotterer, S.~Stieberger and D.~Tsimpis,
	``A recursive method for SYM n-point tree amplitudes,''
	Phys.\ Rev.\ D {\bf 83}, 126012 (2011).
	[arXiv:1012.3981 [hep-th]].
	%%CITATION = arXiv:1012.3981%%
}
\lref\nptTree{
	C.R.~Mafra, O.~Schlotterer and S.~Stieberger,
	``Complete N-Point Superstring Disk Amplitude I. Pure Spinor Computation,''
	Nucl.\ Phys.\ B {\bf 873}, 419 (2013).
	[arXiv:1106.2645 [hep-th]].
	%%CITATION = arXiv:1106.2645%%
\semi
  	C.R.~Mafra, O.~Schlotterer and S.~Stieberger,
	``Complete N-Point Superstring Disk Amplitude II. Amplitude and Hypergeometric Function Structure,''
	Nucl.\ Phys.\ B {\bf 873}, 461 (2013).
	[arXiv:1106.2646 [hep-th]].
	%%CITATION = arXiv:1106.2646%%
}

\lref\partI{
	C.R.~Mafra and O.~Schlotterer,
  	``Cohomology foundations of one-loop amplitudes in pure spinor superspace,''
	[arXiv:1408.3605 [hep-th]].
	%%CITATION = arXiv:1408.3605%%
}
\lref\genredef{
	E.~Bridges and C.R.~Mafra,
	``Algorithmic construction of SYM multiparticle superfields in the BCJ gauge,''
	JHEP \textbf{10}, 022 (2019)
	%doi:10.1007/JHEP10(2019)022
	[arXiv:1906.12252 [hep-th]].
}
\lref\FTlimit{
	C.R.~Mafra,
  	``Berends-Giele recursion for double-color-ordered amplitudes,''
	JHEP {\bf 1607}, 080 (2016).
	[arXiv:1603.09731 [hep-th]].
	%%CITATION = arXiv:1603.09731%%
}
\lref\towards{
	C.R.~Mafra,
	``Towards Field Theory Amplitudes From the Cohomology of Pure Spinor Superspace,''
	JHEP {\bf 1011}, 096 (2010).
	[arXiv:1007.3639 [hep-th]].
	%%CITATION = arXiv:1007.3639%%
}
\lref\Polylogs{
	J.~Broedel, O.~Schlotterer and S.~Stieberger,
	``Polylogarithms, Multiple Zeta Values and Superstring Amplitudes,''
	Fortsch.\ Phys.\  {\bf 61}, 812 (2013).
	[arXiv:1304.7267 [hep-th]].
	%%CITATION = DAMTP-2013-22%%
}
\lref\DPellis{
	F.~Cachazo, S.~He and E.Y.~Yuan,
	``Scattering of Massless Particles: Scalars, Gluons and Gravitons,''
	JHEP {\bf 1407}, 033 (2014).
	[arXiv:1309.0885 [hep-th]].
	%%CITATION = arXiv:1309.0885%%
}
\lref\PScomb{
	C.R.~Mafra,
	``Planar binary trees in scattering amplitudes.''
	Algebraic Combinatorics, Resurgence, Moulds and Applications (CARMA) (2020): 349-365.
%	doi:10.4171/205-1/6
	[arXiv:2011.14413 [math.CO]].
}
\lref\Reutenauer{
	C.~Reutenauer,
	``Free Lie Algebras,''
	London Mathematical Society Monographs, 1993
}
\lref\BCJloop{
	Z.~Bern, J.J.M.~Carrasco and H.~Johansson,
	``Perturbative Quantum Gravity as a Double Copy of Gauge Theory,''
	Phys.\ Rev.\ Lett.\  {\bf 105}, 061602 (2010).
	[arXiv:1004.0476 [hep-th]].
	%%CITATION = arXiv:1004.0476%%
}

\lref\BernUF{
	Z.~Bern, J.~J.~M.~Carrasco, L.~J.~Dixon, H.~Johansson and R.~Roiban,
  	``Simplifying Multiloop Integrands and Ultraviolet Divergences of Gauge Theory and Gravity Amplitudes,''
	Phys.\ Rev.\ D {\bf 85}, 105014 (2012).
	[arXiv:1201.5366 [hep-th]].
	%%CITATION = arXiv:1201.5366%%
\semi
	Z.~Bern, S.~Davies, T.~Dennen and Y.~t.~Huang,
  	``Absence of Three-Loop Four-Point Divergences in N=4 Supergravity,''
	Phys.\ Rev.\ Lett.\  {\bf 108}, 201301 (2012).
	[arXiv:1202.3423 [hep-th]].
	%%CITATION = UCLA-12-TEP-101%%
\semi
	Z.~Bern, S.~Davies, T.~Dennen and Y.~t.~Huang,
  	``Ultraviolet Cancellations in Half-Maximal Supergravity as a Consequence of the Double-Copy Structure,''
	Phys.\ Rev.\ D {\bf 86}, 105014 (2012).
	[arXiv:1209.2472 [hep-th]].
	%%CITATION = UCLA-12-TEP-106%%
\semi
 	Z.~Bern, S.~Davies, T.~Dennen, A.~V.~Smirnov and V.~A.~Smirnov,
  	``Ultraviolet Properties of N=4 Supergravity at Four Loops,''
	Phys.\ Rev.\ Lett.\  {\bf 111}, no. 23, 231302 (2013).
	[arXiv:1309.2498 [hep-th]].
	%%CITATION = arXiv:1309.2498%%
\semi
	Z.~Bern, S.~Davies and T.~Dennen,
  	``Enhanced Ultraviolet Cancellations in N = 5 Supergravity at Four Loop,''
	[arXiv:1409.3089 [hep-th]].
	%%CITATION = UCLA-14-TEP-106%%
}
\lref\Gauge{
	S.~Lee, C.R.~Mafra and O.~Schlotterer,
  	``Non-linear gauge transformations in $D=10$ SYM theory and the BCJ duality,''
	JHEP {\bf 1603}, 090 (2016).
	[arXiv:1510.08843 [hep-th]].
	%%CITATION = DAMTP-2015-68%%
}
\lref\CJfive{
	J.J.~Carrasco and H.~Johansson,
  `	`Five-Point Amplitudes in N=4 Super-Yang-Mills Theory and N=8 Supergravity,''
	Phys.\ Rev.\ D {\bf 85}, 025006 (2012).
	[arXiv:1106.4711 [hep-th]].
	%%CITATION = arXiv:1106.4711%%
}

\lref\yuansix{
	E.~Y.~Yuan,
	``Virtual Color-Kinematics Duality: 6-pt 1-Loop MHV Amplitudes,''
	JHEP \textbf{05}, 070 (2013)
%	doi:10.1007/JHEP05(2013)070
	[arXiv:1210.1816 [hep-th]].
}

\lref\olipartial{
	S.~He and O.~Schlotterer,
	``New Relations for Gauge-Theory and Gravity Amplitudes at Loop Level,''
	Phys. Rev. Lett. \textbf{118}, no.16, 161601 (2017)
	%doi:10.1103/PhysRevLett.118.161601
	[arXiv:1612.00417 [hep-th]].
}
\lref\eulermaster{
	K.~Dilcher ``Sums of products of Bernoulli numbers''. Journal of Number Theory. 1996 Sep 1;60(1):23-41.
}
\lref\GSWII{
	M.B.~Green, J.H.~Schwarz and E.~Witten,
  	``Superstring Theory. Vol. 2: Loop Amplitudes, Anomalies And Phenomenology,''
	Cambridge, UK: Univ.~Pr.~(1987) 596 P. (Cambridge Monographs On Mathematical Physics).
}
\lref\GreenFT{
	M.B.~Green, J.H.~Schwarz and L.~Brink,
  	``N=4 Yang-Mills and N=8 Supergravity as Limits of String Theories,''
	Nucl.\ Phys.\ B {\bf 198}, 474 (1982).
	%%CITATION = CALT-68-880%%
}
\lref\piotr{
	P.~Tourkine,
	``Tropical Amplitudes,''
	Annales Henri Poincare \textbf{18}, no.6, 2199-2249 (2017)
	%doi:10.1007/s00023-017-0560-7
	[arXiv:1309.3551 [hep-th]].
}
\lref\FramptonAnomaly{
	P.~H.~Frampton and T.~W.~Kephart,
  	``Explicit Evaluation of Anomalies in Higher Dimensions,''
	Phys.\ Rev.\ Lett.\  {\bf 50}, 1343 (1983), [Erratum-ibid.\  {\bf 51}, 232 (1983)].
	%%CITATION = IFP-192-UNC%%
\semi
	P.H.~Frampton and T.~W.~Kephart,
  	``The Analysis of Anomalies in Higher Space-time Dimensions,''
	Phys.\ Rev.\ D {\bf 28}, 1010 (1983).
	%%CITATION = IFP-193-UNC%%
}
\lref\GSanomaly{
	M.B.~Green and J.H.~Schwarz,
	``Anomaly Cancellation in Supersymmetric D=10 Gauge Theory and Superstring Theory,''
	Phys.\ Lett.\ B {\bf 149}, 117 (1984).
	%%CITATION = CALT-68-1182%%
\semi
	M.B.~Green and J.H.~Schwarz,
	``The Hexagon Gauge Anomaly in Type I Superstring Theory,''
	Nucl.\ Phys.\ B {\bf 255}, 93 (1985).
	%%CITATION = CALT-68-1224%%
}
\lref\ChenEVA{
	W.M.~Chen, Y.~t.~Huang and D.~A.~McGady,
  	``Anomalies without an action,''
	[arXiv:1402.7062 [hep-th]].
	%%CITATION = PUPT-2459%%
}
\lref\veronneu{
	Z.~Bern, C.~Boucher-Veronneau and H.~Johansson,
	``N >= 4 Supergravity Amplitudes from Gauge Theory at One Loop,''
	Phys. Rev. D \textbf{84}, 105035 (2011)
	%doi:10.1103/PhysRevD.84.105035
	[arXiv:1107.1935 [hep-th]].
}
\lref\KKLance{
	V.~Del Duca, L.J.~Dixon and F.~Maltoni,
	``New color decompositions for gauge amplitudes at tree and loop level,''
	Nucl.\ Phys.\ B {\bf 571}, 51 (2000).
	[hep-ph/9910563].
	%%CITATION = hep-ph/9910563%%
}
\lref\website{
	http://www.southampton.ac.uk/\~{}crm1n16/1loopIntegrands.html
}
\lref\stringinspired{
	S.~He, R.~Monteiro and O.~Schlotterer,
	``String-inspired BCJ numerators for one-loop MHV amplitudes,''
	JHEP \textbf{01}, 171 (2016)
%	doi:10.1007/JHEP01(2016)171
	[arXiv:1507.06288 [hep-th]].
}
\lref\MonteiroOx{
	N.E.J.~Bjerrum-Bohr, T.~Dennen, R.~Monteiro and D.~O'Connell,
  	``Integrand Oxidation and One-Loop Colour-Dual Numerators in N=4 Gauge Theory,''
  	JHEP {\bf 1307}, 092 (2013).
	[arXiv:1303.2913 [hep-th]].
	%%CITATION = arXiv:1303.2913%%
}
\lref\FunctionsBook{
	I. S. Gradshteyn, I. M. Ryzhik,
	``Table of Integrals, Series, and Product,''
	Academic press, 2014.
}
\lref\genvanhove{
	N.~E.~J.~Bjerrum-Bohr, P.~H.~Damgaard, T.~Sondergaard and P.~Vanhove,
	``Monodromy and Jacobi-like Relations for Color-Ordered Amplitudes,''
	JHEP \textbf{06}, 003 (2010)
%doi:10.1007/JHEP06(2010)003
	[arXiv:1003.2403 [hep-th]].
}

\Date{February 2021}

\listtoc
\writetoc
\filbreak

%*******************
\newsec Description of the problem and its solution

This paper aims to answer a question left over from the pure spinor construction of one-loop integrands of super-Yang--Mills
(SYM) using locality and BRST invariance \towardsOne. Can one find a set of local and supersymmetric numerators for
ten-dimensional SYM one-loop integrands at six points satisfying the Bern-Carrasco-Johansson (BCJ)\foot{A brief review of the
BCJ color-kinematics duality sufficient for our purposes will be given below in section~\BCJintro\ but a much more in-depth
review is contained in \bcjreview.} color-kinematics duality? We will see below that the answer is {\it yes}, and we will also
outline the solution for seven-point integrands.

The one-loop integrands of SYM in ten dimensions for five and six points
were constructed in \towardsOne, where it was shown that the numerators for the five-point amplitude
satisfied the color-kinematics duality while those at six points did not. The proposal of \towardsOne\
was based on two main ingredients: locality and BRST invariance. Using the
multiparticle superfields in pure spinor superspace developed in \EOMbbs, these requirements together
with a basic understanding of the zero-mode saturation rules of the pure spinor formalism \refs{\MPS,\psf} led
to intuitive rules mapping one-loop cubic graphs to superspace numerators, see \figsamp. By assembling the numerators
of the cubic graphs for all $p$-gons of a $n$-point amplitude such that their sum is in the pure spinor BRST cohomology (up to
anomalous terms of the form discussed in \refs{\anomalypaper,\anomalysix}),
the amplitudes of the color-ordered five and six-point amplitudes for the canonical color ordering were constructed. The six-point
integrand was later successfully used in \yutin, passing some consistency checks.

\ifig\figsamp{The multiparticle superfields and pure spinor one-loop building blocks lead to intuitive
mappings between one-loop cubic graphs and pure spinor superspace expressions encoding the polarization dependence
of ten-dimensional supersymmetric Yang--Mills states \towardsOne.}
{\epsfxsize=0.2\hsize\epsfbox{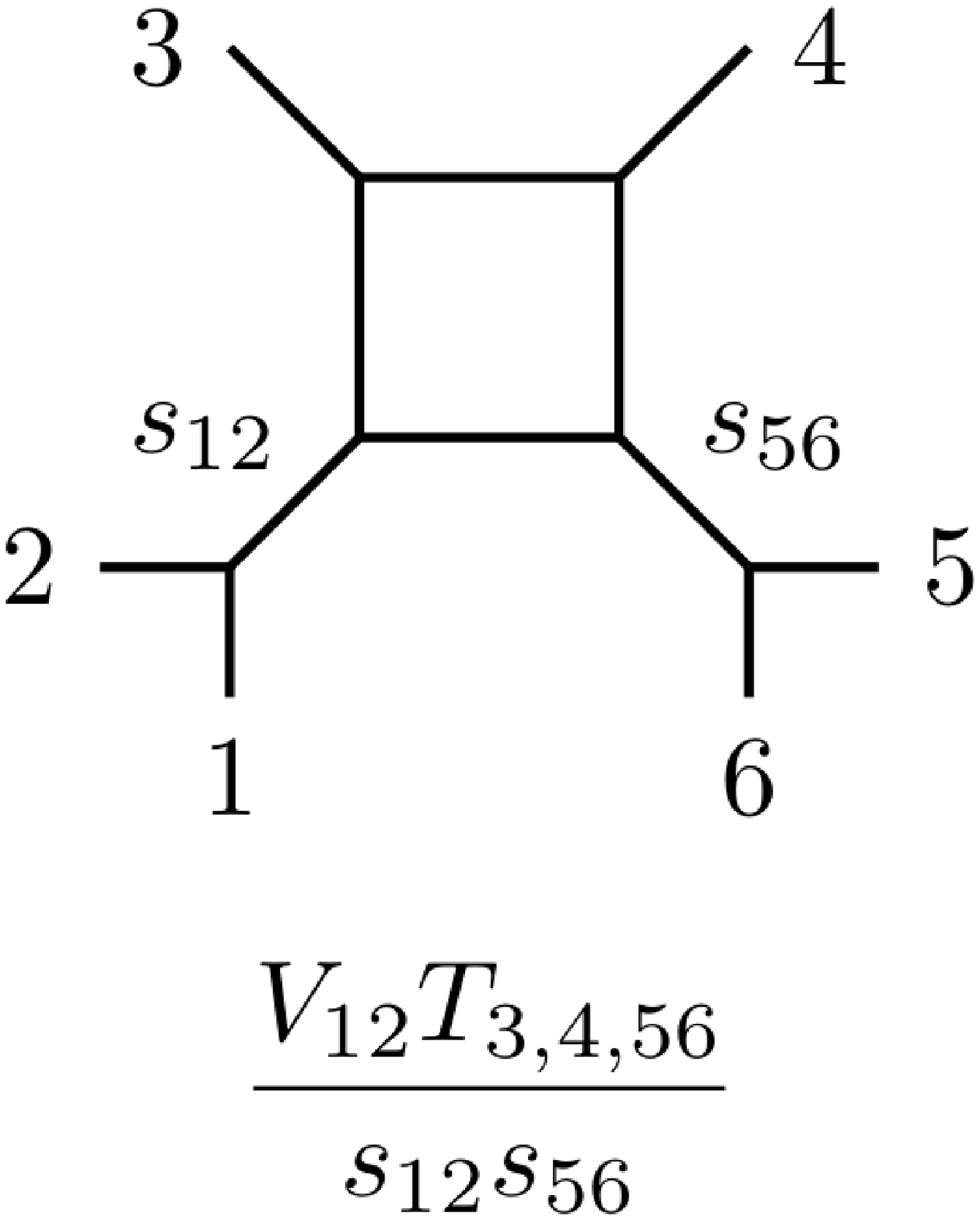}}

%*************************************************************************************
\newsubsec\introcorr Genus-one open superstring correlators in  pure spinor superspace

In this paper we will also use the same formalism of multiparticle superfields in pure spinor superspace to
present local representations of the five-, six- and seven-point amplitudes that do obey the color-kinematics
duality. Since we are using the same superfield language,
it is therefore important to highlight the differences with respect to the previous analysis of \towardsOne.
The difference stems from
the knowledge of the open-string one-loop {\it correlators} recently obtained in \refs{\oneloopI\oneloopII{--}\oneloopIII} up to seven
points. They are given by
\eqnn\FourPointCorr
\eqnn\FivePointCorr
\eqnn\SixPointCorr
\eqnn\SevenPointCorr
$$\eqalignno{
{\cal K}_4(\ell)&=V_1T_{2,3,4}{\cal Z}_{1,2,3,4}\,,&\FourPointCorr\cr
{\cal K}_5(\ell)&=V_1T_{2,3,4,5}^m{\cal Z}_{1,2,3,4,5}^m&\FivePointCorr\cr
&+\big(V_AT_{B,C,D}{\cal Z}_{A,B,C,D}+[A,B,C,D|12345]\big)\,,\cr
{\cal K}_6(\ell)&=\frac{1}{2}V_1T_{2,3,4,5,6}^{mn}{\cal Z}_{1,2,3,4,5,6}^{mn}&\SixPointCorr\cr
&+\big(V_AT_{B,C,D,E}^m{\cal Z}^m_{A,B,C,D,E}+[A,B,C,D,E|123456]\big)\cr
&+\big(V_AT_{B,C,D}{\cal Z}_{A,B,C,D}+[A,B,C,D|123456]\big)\,,\cr
{\cal K}_7(\ell)&=\frac{1}{6}V_1T_{2,3,4,5,6,7}^{mnp}{\cal Z}_{1,2,3,4,5,6,7}^{mnp}&\SevenPointCorr\cr
&+\frac{1}{2}\big(V_AT_{B,C,D,E,F}^{mn}{\cal Z}^{mn}_{A,B,C,D,E,F}+[A,B,C,D,E,F|1234567]\big)\cr
&+\big(V_AT_{B,C,D,E}^m{\cal Z}^m_{A,B,C,D,E}+[A,B,C,D,E|1234567]\big)\cr
&+\big(V_AT_{B,C,D}{\cal Z}_{A,B,C,D}+[A,B,C,D|1234567]\big)\cr
&-\big(V_1J_{2|3,4,5,6,7}^m{\cal Z}_{2|1,3,4,5,6,7}^m+(2\leftrightarrow 3,4,5,6,7)\big)\cr
&-\big(\big(V_AJ_{B|C,D,E,F}{\cal Z}_{B|A,C,D,E,F}+(B\leftrightarrow C,D,E,F)\big)+[A,B,C,D,E,F|1234567]\big)\cr
&-\big(\Delta_{1|2|3,4,5,6,7}{\cal Z}_{12|3,4,5,6,7}+(2\leftrightarrow 3,4,5,6,7)\big)\,,\cr
}$$
where the $[A_1,...,A_m | 12...n]$ notation is used to denote a sum over
Stirling cycles \oneloopIII, see the appendix~\Convapp\ for more details.\foot{These sums can also be described
by all
the ways in which $12...n$ can be completely decomposed into $m$ Lyndon words, with every letter
appearing in precisely one such word.}
The supersymmetric polarizations of ten-dimensional gluons and gluinos
are encoded in the pure spinor multiparticle building blocks $V_A T^{m \ldots}_{B,C,D \ldots}$ reviewed in \oneloopI. The
various $\cZ^{m \ldots}_{A,B,C, \ldots}$ are
worldsheet functions elaborated in \oneloopII\ and they
depend on the insertion points of the vertices on the Riemann
surface and on the loop momentum $\ell^m$.

The open string amplitudes for supersymmetric states are obtained
from these correlators after integration over the vertex insertion points, over the loop momentum, and over the modulus $\tau$ of the
genus-one Riemann surfaces
\eqn\theamp{
{\cal A}_n  =
\sum_{\rm top} C_{\rm top} \int_{D_{\rm top}}\!\!\!\!
d\tau \, dz_2 \, d z_3 \, \ldots \, d z_{n} \, \int d^{D} \ell \ |{\cal I}_n(\ell)| \,
\langle {\cal K}_n(\ell)  \rangle \, ,
}
where ${\cal I}_n(\ell)$ denotes the Koba-Nielsen factor, $D_{\rm top}$ denotes an ordered region of
integration over the insertion points $z_i$, and $C_{\rm top}$ denotes a group-theory factor which depends
on the topology of the genus-one surface (cylinder, M\"obius strip or non-planar cylinder) \GSWII. For simplicity
we will consider only the planar cylinder topology in the following. For more details on this setup, see section~2 of \oneloopI.

To gain intuition why the one-loop open-string correlators lead to a representation of one-loop SYM
numerators that satisfy the color-kinematics duality
it will be illustrative to review the quest for local BCJ-satisfying ten-dimensional supersymmetric
numerators at tree level, solved in pure spinor superspace in \BCJtree.

%**********************************************************************************************
\newsubsec\introcorrBCJtree BCJ-satisfying local numerators at tree level from string disk correlators

\newsubsubsec\fivebcjfail Cohomology analysis: Five-point tree numerators from relabeling

When the tree-level
color-ordered amplitudes
were first proposed in \nptMethod, the construction was based on the principles of locality and BRST invariance
of pure spinor superspace expressions using multiparticle superfields. These same principles were later used
when proposing SYM one-loop integrands in \towardsOne. The difference between the expressions in \nptMethod\ and \towardsOne\
originates from the differences in the pure spinor amplitude prescriptions at tree level \psf\ and one loop \MPS.
The $n$-point tree-level numerators of \nptMethod\ had to be built from three unintegrated (multiparticle) vertices $V$ following the
OPE contractions with $(n-3)$ integrated vertices $U(z)$.
For example, at tree level the five-point SYM amplitude in the canonical color
ordering was obtained as\foot{For convenience we shall frequently omit from amplitudes such as \fivept\ 
the pure spinor brackets $\langle \ldots\rangle$ that extract the top element
$(\l\g^m\t)(\l\g^n\t)(\l\g^p\t)(\t\g_{mnp}\t)$ in the cohomology of the pure spinor BRST operator \psf.
The component evaluation of ghost-number three expressions uses the identities from the appendix of \anomalypaper.}
\eqn\fivept{
A^{\rm SYM}(1,2,3,4,5) =
{V_{[12,3]}V_4V_5\over s_{12}s_{123}}
+{V_{[1,23]}V_4V_5\over s_{23}s_{123}}
+ {V_{[1,2]}V_{[3,4]}V_5\over s_{12}s_{34}}
+ {V_1V_{[23,4]}V_5\over s_{23}s_{234}}
+ {V_1V_{[2,34]}V_5\over s_{34}s_{234}}
}
where $V_{[A,B]}$ denotes the multiparticle unintegrated vertex operator in the BCJ gauge, see the review on multiparticle
superfields in section~3 of \oneloopI\ and section~4.3 of \genredef.
The expression \fivept\ correctly reproduces the five-point tree amplitude of SYM in the canonical color ordering. The next
task is to check whether this representation leads to numerators that satisfy the color-kinematics duality, this is where a
subtle point arises.

A triplet of numerators participating in a kinematic Jacobi identity necessarily involves numerators from
amplitudes with different color orderings, but the naive relabeling of the amplitude \fivept\ does not lead to a representation
satisfying the BCJ color-kinematics duality.
Let us illustrate this point with an example. Using the
parameterization of numerators from \BCJ\ where
\eqnn\bcjpar
$$\eqalignno{
A^{\rm SYM}(1,2,3,4,5) &=
{n_1\over s_{12}s_{45}}
+ {n_2\over s_{23}s_{51}}
+ {n_3\over s_{34}s_{12}}
+ {n_4\over s_{45}s_{23}}
+ {n_5\over s_{51}s_{34}}&\bcjpar\cr
A^{\rm SYM}(1,4,3,2,5) &=
{n_6\over s_{14}s_{25}}
+ {n_5\over s_{51}s_{34}}
+ {n_7\over s_{32}s_{14}}
+ {n_8\over s_{25}s_{34}}
+ {n_2\over s_{51}s_{23}}
}$$
in order
to check whether the numerators $n_3,n_5$ and $n_8$ satisfy the kinematic Jacobi identity $n_3-n_5+n_8=0$
one needs to extract the numerator $n_8$ of the pole in $1/(s_{25}s_{34})$, $n_3$ of $1/(s_{34}s_{12})$ and $n_5$ of
$1/(s_{51}s_{34})$. While $n_3$ and $n_5$
can be read off from the amplitude $A(1,2,3,4,5)$ in \fivept, the numerator $n_8$ is found in the different color
ordering $A(1,4,3,2,5)$. If we assume that
this color ordering is given by the relabeling of \fivept\ the kinematic Jacobi relating these three numerators is not
satisfied,
\eqn\failure{
n_3 - n_5 + n_8 = V_{[1,2]}V_{[3,4]}V_5 - V_1V_{[2,34]}V_5 + V_1V_{[43,2]}V_5 \neq 0\,,
}
where we used $n_8=V_1V_{[43,2]}V_5$ obtained from $n_4=V_1V_{[23,4]}V_5$ via the relabeling $2\leftrightarrow 4$.

%*********************************************************************************************
\newsubsubsec\FTsec Open superstring: Five-point tree numerators from the field-theory limit

The solution to the above problem was found in \BCJtree\ by utilizing the $n$-point string disk correlator of \nptTree\
to generate
different color orderings in its field-theory limit. These orderings follow from
the various integration regions over the insertion points $z_i$ ordered along the
boundary of a disk.
For five points the superstring tree-level correlator is
\eqn\fiveptcorr{
{\cal K}_5(z_1, \ldots,z_5) =
{V_{123}V_{4}V_5\over z_{12}z_{23}} +
{V_1V_{432}V_5\over z_{43}z_{32}} +
{V_{12}V_{43}V_5\over z_{12}z_{43}} + (2\leftrightarrow3)\,.
}
The string tree-level amplitudes with different color orderings are obtained by the different integration regions of the vertex
insertion points relative to each other. The corresponding color-ordered SYM amplitudes follow from the field-theory limit
$\ap\to0$ of the disk integrals, encoded in the biadjoint scalar amplitudes \DPellis\ (see also \Polylogs).
More precisely, one can express the field-theory limit of the
string correlator \fiveptcorr\ as follows \refs{\stringinspired,\FTlimit}
\eqn\SYMstring{
A^{\rm SYM}(\Sigma) =\!\!\!\! \sum_{XY=23}\!\!\! V_{1X} V_{(n-1)\tilde Y}V_n\,
m(\Sigma|1,X,n,Y,n-1)(-1)^{\len{Y}+1} + (2\leftrightarrow3)\,,
}
where $m(\Sigma|\Omega)$ denotes the biadjoint tree amplitudes,
\eqn\biad{
m(P,n|Q,n) = s_P \phi_{P|Q}
}
and $\phi_{P|Q}$ are the Berends-Giele double currents \FTlimit. They can be computed recursively
\eqn\BGdoubleDef{
\phi_{P|Q}=\frac{1}{s_P}\sum_{XY=P}\sum_{AB=Q}\big(\phi_{X|A}\phi_{Y|B}-(X\leftrightarrow Y)\big)\,,
\quad{}\phi_{P|Q}=0\hbox{ if } P\setminus Q\neq\emptyset\,.
}
in terms
of generalized Mandelstam invariants $s_P=\half k_P\cdot k_P$ where $k_P$ is a multiparticle momentum 
defined by $k_P=k_{p_1}+k_{p_2}+ \cdots$ (for example $k_{123} = k_1+k_2+k_3$).

Extracting the field-theory limit of the string disk integrals computed
in the ordering $z_1 \le z_4\le z_3\le z_2\le z_5$ -- corresponding to $\Sigma=14325$ in \SYMstring\ --
leads to the following color-ordered amplitude
\eqnn\onefour
$$\eqalignno{
A^{\rm SYM}(1,4,3,2,5) &= 
% from BGdoublecurrent/ftlimit.frm (must swap sign)
{1\over s_{14}s_{25}} (V_1V_{432} + V_{12}V_{43} + V_{13}V_{42} + V_{132}V_4)V_5 &\onefour\cr
	% + Is(1,3,4)*Is(1,4) * (
        %   - V(1)*V(4,3,2)*V(5)
        %   - V(1,2)*V(4,3)*V(5)
        %   - V(1,3)*V(4,2)*V(5)
        %   - V(1,3,2)*V(4)*V(5)
        %   )
&+{1\over s_{51}s_{34}} V_1V_{432}V_5
       % + Is(2,3,4)*Is(3,4) * (
        %   - V(1)*V(4,3,2)*V(5)
        %   );
-{1\over s_{23}s_{14}} (V_1V_{[4,23]} + V_{[1,23]}V_4)V_5\cr
       % + Is(1,4)*Is(2,3) * (
        %   + V(1)*V(4,2,3)*V(5)
        %   - V(1)*V(4,3,2)*V(5)
        %   + V(1,2,3)*V(4)*V(5)
        %   - V(1,3,2)*V(4)*V(5)
        %   )
&+{1\over s_{25}s_{34}} (V_1V_{432} + V_{12}V_{43})V_5
       % + Is(1,3,4)*Is(3,4) * (
        %   - V(1)*V(4,3,2)*V(5)
        %   - V(1,2)*V(4,3)*V(5)
        %   )
-{1\over s_{51}s_{23}} V_1V_{[4,23]}V_5\,.
       % + Is(2,3)*Is(2,3,4) * (
        %   + V(1)*V(4,2,3)*V(5)
        %   - V(1)*V(4,3,2)*V(5)
        %   )
}$$
One can now read off the numerator $n_8=V_1V_{432}V_5+V_{12}V_{43}V_5$ and verify that 
the BCJ identity $n_3-n_5+n_8=0$ is
identically\foot{{\it Identically} means that no BRST cohomology identity (of the type discussed in \partI) is required to
verify the vanishing of the triplet of numerators; it vanishes at the superfield level.} satisfied \BCJtree
\eqn\success{
n_3 - n_5 + n_8 = V_{[1,2]}V_{[3,4]}V_5 - V_1V_{[2,34]}V_5 + (V_1V_{432}V_5+V_{12}V_{43}V_5) = 0\,,
}
where the bracket notation reviewed in section~3.4.3 of \oneloopI\ implies
$V_{432}=V_{[23,4]}$ and $V_{43}=-V_{[3,4]}$.

The field-theory tree-level SYM numerators are extracted from the knowledge of the singular behavior of
the correlator as vertex operators collide as encoded in the biadjoint Berends-Giele currents. But we know
that these limits constitute a local property of the Riemann surface and therefore must be
independent of its genus.
These results together with the analysis of \mizera\ lead to the following expectation:
\smallskip
{\narrower\smallskip\noindent {\it
The field-theory limit of the one-loop string correlators integrated along different vertex insertion orderings should give rise
to a local representation for SYM one-loop integrands that satisfy the BCJ color-kinematics duality.}\smallskip}

\noindent  As an illustration of this method -- to be fully developed in the next sections --
let us apply it in the simplest case of the five-point SYM integrand/amplitude
following from the string correlator \FivePointCorr.

%****************************************************************************************************
\newsubsec\introcorrBCJ BCJ-satisfying local numerators at one loop from string genus-one correlators

%*********************************************************************************
\newsubsubsec\fivebcjdiff Cohomology analysis: Five-point pentagon from relabeling

As mentioned above, the five-point SYM integrand was proposed based on a few constraints such as
locality and BRST invariance. The pentagon for the color order $A(1,2,3,4,5)$ was given as \towardsOne
\eqn\pentfive{
N^{(5)}_{1|2,3,4,5}(\ell) = \ell^m V_1 T^m_{2,3,4,5} + \half\big[V_{12}T_{3,4,5}+(2\leftrightarrow3,4,5)\big]
+\half\big[V_1 T_{23,4,5} + (2,3|2,3,4,5)\big]
}
where the notation $+(i,j|2,3,4,5)$ denotes a sum over all possible ways to choose two elements $i$ and $j$ from the set
$\{2,3,4,5\}$ while keeping the same order of $i$ and $j$ within the set. The cubic graph associated to this pentagon is
displayed in \figFivePent. Note the convention of assigning the loop momentum $\ell$ to the edge between $5$ and $1$.

\ifig\figFivePent{The cubic graph associated to the pentagon $N^{(5)}_{1|2,3,4,5}(\ell)$ from equation \pentfive. The convention
for the loop momentum $\ell$ is to run from the last argument of the numerator to the first.}
{\epsfxsize=0.18\hsize\epsfbox{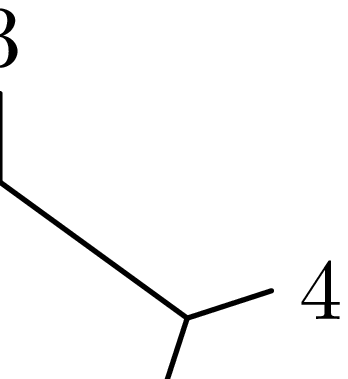}}

Using BRST cohomology arguments the box and pentagon numerators following from relabelings (while respecting
the loop momentum assignment convention and the constraint that leg $1$ is contained in a multiparticle unintegrated vertex $V$)
were proposed in \towardsOne,
\eqnn\genBox
\eqnn\genPent
$$\eqalignno{
N^{(4)}_{A|B,C,D}(\ell) &= V_AT_{B,C,D}&\genBox\cr
N^{(5)}_{A|B,C,D,E}(\ell) &=
\ell_m  V_A T^m_{B,C,D,E} + {1\over 2} \big[V_{[A,B]} T_{C,D,E} + (B\leftrightarrow C,D,E) \big] &\genPent\cr
&\quad{} + {1\over 2} \big[V_{A} T_{[B,C],D,E} + (B,C|B,C,D,E) \big]\,.
}$$
For a cubic-graph parameterization of the five-point integrand
to obey the BCJ color-kinematics duality the antisymmetric combination of two pentagons
in the legs $1$ and $2$ must give rise to a box \CJfive
\medskip
\centerline{\hbox{
\raise0pt\hbox{{\epsfxsize=0.18\hsize\epsfbox{cubic_graphs-10.eps}}}%
\hbox to 10pt{}\raise30pt\hbox{$\ - \ $}\hbox to 10pt{}%
\raise0pt\hbox{{\epsfxsize=0.2\hsize\epsfbox{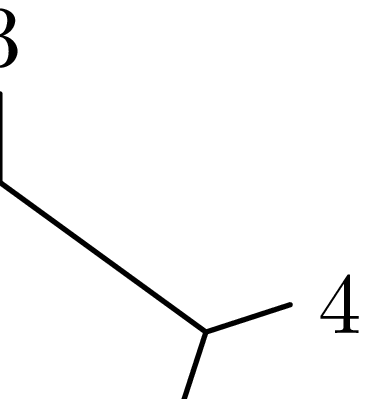}}}
\hbox to 5pt{}\raise30pt\hbox{$\ - \ $}\hbox to 5pt{}
\raise-5pt\hbox{{\epsfxsize=0.15\hsize\epsfbox{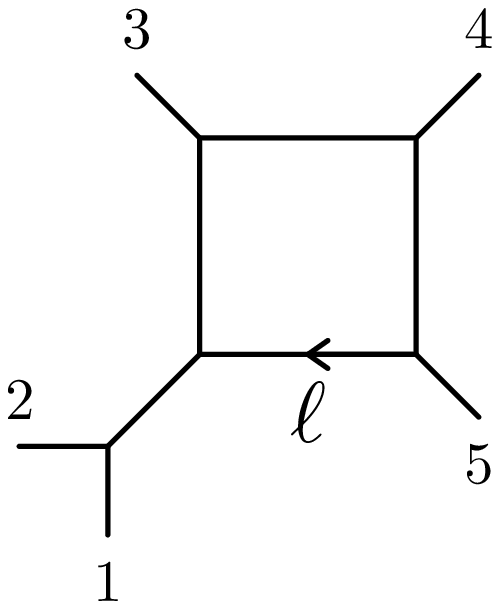}}}
\hbox to 5pt{}\raise30pt\hbox{$\ = \ 0$}\hbox to 5pt{}
}}
\smallskip
\noindent From the figure above we see that the pentagon in the middle must come from
the color ordering $A(2,1,3,4,5)$ so as to keep the momenta in the common edges of the
participating cubic graphs the same while respecting the loop momentum convention mentioned above.
However, the generic expression \genPent\ has to ensure that the leg $1$ appears in $A$, so the solution
proposed in \towardsOne\ satisfying both constraints was to assign the pentagon numerator
$N_{1|3,4,5,2}(\ell-k_2)$ to the middle diagram, with a shift in the loop momentum.
Using that
the $12$-box numerator is $V_{12}T_{3,4,5}$, the
expression \genPent\ implies that the numerator translation of the diagrams above
is given by
\eqnn\bcjtriv
\eqnn\bcjtrivv
$$\displaylines{
\langle N^{(5)}_{1|2,3,4,5}(\ell) - N^{(5)}_{1|3,4,5,2}(\ell-k_2) - N^{(4)}_{12|3,4,5}\rangle =\hfil\bcjtriv\hfilneg\cr
\langle k^2_m V_{1}T^m_{2,3,4,5} + V_{21}T_{3,4,5} + V_1 T_{23,4,5} + V_1 T_{24,3,5} + V_1
T_{25,3,4}\rangle=0.\hfil\bcjtrivv\hfilneg
}$$
The BCJ color-kinematic identity relating two pentagons with a box
is satisfied, but only up to BRST-exact terms in pure spinor superspace that are annihilated by the pure spinor cohomology bracket
$\langle \ldots\rangle$. The BRST exactness of the second line was shown in \partI.

%*********************************************************************************************
\newsubsubsec\newbcjcorr The BCJ pentagon from the field-theory limit of the string correlator

The five-point analysis of \towardsOne\ was primarily based on the BRST cohomology properties of the integrands,
and as we reviewed above this was enough to obtain a BCJ-satisfying parameterization up to BRST-exact terms. However,
using the field-theory limit of the string correlator the resulting numerators for the pentagons improve the BCJ identity
to be satisfied identically at the superspace level, requiring no cohomology manipulations.

To see this we consider the five-point correlator \FivePointCorr\ written in terms
of the Eisenstein-Kronecker coefficient functions $g^{(1)}_{ij}$ of \oneloopII, namely $\cZ^m_{1,2,3,4,5} = \ell^m$ and
$\cZ_{12,3,4,5} = g^{(1)}_{12}$
\eqn\FivePtCorr{
{\cal K}_5(\ell)= V_{1} T^m_{2,3,4,5}  \ell^m
+\big[ V_{12}T_{3,4,5}  g^{(1)}_{12} + (2\leftrightarrow 3,4,5) \big]
+\big[ V_{1}T_{23,4,5}  g^{(1)}_{23} + (2,3|2,3,4,5) \big]\,.
}
In the string-based formalism \worldline, the field-theory limit of the string propagator (in our case the $g^{(1)}_{ij}$
functions) depends on
the relative ordering of how the vertex insertion points are integrated by a term proportional to $\sgn_{ij}$. More precisely,
if the color ordering of the resulting SYM integrand
is $P$, the field-theory limit of $g^{(1)}_{ij}$ contains a term $\half{\rm sgn}_{ij}^P$, where ${\rm sgn}_{ij}^P$ is defined in \sgnDef.
Therefore the pentagons of the integrands in the $A(1,2,3,4,5)$ and $A(2,1,3,4,5)$ orderings differ by a sign in the term coming
from $g^{(1)}_{12}$. This gives rise to the following pentagons:
\eqnn\pentcan
\eqnn\pentncan
$$\eqalignno{
N_{1|2,3,4,5}(\ell) &= \ell^m V_1 T^m_{2,3,4,5}
+ \half V_{12}T_{3,4,5}
+ \half V_{13}T_{2,4,5}
+ \half V_{14}T_{2,3,5}
+ \half V_{15}T_{2,3,4}&\pentcan\cr
&+\half V_1 T_{23,4,5}
+\half V_1 T_{24,3,5}
+\half V_1 T_{25,3,4}
+\half V_1 T_{34,2,5}
+\half V_1 T_{35,2,4}
+\half V_1 T_{45,2,3}\cr
N_{2|1,3,4,5}(\ell) &= \ell^m V_1 T^m_{2,3,4,5}
- \half V_{12}T_{3,4,5}
+ \half V_{13}T_{2,4,5}
+ \half V_{14}T_{2,3,5}
+ \half V_{15}T_{2,3,4}&\pentncan\cr
&+\half V_1 T_{23,4,5}
+\half V_1 T_{24,3,5}
+\half V_1 T_{25,3,4}
+\half V_1 T_{34,2,5}
+\half V_1 T_{35,2,4}
+\half V_1 T_{45,2,3}\cr
}$$
where we note that the constraint that leg $1$ is within $V$ is automatically satisfied because
the correlator \FivePtCorr\ is always the same, what changes is the relative ordering of integration
of the vertex positions.

It is easy to see that the numerators \pentcan\ and \pentncan\ imply that the
BCJ identity is identically satisfied at the superfield level,
\eqn\bcjs{
N_{1|2,3,4,5}(\ell) - N_{2|1,3,4,5}(\ell) - N_{12|3,4,5}(\ell) = 0\,,
}
where $N_{12|3,4,5}(\ell) = V_{12}T_{3,4,5}$.
We thus see that the derivation of $n$-gon numerators from the field-theory limit of the open superstring correlator
evaluated at different regions of integration implies that
the associated BCJ identity is satisfied even before applying the pure spinor cohomology bracket to extract the polarization content
of the superfields, unlike the case \bcjtriv\ obtained from relabeling.
For five points this difference is immaterial as both approaches eventually satisfy the color-kinematics duality
in the cohomology. However,
we will see below that the field-theory limit technique leads to a six-point representation that
satisfies the color-kinematics duality in contrast to the representation of \towardsOne.

%%%%%%%%%%%%%%%%%%%%%%%%%%%%%%%%%%%%%%%%%%%%%%%%%%%%%
%%%%%%%%%%%%%%%%%%%%%%%%%%%%%%%%%%%%%%%%%%%%%%%%%%%%%

\newnewsec\FTLimitGeneral SYM one-loop integrands from string correlators

The field-theory limit of the one-loop string correlators is obtained by shrinking the strings to
points with $\ap\to0$
while degenerating the genus-one surface with modular parameter $\tau$ to point-particle worldline diagrams
with $\Im(\tau)\to\infty$ \GreenFT.
In principle this can be done using the tropical limit
techniques of \piotr\ or the string-based formalism \worldline, although
the explicit form of the Kronecker-Eisenstein coefficient functions $g^{(n)}(z,\tau)$ lead to subtleties
arising from the regular functions with $n\ge 2$. Alternatively, one
can combine the strengths of these approaches with the requirement
that the field-theory integrands for different color orderings and loop-momentum parameterizations
obtained from the string correlators are in the BRST cohomology of the pure spinor
BRST charge. Some trial and error led to the combinatorial rules described below.

\newsubsec\BGphi Kinematic poles and biadjoint Berends-Giele currents

The kinematic poles arise when the insertion points of the vertex operators
approach each other $z_i\to z_j$ on the Riemann surface. The short-distance behavior
of the Koba-Nielsen factor and the OPE propagator is independent
of the genus of the Riemann surface.
This means that the
pole structure of the genus-one string correlators can be described by the same combinatorics of
tree-level poles, given by the biadjoint scalar amplitudes \biad.
These amplitudes are efficiently
computed using the Berends-Giele double currents $\phi_{P|Q}$ of explicit form given in \BGdoubleDef\ where
the words $P$ and $Q$ encode the integration region and integrand.

In the one-loop case however, in
addition to the tree-level kinematic poles in Mandelstam invariants
the field-theory limit of the genus-one string correlators also
yield Feynman loop momentum integrands
\eqn\Feynman{
I_{A_{n+1}1A_1,A_2, \ldots,A_n}(\ell) = {1\over (\ell-k_{A_1})^2(\ell-k_{A_1A_2})^2 \cdots (\ell-k_{A_1A_2 \ldots A_n})^2}
}
to be integrated over a $D$-dimensional loop momentum $\ell$ with $\int d^D \ell$. Note the special role
played by the label $1$ in the above definition; this handling fixes the freedom
to shift the loop momentum and is useful in obtaining BRST-closed SYM integrands \towardsOne.

In summary, the field-theory limit of genus-one open string correlators will be described by poles in Mandelstam
invariants encoded in Berends-Giele double currents multiplied by Feynman loop momentum integrals.

\newsubsubsec\orderingsec Encoding different integration regions

In the same way as in the tree-level case, the color ordering of the resulting SYM integrand from the field-theory limit
of the genus-one open string correlator is associated to the relative ordering of the $z_i$ variables among each other on
the boundary of the Riemann surface. For
example, the ordering $z_1\le z_3\le z_5 \le z_4 \le z_2$ yields an integrand with color ordering $\sigma=13542$.

The presence or absence of kinematic poles depend crucially on the region of integration relative to the ordering of the
$z_{ij}$ variables being integrated. To encode this information we define a map $\Ord_A(B)$ acting on two words $A$ and $B$ that
crops the word $A$
while maintaining the letters it shares with $B$.
That is, we take the word $B$ and return the smallest sequence of consecutive
letters in the cyclic-symmetric object $A$ containing every letter in $B$.
For example,
\eqnn\OrdExamples
$$\eqalignno{
\Ord_{123456}(32)=23\,,\quad{}
\Ord_{123456}(13)&=123\,,\quad{}
\Ord_{123456}(15)=561\,,\quad{}&\OrdExamples\cr
\Ord_{24856317}(58)=85\,,\quad{}
\Ord_{24856317}(465)&=4856\,,\quad{}
\Ord_{24856317}(78)=7248\,.
}$$
This map can be defined algebraically by
\eqnn\OrdDef
$$\eqalignno{
\Ord_A(B)&=\cases{
A_iA_{i+1}...A_{j-1}A_j&if $A_i,A_j\in B,\;\; B\subseteq A_i...A_j,\;\;j-i\leq\frac{|A|}{2}$\cr
A_jA_{j+1}...A_{|A|}A_1A_2...A_i&if $A_i,A_j\in B,\;\; B\subseteq A_i...A_j,\;\;j-i>\frac{|A|}{2}$\cr
0&else$\;\;\;\qquad{}\qquad{}\qquad{}\qquad{}\qquad{}\qquad{}\qquad{}\qquad{}\OrdDef$}&
}$$
This map will be used with the Berends-Giele double current to correctly generate kinematic poles for each integration region
$\sigma$. It will be convenient to introduce the notation: 
\eqnn\phiAbbrev
$$\eqalignno{
\hat\phi(\sigma|A)&\equiv \phi_{\Ord_\sigma(A)|A}\,,&\phiAbbrev
}$$
for an amplitude with color ordering $\sigma$.

%****************************************************
\newsubsec\introInts $p$-gon loop momentum integrands

Frequently we will need the Feynman loop momentum integrands \Feynman\ with a general shift in the loop momentum
$\ell\to\ell + a^i k_i$. This will be indicated by superscripts
\eqn\Feynmanshift{
I^{a_1,a_2,...,a_m}_{A_{n+1}1A_1,A_2,...,A_n}(\ell) = I_{A_{n+1}1A_1,A_2,...,A_n}(\ell+a_1k_1+a_2k_2+ \cdots+a_mk_m)
}
Explicitly we have
\eqn\DefI{
I^{a_1,a_2,...,a_m}_{A_{n+1}1A_1,A_2,...,A_n}=\frac{1}{(\ell+f_{a_1...a_m}-k_{A_1})^2...(\ell+f_{a_1...a_m}-k_{A_1A_2...A_n})^2}\,,
}
where we defined for convenience
\eqn\DefII{
f_{a_1,...,a_m}=a_1k_1+a_2k_2+...+a_mk_m\,.
}
In the event of an $a_i$ being zero, we will omit it from the notation. Note that
the words characterizing the integrands \DefI\ are totally symmetric e.g. $I_{1,342,5,6}=I_{1,234,5,6}$.

We will sometimes simplify the notation for the loop momentum integrands
by dropping all indices which are single letters, and dropping
the shifts in the loop momentum. When this is done it should always be clear the
color ordering of the amplitude.
For example, in the canonical ordering
$A(1,2,...,n;\ell)$ we have
\eqnn\LEgs
$$\eqalignno{
I_\emptyset=I&=I^{a_1,...,a_n}_{1,2,...,n}\,,\qquad{}\qquad{}\qquad{}
I_{234}=I^{a_1,...,a_n}_{1,234,5,6,...,n}\,,&\LEgs\cr
I_{23,56}&=I^{a_1,...,a_n}_{1,23,4,56,7,8,...,n}\,,\qquad{}
I_{n1,34}=I^{a_1,...,a_n}_{n1,2,34,5,6,...,n-1}\,.
}$$
In a few instances, we may wish to use this notation when it is not
immediately clear what the underlying color ordering is.
In these circumstances we will include it as a superscript in the $I$. So, for example
\eqnn\LEgsTwo
$$\eqalignno{
&I^{235416}_\emptyset=I^{235416}=I_{2,3,5,4,1,6}\,,\quad{}
I^{235416}_{53}=I_{2,35,4,1,6}\,,\quad{}
I^{235416}_{612}=I_{162,3,5,4}\,.&\LEgsTwo
}$$

%**********************************************************************
\newsubsec\FTgs Field-theory limit of Kronecker-Eisenstein coefficients

We are now ready to give the field theory limits. These are:
\eqnn\gGenLimits
\eqnn\gGenLimitsTwo
\eqnn\ThreeKEterms
$$\eqalignno{
g^{(p)}_{ij}&\rightarrow 
b^{(p)}_{ij}P
+c^{(p)}_{ij}P(ij)&\gGenLimits\cr
g^{(p)}_{ij}
g^{(q)}_{kl}&\rightarrow
b^{(p)}_{ij}b^{(q)}_{kl}P
+b^{(p)}_{ij}c^{(q)}_{kl}P(kl)
+c^{(p)}_{ij}b^{(p)}_{kl}P(ij)
+c^{(p)}_{ij}c^{(q)}_{kl}
P(ij,kl)&\gGenLimitsTwo\cr
g^{(p_1)}_{{i_1}{j_1}}
g^{(p_2)}_{{i_2}{j_2}}
g^{(p_3)}_{{i_3}{j_3}}&\rightarrow 
%%%%
b^{(p_1)}_{{i_1}{j_1}}b^{(p_2)}_{{i_2}{j_2}}b^{(p_3)}_{{i_3}{j_3}}P
%%%%
+b^{(p_1)}_{{i_1}{j_1}}b^{(p_2)}_{{i_2}{j_2}}c^{(p_3)}_{{i_3}{j_3}}P({i_3}{j_3})&\ThreeKEterms\cr
&+b^{(p_1)}_{{i_1}{j_1}}c^{(p_2)}_{{i_2}{j_2}}b^{(p_3)}_{{i_3}{j_3}}P({i_2}{j_2})
+c^{(p_1)}_{{i_1}{j_1}}b^{(p_2)}_{{i_2}{j_2}}b^{(p_3)}_{{i_3}{j_3}}P({i_1}{j_1})\cr
%%%%
&+b^{(p_1)}_{{i_1}{j_1}}c^{(p_2)}_{{i_2}{j_2}}c^{(p_3)}_{{i_3}{j_3}}P({i_2}{j_2},{i_3}{j_3})
+c^{(p_1)}_{{i_1}{j_1}}b^{(p_2)}_{{i_2}{j_2}}c^{(p_3)}_{{i_3}{j_3}}P({i_1}{j_1},{i_3}{j_3})\cr
&+c^{(p_1)}_{{i_1}{j_1}}c^{(p_2)}_{{i_2}{j_2}}b^{(p_3)}_{{i_3}{j_3}}P({i_1}{j_1},{i_2}{j_2})
+c^{(p_1)}_{{i_1}{j_1}}c^{(p_2)}_{{i_2}{j_2}}c^{(p_3)}_{{i_3}{j_3}}P({i_1}{j_1},{i_2}{j_2},{i_3}{j_3})
}$$
These limits always have the same form; we take the subscripts of the $g_{ij}^{(p)}$,
and sum over the possible ways to assign these to either a $b^{(p)}$ or a $c^{(p)}$ (to be defined below), and
whenever we assign them to a $c^{(p)}$ they are also entered into the $P$ function. In turn these
are defined by
\eqnn\Pdef
$$\eqalignno{
P&=I&\Pdef\cr
P(ij)&=\hat\phi(\sigma|ij)I_{ij}\cr
P(ij,kl)&=\cases{
\hat\phi(\sigma|ijl)I_{ijl}& \quad{}\qquad{}$\;\;$if $j=k$\cr
\hat\phi(\sigma|ij)\hat\phi(\sigma|kl)I_{ij,kl}& \quad{}\qquad{}$\;\;$if all $i$ unique
}\cr
P(ij,kl,mn)&=\cases{
\hat\phi(\sigma|ijln)I_{ijln}& if $j=k$, $\;\;l=m$\cr
\hat\phi(\sigma|ijl)\hat\phi(\sigma|mn)I_{ijl,\,mn}& if $j=k$, $\;\;m,n\notin\{i,j,k,l\}$\cr
\hat\phi(\sigma|ijn)\hat\phi(\sigma|kl)I_{ijn,\,kl}& if $j=m$, $\;\;k,l\notin \{i,j,m,n\}$\cr
\hat\phi(\sigma|ij)\hat\phi(\sigma|kln)I_{ij,\,kln}& if $l=m$, $\;\;i,j\notin\{k,l,m,n\}$\cr
\hat\phi(\sigma|ij)\hat\phi(\sigma|kl)\hat\phi(\sigma|mn)I_{ij,\,kl,mn}& if all $i$ unique
}
}$$
where we used the notation \phiAbbrev.
The cases provided above will be sufficient for our purposes.

Finally, the coefficients $b^{(p)}$ and $c^{(p)}$
for an integrand
$A(\sigma;\ell+\sum_{i=1}^na_ik_i)$ are given by
\eqnn\xDef
\eqnn\yDef
$$\eqalignno{
b^{(p)}_{ij}&=\sum_{m=0}^p\left(\sgn^\sigma_{ij}\right)^{m}\frac{B_{m}(a_j-a_i)^{p-m}}{m!(p-m)!}&\xDef\cr
c^{(p)}_{ij}&=\frac{1}{2(p-1)!}\big((a_j-a_i)+\sgn^\sigma_{ij}\dist_4^{\sigma}(i,j)\big)^{p-1}&\yDef
}$$
where
$B_n$ denotes the $n^{th}$ Bernoulli number\foot{The
amplitudes up to seven points require up to $B_3$:
$B_0=1\,,B_1=\frac{1}{2}\,,B_2=\frac{1}{6}\,,B_3=0$.
}
and
\eqn\sgnDef{
\sgn_{ij}^B=\cases{
+1&: $i$ is left of $j$ in $B$\cr
-1&: $i$ is right of $j$ in $B$
}}
The function $\dist_a^B(i,j)$ measures the distance
between $i$ and $j$ in the word $B$ and returns $+1$ if it is larger than
$a$ and $0$ otherwise,
\eqn\distDef{
\dist_a^B(i,j)=\cases{
+1 &: if $i$ is $a$ or more letters to the left or right of $j$ in $B$\cr
0 &: if $i$ is fewer than $a$ letters to the left or right of $j$ in $B$
}}
Note that when $a_i=0\;\forall\,i$, we must take $0^0=1$ in the above.

%****************************************************
\newsubsubsec\ExampleOfGenLimit A seven-point example

The field-theory limit of the term
$g^{(1)}_{25}g^{(1)}_{57}g^{(1)}_{76}V_1T_{2576,3,4}$
in the seven-point string correlator \SevenPointCorr\
for the SYM integrand with color ordering
$A(1,2,3,4,5,6,7;\ell+4k_4-6k_5)$
follows from \ThreeKEterms\ with $a_4=4$ and $a_5=-6$,
\eqnn\gLimEgPtwo
$$\eqalignno{
g^{(1)}_{25}
g^{(1)}_{57}
g^{(1)}_{76} &\rightarrow
%%%%
b^{(1)}_{25}b^{(1)}_{57}b^{(1)}_{76}P
%%%%
+b^{(1)}_{25}b^{(1)}_{57}c^{(1)}_{76}P(76)\cr
& +b^{(1)}_{25}c^{(1)}_{57}b^{(1)}_{76}P(57)
+c^{(1)}_{25}b^{(1)}_{57}b^{(1)}_{76}P(25)&\gLimEgPtwo\cr
%%%%
& +b^{(1)}_{25}c^{(1)}_{57}c^{(1)}_{76}P(57,76)
+c^{(1)}_{25}b^{(1)}_{57}c^{(1)}_{76}P(25,76)\cr
& +c^{(1)}_{25}c^{(1)}_{57}b^{(1)}_{76}P(25,57)
+c^{(1)}_{25}c^{(1)}_{57}c^{(1)}_{76}P(25,57,76)\,.
}$$
Many of these terms vanish. For instance using \Pdef\ the factor $P(57)$ is
proportional to
$\hat\phi(1234567|57)=\phi_{57|\Ord_{1234567}(57)}=\phi_{57|567}=0$.
Similarly, we find
\eqn\phiDissappearEgs{
P(25)=P(25,76)=P(25,57)=P(25,57,76)=0\,.
}
The non-zero terms are then given by
\eqnn\NonZeroPs
$$\eqalignno{
P&=I=I^{a_4,a_5}_{1,2,3,4,5,6,7}&\NonZeroPs\cr
P(76)&=\hat\phi(1234567|76)I_{76}=\phi_{76|67}I_{76}=-\frac{1}{s_{67}}I^{a_4,a_5}_{1,2,3,4,5,76}\cr
P(57,76)&=\hat\phi(1234567|576)I_{576}=\phi_{576|567}I_{1,2,3,4,576}=-\frac{1}{s_{67}s_{567}}I^{a_4,a_5}_{1,2,3,4,576}
}$$
The various $b^{(1)}_{ij}$ and $c^{(1)}_{ij}$ terms are given by \xDef\ and \yDef. In the
$g_{25}^{(1)}$ case, these are given by (recall that $a_4=4, a_5=-6$)
\eqnn\TwoFiveEgs
$$\eqalignno{
b^{(1)}_{25}&=\frac{B_{0}(-6)^1}{0!\,1!}+\frac{B_{1}(-6)^0}{1!\,0!}=-6+\frac{1}{2}=-\frac{11}{2}&\TwoFiveEgs\cr
c^{(1)}_{25}&=\frac{\big(a_5-a_2+\sgn_{25}^{1234567}\dist_4^{1234567}(2,5)\big)^{1-1}}{2(1-1)!}
=\frac{\big(-6+(-1)^0\times 0\big)^{0}}{2}=\frac{1}{2}
}$$
The others are given by
\eqn\RestOfEgs{
b^{(1)}_{57}=\frac{13}{2}\,,\quad
c^{(1)}_{57}=\frac{1}{2}\,,\quad
b^{(1)}_{76}=-\frac{1}{2}\,,\quad
c^{(1)}_{76}=\frac{1}{2}\,.
}
Putting everything together, we see that the limit is given by
\eqn\EndOfEg{
g^{(1)}_{25}
g^{(1)}_{57}
g^{(1)}_{76}%V_1T_{2576,3,4}
\rightarrow
%V_1T_{2576,3,4}
\frac{143}{8}I^{a_4,a_5}_{1,2,3,4,5,6,7}
+\frac{143}{8}\frac{1}{s_{67}}I^{a_4,a_5}_{1,2,3,4,5,67}
+\frac{11}{8}\frac{1}{s_{67}s_{567}}I^{a_4,a_5}_{1,2,3,4,567}
}
Doing this analysis for the full seven point correlator leads to a BRST closed expression up to anomalous terms (the
explicit expression is available to download from \website).
In the appendix~\sevAp\ the BRST variation of this numerator is worked out and shown to have the desired
property of canceling propagators.

%*********************************************************
\newsubsec\FTpara The one-loop SYM field-theory integrands

The one-loop correlators of the open superstring are integrated over the vertex insertions $z_i$ ordered along
the boundary of a genus one surface. After taking the field-theory limit, the color ordering of the resulting SYM integrand
corresponds to the ordering of the insertions $z_i$. As alluded to in section~\BGphi, the
field-theory limit of the open-string correlators will be written as a field-theory {\it integrand} depending on the loop
momentum $\ell^m$. The parameterization of the one-loop graphs by Feynman loop integrals
is notoriously plagued with the ``labelling problem'':
arbitrary shifts of the loop momentum must not affect the integrated amplitude. This
will be indicated by labelling a color-ordered SYM integrand with the explicit parameterization of the loop momentum
as
\eqn\AmpNot{
A(1,2,...,n;\ell+a_{1}k_{1}+\cdots +a_{n}k_{n})
}
This refers to the amplitude with color ordering $1,2,...,n$, constructed
such that the momentum going from the $n$th leg to the $1$st leg is
$\ell+a_{1}k_{1}+...+a_{n}k_{n}$.
For example, the field-theory limit of the five-point correlator with insertion points ordered
according to $z_1\le z_3\le z_5\le z_2\le z_4$ and loop momentum $\ell$ running between legs $4$ and $1$ is represented by
the SYM integrand\foot{For simplicity we will consider only the planar topology.} $A(1,3,5,2,4;\ell)$.
The statement of cyclicity -- proven in the appendix~\SixPointGeneral\ -- in the color ordering becomes
\eqn\CycNotationSym{
A(1,2,...,n;\ell+a_{1}k_{1}+...+a_{n}k_{n})
=A(2,3,...,n,1;\ell+(a_{1}-1)k_{1}+a_{2}k_{2}...+a_{n}k_{n})
}
Using this, one can always choose to fix the color ordering of the SYM integrand
to start with a leading $1$.

%************************************************
\newsubsubsec\introNs The field-theory numerators

The field-theory limit of the open superstring $n$-point correlator for will be parameterized by
a sum over $p$-gon cubic graphs ranging from $p=4$ (boxes) to $p=n$:
\eqn\integrand{
A(i_1i_2 \ldots i_n;\ell+a^jk_j) = \sum_{p=4}^n
\sum_{A_1 \ldots A_{p+1}=i_2 \ldots i_n}
{\cal N}^{a_1,a_2,\ldots ,a_n}_{A_{p+1}i_1 A_1\vert A_2, \ldots,A_p}(\ell)\;I^{a_1,a_2, \ldots,a_n}_{i_1A_1, A_2, \ldots,A_p}
}
where $\cN^{a_1,a_2, \ldots,a_n}_{A_1|A_2, \ldots,A_p}(\ell)$ denotes the kinematic Berends-Giele
numerator of a $p$-gon constructed as described in the appendix~\Convapp\ and
$I^{a_1,a_2,...,a_n}_{A_1,A_2,...,A_p}$ represents the $p$-gon integrand. We note that in extracting
a local numerator $N_{ \ldots}$ from \integrand\ there will be a factor of $1/2$ for each inverse
Mandelstam invariant, see the definition \bMap.

%********************************
\newsubsubsec\fourSYM Four points

The extraction of the field theory limit at four points is trivial as there is no
propagator function \MPS. The only limit to consider is the Koba-Nielsen factor and we
get
\eqn\FourPtAmp{
A(\sigma_1,\sigma_2,\sigma_3,\sigma_4|\ell+a_1k_{\sigma_1}+...+a_4k_{\sigma_4})=V_1T_{2,3,4}I_{1,2,3,4}\,.
}

%********************************************
\newsubsubsec\GenProcedureFivePts Five points

The five-point genus-one superstring correlator is given by \oneloopIII
\eqnn\fiveptcorrloop
$$\eqalignno{
{\cal K}_5(\ell)&=V_1T_{2,3,4,5}^m{\cal Z}_{1,2,3,4,5}^m
+ \big[V_{12}T_{3,4,5}\cZ_{12,3,4,5} + (2\leftrightarrow3,4,5)\big]&\fiveptcorrloop\cr
&+ \big[V_{1}T_{23,4,5}\cZ_{23,1,4,5} + (2,3|2,3,4,5)\big]
%&+\big(V_AT_{B,C,D}{\cal Z}_{A,B,C,D}+[A,B,C,D|12345]\big)\,,
}$$
with the worldsheet functions \oneloopII
\eqn\czsfive{
\cZ^m_{1,2,3,4,5} = \ell^m\,,\qquad \cZ_{12,3,4,5} = g^{(1)}_{12}\,.
}
This correlator gives rise to
five terms with non-vanishing poles in the canonical color ordering, namely
$g_{12}^{(1)},
g_{23}^{(1)},
g_{34}^{(1)},
g_{45}^{(1)}$, and
$g_{51}^{(1)}$.
The parameterization of the integrand $A(1,2,3,4,5;\ell+a^ik_i)$ from \integrand\ is given by
\eqnn\fivexcB
$$\eqalignno{
A(1,2,3,4,5;\ell+a^ik_i) &=
N_{1|2,3,4,5}(\ell)I^{a_1, \ldots,a_5}_{1,2,3,4,5}(\ell)  &\fivexcB\cr
&+ {1\over 2 s_{12}}N_{12|3,4,5}(\ell)I^{a_1, \ldots,a_5}_{12,3,4,5}(\ell)
+ {1\over 2 s_{23}}N_{1|23,4,5}(\ell)I^{a_1, \ldots,a_5}_{1,23,4,5}(\ell)\cr
& + {1\over 2 s_{34}}N_{1|2,34,5}(\ell)I^{a_1, \ldots,a_5}_{1,2,34,5}(\ell)
+ {1\over 2 s_{45}}N_{1|2,3,45}(\ell)I^{a_1, \ldots,a_5}_{1,2,3,45}(\ell)\cr
&+ {1\over 2 s_{51}}N'_{51|2,3,4}(\ell)I^{a_1, \ldots,a_5}_{1,2,3,4}(\ell)\,.
}$$
Since the field-theory limit rules behave differently for labels at the extremities of the color
ordering,
the $51$-pentagon numerator is denoted $N'_{51|2,3,4}(\ell)$.
Using the field-theory limit \gGenLimits\ and comparing the outcome with \fivexcB\ we can read off the
box numerators. They are independent of the loop momentum and are uniformly described by
\eqn\BoxDef{
N_{A|B,C,D}= V_AT_{B,C,D}\,.
}
In particular, $N'_{51|2,3,4}=N_{51|2,3,4}= V_{51}T_{2,3,4}$. This result agrees with the analysis of \towardsOne.

The pentagon $I^{a_1, \ldots,a_5}_{1,2,3,4,5}(\ell)$ arises from the pieces with no kinematic poles in \gGenLimits\ and collecting
its associated superfields yields the numerator
\eqnn\FivePtGeneralCorr
$$\eqalignno{
N_{1|2,3,4,5}^{a_1,...,a_5}(\ell)=&V_1T_{2,3,4,5}^m\ell^m+\Big[V_{12}T_{3,4,5}\big(a_2-a_1+\frac{1}{2}\big)+(2\leftrightarrow
3,4,5)\Big]
&\FivePtGeneralCorr\cr
&+\Big[V_1T_{23,4,5}\big(a_3-a_2+\frac{1}{2}\big)+(2,3|2,3,4,5)\Big].
}$$
A straightforward but tedious calculation shows that
\eqnn\GenEllRewritting
$$\eqalignno{
QN^{a_1,...,a_5}_{1|2,3,4,5}(\ell)&=
\frac{1}{2}V_1V_2T_{3,4,5}((\ell+f_{a_1...a_5}-k_{12})^2-(\ell+f_{a_1...a_5}-k_1)^2)&\GenEllRewritting\cr
&+\frac{1}{2}V_1V_3T_{2,4,5}((\ell+f_{a_1...a_5}-k_{123})^2-(\ell+f_{a_1...a_5}-k_{12})^2)\cr
&+\frac{1}{2}V_1V_4T_{2,3,5}((\ell+f_{a_1...a_5}-k_{1234})^2-(\ell+f_{a_1...a_5}-k_{123})^2)\cr
&+\frac{1}{2}V_1V_5T_{2,3,4}((\ell+f_{a_1...a_5}-k_{12345})^2-(\ell+f_{a_1...a_5}-k_{1234})^2)\cr
}$$
with the $f_{a_1...a_5}$ defined as in \DefII.
It is then not hard to check that the above cancels the BRST variation of the
box terms. For example, the terms
proportional to
$(\ell+f_{a_1...a_5}-k_{123})^2$ are given by
\eqn\VarCancelEg{
\frac{1}{2}(V_{1}V_3T_{2,4,5}-V_1V_4T_{2,3,5})=-\frac{1}{2s_{34}}QV_1T_{2,34,5}
}
and cancel the BRST variation of the $34$-box in \fivexcB\ since
\eqn\canI{
(\ell+f_{a_1...a_5}-k_{123})^2 I^{a_1, \ldots,a_5}_{1,2,3,4,5}(\ell) = I^{a_1, \ldots,a_5}_{1,2,34,5}(\ell)\,.
}
Similar calculations show that $QN_{1|2,3,4,5}^{a_1, \ldots,a_5}(\ell)I_{1,2,3,4,5}^{a_1, \ldots,a_5}=-QA_{\rm box}(1,2,3,4,5)$
and therefore the five-point SYM integrand \fivexcB\ is BRST invariant.

The BRST cohomology identities \partI
\eqnn\Tmone
$$\eqalignno{
\langle V_1 k^1_m T^m_{2,3,4,5} \rangle &= \langle -V_{12} T_{3,4,5} + (2\leftrightarrow 3,4,5) \rangle &\Tmone\cr
\langle V_1 k^2_m T^m_{2,3,4,5} \rangle&= \langle V_{12} T_{3,4,5} +  \big[ - V_1 T_{23,4,5}+(3\leftrightarrow 4,5) \big] \rangle
}$$
can be used to show that
\eqn\pentas{
\langle N^{(5)}_{1|2,3,4,5}(\ell+a^i k_i)\rangle = \langle N_{1|2,3,4,5}^{a_1,...,a_5}(\ell)\rangle
}
where $N^{(5)}_{1|2,3,4,5}(\ell)$ is given by \pentfive\ and $I_{1,2,3,4,5}^{a_1, \ldots,a_5}(\ell) = I_{1,2,3,4,5}(\ell+a^i k_i)$.
This is an important consistency check on the field-theory rules
spelled out in section~\FTgs.

All color ordering permutations of the five-point SYM integrand is available to download from \website.

%*********************************
\newsubsubsec\SixPtEgVar Six points

The six-point genus-one superstring correlator is given by \oneloopIII
\eqnn\sixLie
$$\eqalignno{
\cK_6(\ell) &=
\half V_{A_1} T^{mn}_{A_2, \ldots,A_6} \cZ^{mn}_{A_1, \ldots,A_6}
+\big[123456|A_1, \ldots,A_6\big] &\sixLie\cr
&\hskip8.5pt{}+ V_{A_1} T^m_{A_2, \ldots,A_5} \cZ^m_{A_1, \ldots,A_5}
+ \big[123456|A_1, \ldots,A_5\big]\cr
&\hskip8.5pt{}+ V_{A_1} T_{A_2, \ldots,A_4} \cZ_{A_1, \ldots,A_4}
+ \big[123456|A_1, \ldots,A_4\big]\,,
}$$
with the worldsheet functions \oneloopII,
\eqnn\newgsAgain
$$\eqalignno{
\cZ_{123,4,5,6}&= g^{(1)}_{12}g^{(1)}_{23} + g^{(2)}_{12} +
g^{(2)}_{23} - g^{(2)}_{13}\,, &\newgsAgain\cr
\cZ_{12,34,5,6}&= g^{(1)}_{12}g^{(1)}_{34}
+ g^{(2)}_{13} + g^{(2)}_{24}
- g^{(2)}_{14} - g^{(2)}_{23}\,,\cr
\cZ^m_{12,3,4,5,6}&= \ell^m g^{(1)}_{12} 
+ (k_2^m - k_1^m)g^{(2)}_{12}
+ \big[ k_3^m (g^{(2)}_{13} - g^{(2)}_{23}) + (3\leftrightarrow
4,5,6)\big]\,,\cr
\cZ^{mn}_{1,2,3,4,5,6}&= \ell^m\ell^n +
\bigl[( k_1^{m}k_2^{n}+k_1^{n}k_2^{m}) g^{(2)}_{12} + (1,2|1,2,3,4,5,6)
\bigr]\,.
}$$
To illustrate the field-theory rules of the previous section we will derive the SYM
integrand $A(2,3,4,5,6,1;\ell) = A(1,2,3,4,5,6;\ell+k_1)$.
We begin with the field theory limit rules given by
\gGenLimits\ and \gGenLimitsTwo
\eqnn\gGenConsistency
$$\eqalignno{
g^{(1)}_{ij}&\rightarrow \frac{1}{2}\sgn_{ij}^{234561}I^{234561}
+\frac{1}{2}\phi_{ij|\Ord_{234561}(ij)}I^{234561}_{ij}&\gGenConsistency\cr
g^{(2)}_{ij}&\rightarrow \frac{1}{12}I^{234561}+{1\over 2 s_{12}}(-\delta_{1i}\delta_{2j}+\delta_{1j}\delta_{2i})I^{234561}_{12}\cr
g^{(1)}_{ij}g^{(1)}_{kl}&\rightarrow \frac{1}{4}\sgn_{ij}^{234561}\sgn_{kl}^{234561}I^{234561}
+\frac{1}{4}\sgn_{kl}^{234561}\phi_{ij|\Ord_{234561}(ij)}I^{234561}_{ij}\cr
&\;\,+\frac{1}{4}\sgn_{ij}^{234561}\phi_{kl|\Ord_{234561}(kl)}I^{234561}_{kl}+\frac{1}{4}P(ij,kl)
}$$
where
\eqnn\PTwoEntry
$$\eqalignno{
&P(ij,kl)=\cases{
\phi_{ijl|Ord_{234561}(ijl)}I^{234561}_{ijl}&if $j=k$\cr
-\phi_{ijk|Ord_{234561}(ijk)}I^{234561}_{ijk}&if $j=l$\cr
-\phi_{jil|Ord_{234561}(ijl)}I^{234561}_{jil}&if $i=k$\cr
\phi_{kij|Ord_{234561}(kij)}I^{234561}_{kij}&if $i=l$\cr
\phi_{ij|Ord_{234561}(ij)}\phi_{kl|Ord_{234561}(kl)}I^{234561}_{ij,kl}&else
}&\PTwoEntry
}$$
Extracting the terms proportional to $I^{234561}=I_{2,3,4,5,6,1}$, we find the
hexagon numerator
\eqnn\SixPtEgHexDef
$$\eqalignno{
N_{2|3,4,5,6,1}(\ell) &=
\frac{1}{2}V_{1}T_{2,3,4,5,6}^{mn}\big(\ell^m\ell^n-\frac{1}{12}[k^1_mk^1_n+(1\leftrightarrow 2,3,4,5,6)]\big)&\SixPtEgHexDef\cr
&+\frac{1}{2}(V_{1}T_{23,4,5,6}^m(\ell^m-\frac{1}{6}k^m_2+\frac{1}{6}k^m_3)+(2,3|2,3,4,5,6))\cr
&-\frac{1}{2}(V_{12}T_{3,4,5,6}^m(\ell^m+\frac{1}{6}k^m_1-\frac{1}{6}k^m_2)+(2\leftrightarrow 3,4,5,6))\cr
&+\frac{1}{6}V_1(T_{[[2,3],4],5,6}+T_{[2,[3,4]],5,6}+(2,3,4|2,3,4,5,6))\cr
&+\frac{1}{4}(V_1T_{23,45,6}+(2,3|4,5|2,3,4,5,6))\cr
&-\frac{1}{4}(V_{12}T_{34,5,6}+(2|3,4|2,3,4,5,6))\cr
&-\frac{1}{6}((V_{123}-2V_{132})T_{4,5,6}+(2,3|2,3,4,5,6))\,.
}$$
We then identify the pentagon numerators, which in all but one case are given by a
generalization of the formulae from \towardsOne , 
\eqnn\SixPtEgPentDef
$$\eqalignno{
N_{A|B,C,D,E1}&=V_{E1}T_{A,B,C,D}^m\ell^m+\frac{1}{2}(V_{[A,E1]}T_{B,C,D}+(A\leftrightarrow B,C,D))&\SixPtEgPentDef\cr
&+\frac{1}{2}(V_{E1}T_{[A,B],C,D}+(A,B|A,B,C,D))
}$$
The exception to the above is the $12$-pentagon, which differs as it has a contribution
from the $g^{(2)}$ terms due to the color ordering $234561$. It is given by the coefficient of $1/2s_{12}I^{a_1=1}_{2,3,4,5,6}$
(note the absence of the label $1$ from the ordering in $I^{a_1=1}_{2,3,4,5,6}$)
\eqnn\SixPtEgPentException
$$\eqalignno{
N'_{21|3,4,5,6}(\ell)=
&-V_1T_{2,3,4,5,6}^{mn}k_2^mk_1^n&\SixPtEgPentException\cr
&-(V_1T_{23,4,5,6}^mk_1^m+(3\leftrightarrow 4,5,6))\cr
&+V_{12}T_{3,4,5,6}^m(\ell^m+k_1^m-k^2_m)\cr
&-(V_{13}T_{2,4,5,6}^m k^m_2+(3\leftrightarrow 4,5,6))\cr
&+\frac{1}{2}(V_{12}T_{34,5,6}+(3,4|3,4,5,6))\cr
&-(V_{13}T_{24,5,6}+(3|4|3,4,5,6))\cr
&+\frac{1}{2}((2V_{132}-V_{123})T_{4,5,6}+(3\leftrightarrow 4,5,6))
}$$
The box numerators have the standard form, with the word containing the
label $1$ assigned to the $V$ superfield, and the other blocks of indices assigned to the $T$
\eqnn\SixPtEgBoxDef
$$\eqalignno{
N_{A|B,C,D1E}&=V_{D1E}T_{A,B,C}\,,\qquad{}
N_{E1A|B,C,D}=V_{E1A}T_{B,C,D}\,.&\SixPtEgBoxDef
}$$
A long calculation shows that the BRST variation of the above integrand is purely anomalous and
given by\foot{See the discussion of \ChenEVA\ as summarized in section~4.5 of \towardsOne\ to understand why \QAwithell\
is not trivially zero due to the cancellation of propagators in the integrand.}
\eqn\QAwithell{
QA^{a_1=1}(1,2,3,4,5,6)=\frac{1}{2}V_1Y_{2,3,4,5,6}(I_{2,3,4,5,6}-\ell^2I_{2,3,4,5,6,1})
}
This is then of a similar form to the $a_1=...=a_6=0$ result found in \towardsOne, and
by an analogous argument to the one presented there one finds the same result for the integrated anomaly
\eqn\QApseudo{
\int d^{10}\ell QA^{a_1=1}(1,2,3,4,5,6)= -\frac{\pi^5}{240}V_1Y_{2,3,4,5,6}\,.
}
Of course the type I superstring theory with gauge group $SO(32)$
is free of gauge anomalies \GSanomaly, but this property does not survive the field-theory
limit of its planar sector and the six-point one-loop SYM amplitude in ten dimensions is anomalous \FramptonAnomaly.
The result \QApseudo\ written in terms of the anomalous building block $Y_{2,3,4,5,6}$ \partI\
is the pure spinor superspace encoding of the field-theory anomaly \refs{\anomalypaper,\anomalysix}.

%**********************************************
\newsubsubsec\GenProcedureSevenPts Seven Points

At seven points, the numerators become far too complex to state here. One example can be found in the appendix~\sevAp.
The derivation of these numerators has one additional complication; as was discussed in
\oneloopIII\ the refined worldsheet functions are given by
\eqnn\cZInitialDef
$$\eqalignno{
&{\cal Z}_{12\vert 3,4,5,6,7}=\partial g^{(2)}_{12}+s_{12}g^{(1)}_{12}g^{(2)}_{12}-3s_{12}g^{(3)}_{12}\,.&\cZInitialDef
}$$
The derivative and the double pole are then removed by using partial integration with the
Koba-Nielsen factor ${\cal I}_7(\ell)$
\eqnn\PartialRearrangement
$$\eqalignno{
(\partial_1 g^{(2)}_{12}){\cal I}_7(\ell)&=
\partial_1(g^{(2)}_{12}{\cal I}_7(\ell))
+ g^{(2)}_{12}\partial_2 {\cal I}_7(\ell)&\PartialRearrangement\cr
&=\partial_1(g^{(2)}_{12}{\cal I}_7(\ell))
+ g^{(2)}_{12}\left((\ell\cdot k_2)+s_{21}g_{21}^{(1)}+s_{23}g_{23}^{(1)}+...+s_{27}g_{27}^{(1)}\right){\cal I}_7(\ell),
}$$
which gives the reformulated expression for \cZInitialDef
\eqnn\cZforJDef
$$\eqalignno{
&{\cal Z}_{12\vert 3,4,5,6,7}=-3s_{12}g^{(3)}_{12}+g^{(2)}_{12}(\ell\cdot k_2
+s_{23}g_{23}^{(1)}+s_{24}g_{24}^{(1)}+...+s_{27}g_{27}^{(1)})&\cZforJDef
}$$
This is the form of the refined worldsheet function we use to extract the numerators and
the computation proceeds analogously as before. And we have
verified the vanishing of the BRST variation of the resulting general expression.

%%%%%%%%%%%%%%%%%%%%%%%%%%%%%%%%%%%%%%%%%%%%%%%%%%%%%
%%%%%%%%%%%%%%%%%%%%%%%%%%%%%%%%%%%%%%%%%%%%%%%%%%%%%
\newnewsec\BCJegs Local BCJ-satisfying numerators

In this section we will obtain the kinematic numerators associated to various one-loop cubic graphs using
the field-theory limit rules of section~\FTgs\ applied to the superstring correlators for six external states as
well as some seven-point numerators.  The results
of this section resolve a puzzle in the analysis of \towardsOne. Namely,
the representation in \towardsOne\ of the six-point integrand did not satisfy the color-kinematics duality by terms
which suspiciously were related to the gauge anomaly. We now show that the six-point integrand representation
arising from the field-theory limit of the string correlator satisfies
all the color-kinematic Jacobi dual relations of Bern-Carrasco-Johansson.

%*********************************************
\newsubsec\BCJintro Color-kinematics duality

The color factors of amplitudes in gauge theory depend on the structure constants
of some gauge group,  $f^{abc}$, that
satisfy the Jacobi identity,
\eqn\JacobiDef{
f^{abe}f^{cde}+f^{bce}f^{ade}+f^{cae}f^{bde}=0\,.
}
The color-kinematics duality conjecture posed by Bern, Carrasco and Johansson (BCJ) states
that the kinematic numerators of cubic-graph diagrams can be chosen to satisfy the same
Jacobi identity relating their color factors \BCJ. That is, if a triplet of diagrams $i,j,k$
whose color factors $c_i,c_j,c_k$ vanish due to the Jacobi identity \JacobiDef, $c_i+c_j+c_k=0$, the
corresponding numerators $N_i,N_j,N_k$ of the diagrams satisfy $N_i(\ell)+N_j(\ell)+N_k(\ell) = 0$
as well. Stated originally at tree-level \BCJ\ (and proven by the field-theory limit of string theory tree amplitudes
\refs{\monodVanhove,\monodStie}) the duality was conjectured at loop-level in \BCJloop, where the kinematic
numerators also depend on loop momenta $\ell$ parameterizing various $n$-gon cubic graphs.
Through this approach, properties of $4\leq {\cal N}\leq 8$ supergravity up to four
loops have been made manifest \BernUF\ (for the five-loop extension see \refs{\genJac,\fivebcj}).

As part of the color-kinematics duality, once the gauge-theory amplitude is written down using
kinematic numerators that satisfy all the kinematic Jacobi identities and automorphism symmetries of the cubic graphs,
the gauge amplitude
can be used to construct a gravity amplitude by replacing the color factors by a second copy
of numerators $c_i\to \tilde N_i(\ell)$ \refs{\BCJ,\BCJloop}. For more details see the review \bcjreview.

We will now show that the numerators extracted from the one-loop string correlators
using the field-theory rules of section~\FTgs\ satisfy all the color-kinematics relations. However,
starting at six points the numerators do not satisfy the required symmetries under shifts
of the loop momentum required by the automorphism symmetries of the cubic graphs (see \CJfive), leading to subtleties in the construction
of the gravity amplitudes which we defer to future work.

The one-loop five-point integrand of SYM in ten dimensions  was already discussed in section~\newbcjcorr\ so we will focus on the
six-point SYM integrand and briefly outline the discussion of the seven-point numerators.

%*****************************
\newsubsec\SixPtsec Six points

The color-kinematics relations are manifestly satisfied within external tree graphs due
to the BCJ gauge used in the multiparticle superfields \refs{\Gauge,\genredef}.
Therefore
we will discuss the kinematic Jacobi identities among $p$-gons with different values of $p$.

%********************************************************************
\newsubsubsec\pentboxsec Kinematic Jacobi between pentagons and a box

The pure spinor superspace expressions of the numerators associated to
the graphs in the following linear combination
\bigskip
\centerline{\hbox{
\raise5pt\hbox{{\epsfxsize=0.18\hsize\epsfbox{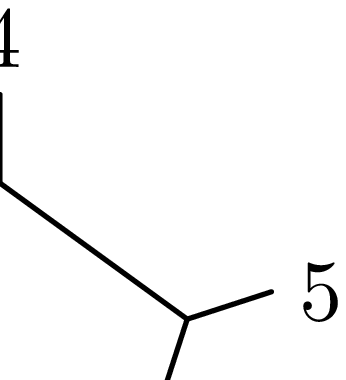}}}%
\hbox to 10pt{}\raise30pt\hbox{$\ - \ $}\hbox to 10pt{}%
\raise0pt\hbox{{\epsfxsize=0.18\hsize\epsfbox{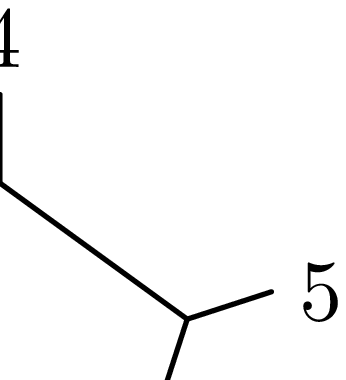}}}
\hbox to 5pt{}\raise30pt\hbox{$\ - \ $}\hbox to 5pt{}
\raise-5pt\hbox{{\epsfxsize=0.15\hsize\epsfbox{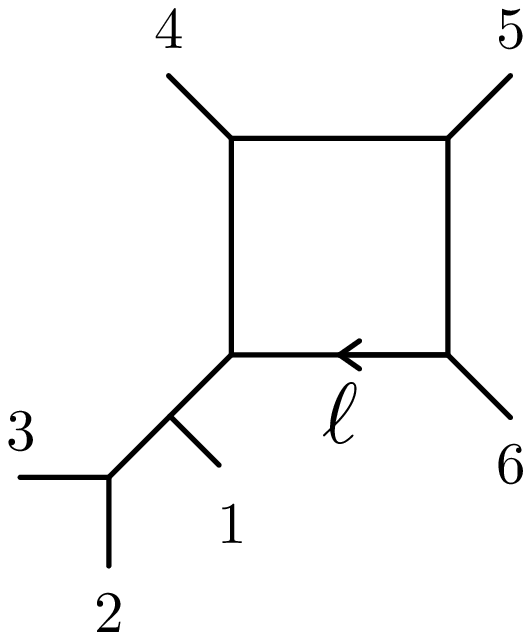}}}
%\hbox to 5pt{}\raise30pt\hbox{$\ = \ 0$}\hbox to 5pt{}
}}
\smallskip
\noindent constitute a good example of how our methods give rise to a
BCJ-satisfying parameterization of the six-point integrand. 

Two of the above graphs are part of the integrand in the canonical color ordering
$A(1,2,3,4,5,6;\ell)$. According to the color-kinematics identity that we are seeking
to show, the loop momentum parameterization of the graphs must have the same momentum
flowing in the edges that are kept the same for all graphs. Therefore the middle graph must have loop momentum
$\ell$ flowing from leg $6$ to the fork $23$. According to the convention shown in \figFivePent\
this is the $23$-pentagon $N_{23|1,4,5,6}(\ell)$ from the integrand
$A(2,3,1,4,5,6;\ell)$ whose expression can be read off from
the field-theory limit rules for this particular ordering.

However the assumption used in the parameterization of \towardsOne\ was that
this pentagon is obtained in a crossing symmetric way as $N_{1|4,5,6,23}(\ell-k_{23})$.
As shown in \towardsOne, using these kinematic numerators the algebraic translation of the BCJ triplet
linear combination above becomes
\eqnn\failb
$$\displaylines{
N_{1|23,4,5,6}(\ell) - N_{1|4,5,6,23}(\ell-k_{23}) - N_{[1,23]|4,5,6}(\ell) = \hfil\failb\hfilneg\cr
k_m^{23} V_1 T^m_{23,4,5,6} + V_{231}T_{4,5,6}
 + \big[ V_1 T_{234,5,6} + (4\leftrightarrow 5,6)\big]
}$$
which is not in the cohomology of the BRST charge and therefore is not vanishing.
In other words, the representation of the six-point integrand
chosen in \towardsOne\ does not satisfy the color-kinematics duality.

In contrast, using the field-theory limit rules of this work the cubic graphs above can be derived
in their native color ordering and they
satisfy the BCJ triplet numerator identity:
\eqn\SixPtBCJegOne{
N_{1|23,4,5,6}(\ell)
%-N^{a_2=a_3=-1}_{1|4,5,6,23}(\ell)
-N_{23|1,4,5,6}(\ell)
-N_{[1,23]|4,5,6}(\ell) = 0\,.
}
To see this vanishing we begin with the box numerator $N_{[1,23]|4,5,6}(\ell)$, the coefficient of
${1\over 4s_{23}s_{123}}I_{123,4,5,6}$ in the integrand $A(1,2,3,4,5,6;\ell)$.
According to \gGenLimitsTwo\ and \Pdef\ the only functions that can generate such a factor are
$g^{(1)}_{12}g^{(1)}_{23}$ and  $g^{(1)}_{13}g^{(1)}_{23}$ via $P(12,23)$ and $P(13,23)$.
There are only two terms featuring these functions in the six-point string correlator \SixPointCorr,
\eqn\BoxEg{
V_{123}T_{4,5,6}g^{(1)}_{12}g^{(1)}_{23} + V_{132}T_{4,5,6}g^{(1)}_{13}g^{(1)}_{32}\,.
}
The field-theory limit of $g^{(1)}_{12}g^{(1)}_{23}$ and  $g^{(1)}_{13}g^{(1)}_{23}$ gives rise to
the box integrand through the $P(ij,jk)$ terms in
\eqnn\DoublePoleEgOne
$$\displaylines{
\frac{1}{4}V_{123}T_{4,5,6}P(12,23) +\frac{1}{4} V_{132}T_{4,5,6}P(13,32)
={1\over4}V_{123}T_{4,5,6}\phi_{123|123}I_{123} + {1\over4}V_{132}T_{4,5,6}\phi_{132|123}I_{123}\cr
={1\over4}V_{123}T_{4,5,6}\left(\frac{1}{s_{12}s_{123}}+\frac{1}{s_{23}s_{123}}\right)I_{123}
+ {1\over4}V_{132}T_{4,5,6}\left(-\frac{1}{s_{23}s_{123}}\right)I_{123}\,.\hfil\DoublePoleEgOne\hfilneg
}$$
The box numerator $N_{[1,23]|4,5,6}(\ell)$ is given by the coefficient of ${1\over4}{1\over s_{23}s_{123}}I_{123}$,
\eqn\BoxEgDef{
N_{[1,23]\vert 4,5,6}=
V_{123}T_{4,5,6} - V_{132}T_{4,5,6}
=V_{[1,23]}T_{4,5,6}
}
The pentagon
$N_{1\vert 23,4,5,6}(\ell)$ is given by the coefficient of
${1\over 2s_{23}}I_{23}$ in the field theory limit of the correlator $\cK_6(\ell)$ for the
color ordering $A(1,2,3,4,5,6;\ell)$. Such factors
arise from any appearance of $g^{(1)}_{23}$ in \SixPointCorr,
\eqnn\PentEgOne
%$$\eqalignno{
$$\displaylines{
V_{1}T^m_{23,4,5,6}\ell^mg^{(1)}_{23}
+\Big[V_{123}T_{4,5,6}g^{(1)}_{12}g^{(1)}_{23}+(2\leftrightarrow 3)\Big]
+\Big[V_{1}T_{234,5,6}g^{(1)}_{23}g^{(1)}_{34}+(4\leftrightarrow 5,6)\Big]\cr
+\Big[V_{14}T_{23,5,6}g^{(1)}_{14}g^{(1)}_{23}+(4\leftrightarrow 5,6)\Big]
+\Big[V_{1}T_{23,45,6}g^{(1)}_{23}g^{(1)}_{45}+(4,5\vert 4,5,6)\Big]\hfil\PentEgOne\hfilneg
}$$
Taking the limits and collecting terms proportional to ${1\over 2s_{23}}I_{23}$ we get
\eqnn\PentEgDefOne
$$\eqalignno{
N_{1\vert 23,4,5,6}(\ell)=V_1T^m_{23,4,5,6}\ell^m &+\half\big[V_{[1,23]}T_{4,5,6}+(23\leftrightarrow
4,5,6)\big]&\PentEgDefOne\cr
&+\half\big[V_1T_{[23,4],5,6}+(23,4\vert 23,4,5,6)\big]\,.
}$$
The numerators \BoxEgDef\ and \PentEgDefOne\ agree with the numerators obtained in \towardsOne.

The middle pentagon in the above figure is the $23$-pentagon in the integrand of $A(2,3,1,4,5,6;\ell)$ since the loop momentum
is running from leg $6$ to $2$. Alternatively, a cyclic rotation as in \CycNotationSym\ yields the integrand
$A(1,4,5,6,2,3;\ell-k_{23})$ whose field-theory limit is computed using the rules of section~\FTgs\ with
with $a_2=a_3=-1$, $a_i=0$ for all other $i$.
The calculation proceeds similarly to
the above. The relevant terms are now\foot{We exploit the total symmetry of the six-point correlator \SixPointCorr\ in
$2,3,4,5,6$}
\eqnn\PentThree
$$\eqalignno{
&V_{1}T^m_{4,5,6,23}\ell^mg^{(1)}_{23}+\Big[V_{123}T_{4,5,6}g^{(1)}_{12}g^{(1)}_{23}+(2\leftrightarrow 3)\Big]&\PentThree\cr
&\qquad{}+\Big[V_{1}T_{423,5,6}g^{(1)}_{42}g^{(1)}_{23}+V_{1}T_{432,5,6}g^{(1)}_{43}g^{(1)}_{32}+(4\leftrightarrow 5,6)\Big]\cr
&\qquad{}+\Big[V_{14}T_{5,6,23}g^{(1)}_{14}g^{(1)}_{23}+(4\leftrightarrow 5,6)\Big]+\frac{1}{2}\Big[V_{1}T_{45,6,23}g^{(1)}_{45}g^{(1)}_{23}+(4,5\vert 4,5,6)\Big].
}$$
Taking the field theory limits and restricting ourselves to the $s_{23}$ single poles,
we see that the numerator is given by %(note that $N_{23|1,4,5,6}(\ell)=N_{1|4,5,6,23}^{a_2=a_3=-1}$)
\eqnn\PentEgDefTwo
$$\eqalignno{
N_{23\vert 1,4,5,6}(\ell)&=V_1T_{4,5,6,23}^m\ell^m-\frac{1}{2}V_{[1,23]}T_{4,5,6}+\frac{1}{2}(V_{[1,4]}T_{5,6,23}+(4\leftrightarrow 5,6))\cr
&+\frac{1}{2}(V_1T_{[23,4],5,6}+(23,4\vert 23,4,5,6)) &\PentEgDefTwo
}$$
This differs from the parameterization of this graph used in \towardsOne, namely $N^{(5)}_{1|4,5,6,23}(\ell-k_{23})$ with the expression
for $N^{(5)}_{A|B,C,D,E}(\ell)$ given in \genPent.
While the representation of \towardsOne\ fails to satisfy the color Jacobi identity, the new representation derived here
obeys the color-kinematics duality. To see this we
plug the superfield expressions of the new field-theory representations of the box \BoxEgDef\ and pentagons
\PentEgDefOne, \PentEgDefTwo\ in the kinematic Jacobi relation
\SixPtBCJegOne\ to obtain
\eqn\sixnotfail{
N_{1|23,4,5,6}(\ell)
-N_{23|1,4,5,6}(\ell)
-N_{[1,23]|4,5,6}(\ell) = 0\,.
}
And we note that the BCJ relation is identically satisfied at the superfield level (i.e. no BRST cohomology identity is
needed). This trivial vanishing for the BCJ triplet at one loop parallels the superfield vanishing of the BCJ triplet
of tree-level numerators obtained from the field-theory of the disk correlators as seen in \success.

%**************************************************************************
\newsubsubsec\SixPthexpent Kinematic Jacobi between hexagons and a pentagon

In a given color ordering, all of the pentagons have a similar structure apart from the $ij$-pentagon
whose labels are cyclically split at the extremities $A(i, \ldots,j;\ell)$.
In this subsection we will demonstrate the validity of a
BCJ relation involving such a numerator. The relation we will show is
\bigskip
\centerline{\hbox{
\raise0pt\hbox{{\epsfxsize=0.17\hsize\epsfbox{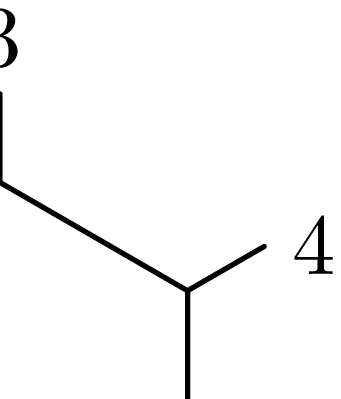}}}%
\hbox to 10pt{}\raise30pt\hbox{$\ - \ $}\hbox to 10pt{}%
\raise0pt\hbox{{\epsfxsize=0.17\hsize\epsfbox{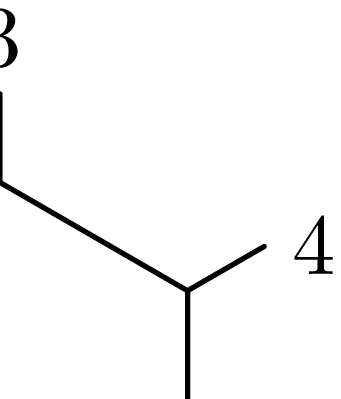}}}
\hbox to 5pt{}\raise30pt\hbox{$\ - \ $}\hbox to 5pt{}
\raise0pt\hbox{{\epsfxsize=0.17\hsize\epsfbox{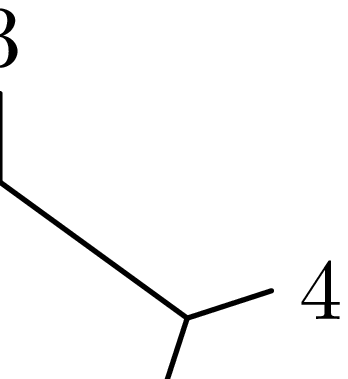}}}
\hbox to 5pt{}\raise30pt\hbox{$\ = \ 0$}\hbox to 5pt{}
}}
\smallskip\noindent which corresponds to
\eqn\SixPtBCJegTwo{
N_{1|2,3,4,5,6}(\ell)-N_{1|6,2,3,4,5}^{a_6=1}(\ell)-N_{61|2,3,4,5}(\ell)=0\,.
}
To find the hexagon numerators, we look at the piece of the field theory limits
proportional to $P=I$. In the first case, this means making the substitution
\eqnn\HexSubEgOne
$$\eqalignno{
&g^{(1)}_{ij}\rightarrow \frac{1}{2}\sgn_{ij}^{123456}I\,,\qquad{}
g^{(1)}_{ij}g^{(1)}_{kl}\rightarrow \frac{1}{4}\sgn_{ij}^{123456}\sgn_{kl}^{123456}I\,,\qquad{}
g^{(2)}_{ij}\rightarrow\frac{1}{12}I\,.&\HexSubEgOne
}$$ 
This then gives the value of the first hexagon numerator as
\eqnn\NSixExample
$$\eqalignno{
N_{1|2,3,4,5,6}(\ell)=
&+ \frac{1}{6}((V_{[[1,2],3]}+V_{[1,[2,3]]})T_{4,5,6}+(2,3|2,3,4,5,6))&\NSixExample\cr
&+ \frac{1}{6}V_1(T_{[[2,3],4],5,6}+T_{[2,[3,4]],5,6}+(2,3,4|2,3,4,5,6))\cr
&+ \frac{1}{4}V_{[1,2]}T_{[3,4],5,6}+(2|3,4|2,3,4,5,6))\cr
&+ \frac{1}{4}V_{1}T_{[2,3],[4,5],6}+(2,3|4,5|2,3,4,5,6))\cr
&+ \frac{1}{2}(V_{[1,2]}T_{3,4,5,6}^m(\ell^m-\frac{1}{6}k^m_1+\frac{1}{6}k^m_2)+(2\leftrightarrow 3,4,5,6))\cr
&+ \frac{1}{2}(V_{1}T_{[2,3],4,5,6}^m(\ell^m-\frac{1}{6}k^m_2+\frac{1}{6}k^m_3)+(2,3|2,3,4,5,6))\cr
&+ \frac{1}{2}V_{1}T_{2,3,4,5,6}^{mn}(\ell^m\ell^n-\frac{1}{12}k^m_1k^n_1-\frac{1}{12}k^m_2k^n_2- \cdots-\frac{1}{12}k^m_6k^n_6)\,.
}$$
For the second hexagon, we consider the field-theory limit of the correlator with the color
ordering $A(1,6,2,3,4,5;\ell+k_1)$. The limits needed now have the form
\eqnn\gLimTwo
$$\eqalignno{
g^{(1)}_{ij}&\rightarrow \half\sgn_{ij}^{162345}+\delta_{j6}-\delta_{i6}\,,&\gLimTwo\cr
g^{(1)}_{ij}g^{(1)}_{kl}&\rightarrow \big(\half\sgn_{ij}^{162345}+\delta_{j6}-\delta_{i6}\big)
\big(\half\sgn_{kl}^{162345} +\delta_{l6}-\delta_{k6}\big)\,,\cr
g^{(2)}_{ij}&\rightarrow \frac{1}{12}+\half\big(\delta_{i6}(1-\sgn_{ij}^{162345})+\delta_{j6}(1+\sgn_{ij}^{162345})\big)\,.
}$$
Using these, the numerator is identified as
\eqnn\NSixShiftedTerm
$$\eqalignno{
N_{1|6,2,3,4,5}^{a_6=1}(\ell)=
&+\frac{1}{2}V_{1}T_{2,3,4,5,6}^{mn}(\ell^m\ell^n+2k_1^mk_6^n-\frac{1}{12}(k^1_mk^1_n+k^2_mk^2_n+ \cdots k^6_m k^6_n))\cr
&+\frac{1}{2}(V_{1}T_{[2,3],4,5,6}^m(\ell^m-\frac{1}{6}k^m_2+\frac{1}{6}k^m_3)+(2,3|2,3,4,5,6))\cr
&- (V_1T_{[2,6],3,4,5}^mk_1^m+(2\leftrightarrow 3,4,5)) &\NSixShiftedTerm\cr
&+\frac{1}{2}(V_{[1,2]}T_{3,4,5,6}^m(\ell^m-\frac{1}{6}k_1^m+\frac{1}{6}k_2^m+2k_6^m)+(2\leftrightarrow 3,4,5))\cr
&+V_{[1,6]}T_{2,3,4,5}^m(\frac{3}{2}\ell^m-\frac{13}{12}k_1^m+\frac{13}{12}k_6^m)\cr
&+\frac{1}{6}V_{1}(T_{[[2,3],4],5,6}+T_{[2,[3,4]],5,6}+(2,3,4|2,3,4,5,6))\cr
&+\frac{1}{6}((V_{[[1,2],3]}+V_{[1,[2,3]]})T_{4,5,6}+(2,3|2,3,4,5))\cr
&-\frac{1}{3}((V_{[[1,2],6]}+V_{[1,[2,6]]})T_{4,5,6}+(2\leftrightarrow 3,4,5))\cr
&+\frac{1}{4}(V_1T_{[2,3],[4,5],6}+(2,3|4,5|2,3,4,5,6))\cr
&+\frac{1}{4}(V_{[1,2]}T_{[3,4],5,6}+(2|3,4|2,3,4,5))\cr
&-\frac{3}{4}(V_{[1,2]}T_{[3,6],4,5}+(2|3|2,3,4,5))\cr
&+\frac{3}{4}(V_{[1,6]}T_{[2,3],4,5}+(2,3|2,3,4,5))
}$$
Finally we have the pentagon term, the superfield coefficient of ${1\over 2s_{16}}I_{61,2,3,4,5}$
from the integrand $A(1,2,3,4,5,6;\ell)$. This can be found to be
\eqnn\NSixOnePentExample
$$\eqalignno{
N'_{61|2,3,4,5}(\ell)=
&+\frac{1}{2}\big[(V_{[[1,2],6]}+V_{[1,[2,6]]})T_{3,4,5}+(2\leftrightarrow 3,4,5)\big]&\NSixOnePentExample\cr
&+\big[V_{[1,2]}T_{[3,6],4,5}+(2|3|2,3,4,5)\big]\cr
&-\frac{1}{2}\big[V_{[1,6]}T_{[2,3],4,5}+(2,3|2,3,4,5)\big]\cr
&-\big[V_{[1,2]}T_{3,4,5,6}^m k_6^m+(2\leftrightarrow 3,4,5)\big]\cr
&+\big[V_1T_{[2,6],3,4,5}^mk_1^m+(2\leftrightarrow 3,4,5)\big]\cr
&-V_{[1,6]}T_{2,3,4,5}^m(\ell^m+k_6^m-k_1^m)\cr
&-V_1T_{2,3,4,5,6}^{mn}k_1^mk_6^n\cr
}$$
It is then simply a matter of plugging the numerators into the identity \SixPtBCJegTwo\ 
to verify its validity.

%**************************************************************************
\newsubsubsec\ijanti Antisymmetry of $ij$-pentagons from $A(i,P,j;\ell)$ in $i\leftrightarrow j$

As mentioned above, the color-kinematics duality relations within external tree
diagrams is manifestly satisfied due to the usage of multiparticle superfields
in the BCJ gauge. For instance, all the boxes and all but one pentagon for an integrand
of arbitrary color ordering $A(P;\ell)$ can be described by
\eqnn\allbox
\eqnn\allpent
$$\eqalignno{
N_{A|B,C,D}(\ell) &= V_{A}T_{B,C,D}(\ell) + (A\leftrightarrow B,C,D)&\allbox\cr
N_{A|B,C,D,E}(\ell) &= \big[V_A T^m_{B,C,D,E}\ell_m + (A\leftrightarrow B,C,D,E)\big]\cr
& + \half\big[V_A T_{[B,C],D,E}+(A|B,C|A,B,C,D,E)\big]\cr
& + \half\big[V_{[A,B]}T_{C,D,E}+(A,B|A,B,C,D,E)\big] &\allpent
}$$
with the additional constraint that $T^{ \ldots}_{ \ldots, A1B, \ldots}=0$ (i.e., setting
to zero all terms in which the label $1$ is not assigned to a multiparticle vertex $V_P$). For example, using \allpent\ we
recover the $23$-pentagon \PentEgDefTwo
\eqnn\exPen
$$\eqalignno{
N_{23\vert 1,4,5,6}(\ell)&=
V_1T_{4,5,6,23}^m\ell^m-\frac{1}{2}V_{[1,23]}T_{4,5,6}+\frac{1}{2}(V_{[1,4]}T_{5,6,23}+(4\leftrightarrow 5,6)) &\exPen\cr
&+\frac{1}{2}(V_1T_{[23,4],5,6}+(23,4\vert 23,4,5,6))
}$$
where we used \genPent\ and the constraint $T^{ \ldots}_{ \ldots, A1B, \ldots}=0$. Since in the BCJ gauge \refs{\Gauge,\genredef} the multiparticle
labels (words) in \allbox\ and \allpent\ satisfy generalized Jacobi identities, the color-kinematics duality
are manifest within those words, with a notable exception.

\ifig\figSixOne{The antisymmetry of the $61$-pentagon from the integrand $A(1,2,3,4,5,6;\ell)$. The momentum running into
the $61$ external tree in the graph on the right is $\ell+k_6$ because in the color ordering $1,2,3,4,5,6$
a momentum $\ell$ must run between $6$ and $1$. Therefore in order to preserve the momentum assignment in the edges between
the two cubic graphs, the pentagon on the left is part of the integrand $A(1,6,2,3,4,5;\ell+k_6)$ with momentum $\ell+k_6$ running
between legs $5$ and $1$ as dictated by the convention \AmpNot. Therefore to extract this pentagon the field-theory rules of
section~\FTgs\ must
be used with $a_6=1$.}
{\epsfxsize=0.53\hsize\epsfbox{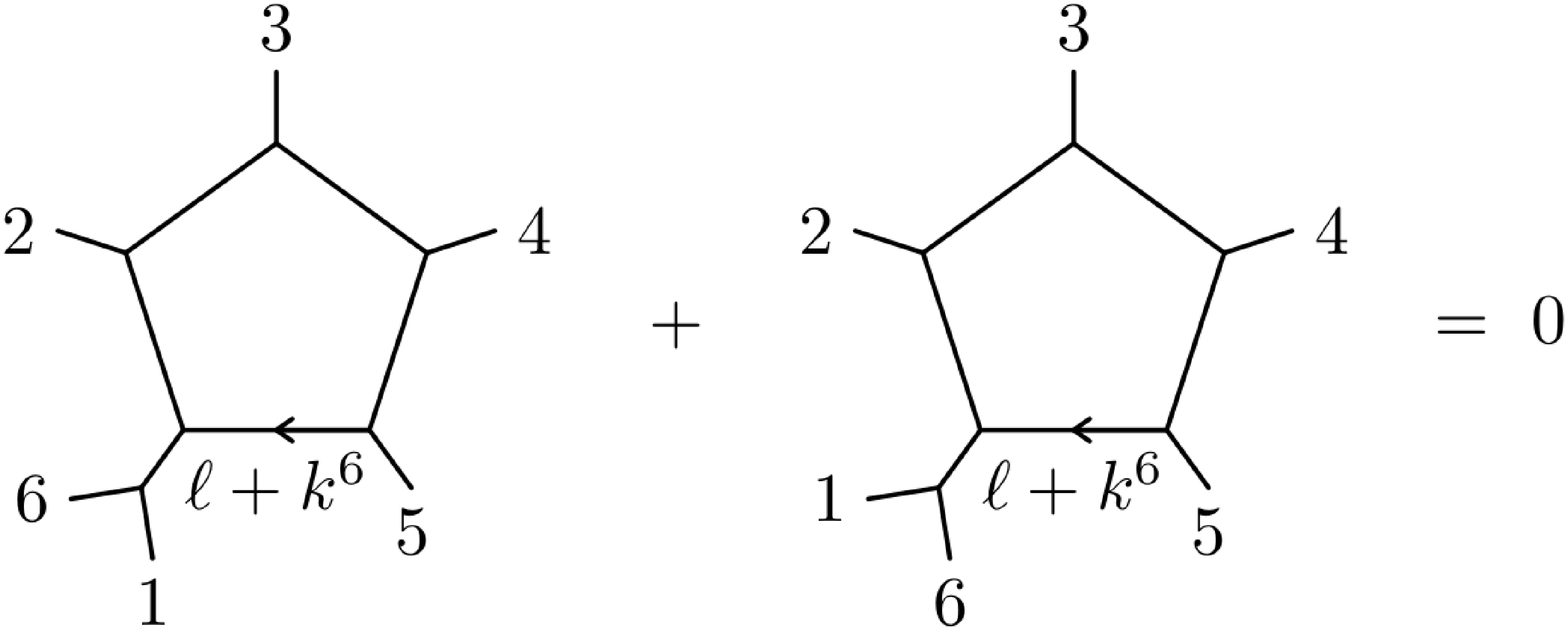}}
%\centerline{\hbox{
%\raise0pt\hbox{{\epsfxsize=0.2\hsize\epsfbox{cubic_graphs-14.eps}}}%
%\hbox to 10pt{}\raise40pt\hbox{$\ + \ $}\hbox to 10pt{}%
%\raise0pt\hbox{{\epsfxsize=0.2\hsize\epsfbox{cubic_graphs-15.eps}}}
%\hbox to 5pt{}\raise40pt\hbox{$\ = \ 0$}\hbox to 5pt{}
%}}

The exception arises for the $ij$-pentagon when the labels $i,j$ are adjacent up to a cyclic rotation, e.g. the $61$-pentagon
in $A(1,2,3,4,5,6;\ell)$ or the $12$-pentagon in $A(2,3,4,5,6,1;\ell)$ do not follow the general formula \allpent, as can be
seen for example in \SixPtEgPentException. The reason this happens is due to a clash between the $ij$ pentagon labels in
$A(j,P,i;\ell)$ and the
convention that the loop momentum $\ell$ runs between $i$ and $j$. So to verify the antisymmetry of the $61$-pentagon from
$A(1,2,3,4,5,6;\ell)$ one needs to compare it to the $16$-pentagon from $A(1,6,2,3,4,5;\ell+k_6)$ using the field-theory rules from section~\FTgs.
The result is
\eqnn\shifk
$$\eqalignno{
N^{a_6=1}_{16|2,3,4,5}(\ell) &=
-\frac{1}{2}\big[(V_{[[1,2],6]}+V_{[1,[2,6]]})T_{3,4,5}+(2\leftrightarrow 3,4,5)\big]&\shifk\cr
&-\big[V_{[1,2]}T_{[3,6],4,5}+(2|3|2,3,4,5)\big]\cr
&+\frac{1}{2}\big[V_{[1,6]}T_{[2,3],4,5}+(2,3|2,3,4,5)\big]\cr
&+\big[V_{[1,2]}T_{3,4,5,6}^m k_6^m - V_1T_{[2,6],3,4,5}^mk_1^m+(2\leftrightarrow 3,4,5)\big]\cr
&+V_{[1,6]}T_{2,3,4,5}^m(\ell^m+k_6^m-k_1^m)\cr
&+V_1T_{2,3,4,5,6}^{mn}k_1^mk_6^n\,.
}$$
Comparing \shifk\ with \NSixOnePentExample\ one immediately verifies the color-kinematics identity
depicted in \figSixOne
\eqn\subsix{
N^{a_6=1}_{16|2,3,4,5}(\ell)+N_{61|2,3,4,5}(\ell)=0\,.
}
It is interesting to observe that the field-theory limit rules yield a
different $16$-pentagon in the in color ordering without a shift in the loop momentum
$A(1,6,2,3,4,5;\ell)$, namely
\eqn\normal{
N_{16|2,3,4,5}(\ell) =
V_{16} T^m_{2,3,4,5} \ell_m
	+ \half \big[V_{16} T_{23,4,5} + (2,3|2,3,4,5)]
          + \half V_{162}T_{3,4,5} + (2\leftrightarrow3,4,5)\big]\,.
}
If we now perform a manual shift $\ell\to\ell+k_6$ in the $16$-pentagon numerator \normal\ and compare it 
with the $16$-pentagon from the shifted amplitude $A(1,6,2,3,4,5;\ell+k_6)$ we find that they are not
BRST equivalent,
\eqn\notb{
Q\big(N^{a_6=1}_{16|2,3,4,5}(\ell) - N_{16|2,3,4,5}(\ell+k_6)\big) = Q(s_{16}V_1 J_{6|2,3,4,5})\,.
}
This shows that the field-theory rules of section~\FTgs\ capture the shifts in the loop momentum
parameterization in a non trivial way, as the limit for  $A(1,6,2,3,4,5;\ell+k_6)$ does not
follow from naively shifting $\ell\to\ell+k_6$ in  $A(1,6,2,3,4,5;\ell)$.

%******************************************
\newsubsubsec\rembcj Remaining BCJ triplets

There are then a number of relations between pentagons and boxes left to show in order
to see that we have a BCJ representation of $A(1,2,3,4,5,6)$, and these
can be seen in the cases a) to d) in the next figure. For each of these in turn we just follow the rules \gGenLimits\ for the following
amplitudes with the following assignments of values for the $a_i$
\medskip
\centerline{\hbox{
\hbox to 10pt{}\raise40pt\hbox{a)}\hbox to 20pt{}%
\raise0pt\hbox{{\epsfxsize=0.2\hsize\epsfbox{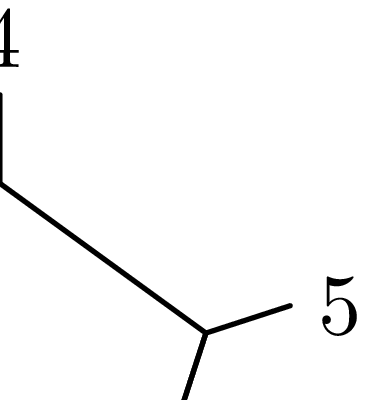}}}%
\hbox to 10pt{}\raise40pt\hbox{$\ - \ $}\hbox to 10pt{}%
\raise0pt\hbox{{\epsfxsize=0.2\hsize\epsfbox{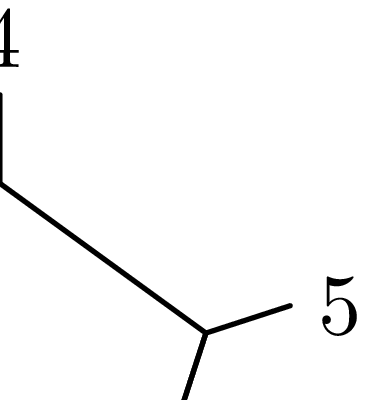}}}
\hbox to 10pt{}\raise40pt\hbox{$\ - \ $}\hbox to 10pt{}%
\raise0pt\hbox{{\epsfxsize=0.15\hsize\epsfbox{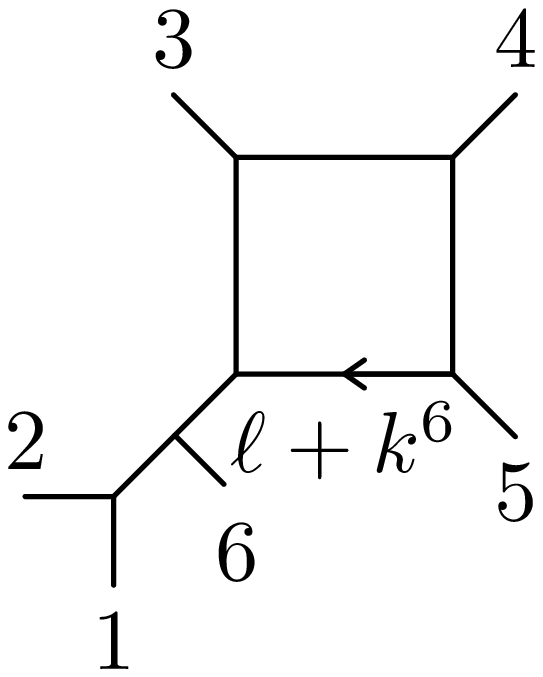}}}
\hbox to 5pt{}\raise40pt\hbox{$\ = \ 0$}\hbox to 5pt{}
}}
\smallskip
\centerline{\hbox{
\hbox to 10pt{}\raise40pt\hbox{b)}\hbox to 20pt{}%
\raise0pt\hbox{{\epsfxsize=0.2\hsize\epsfbox{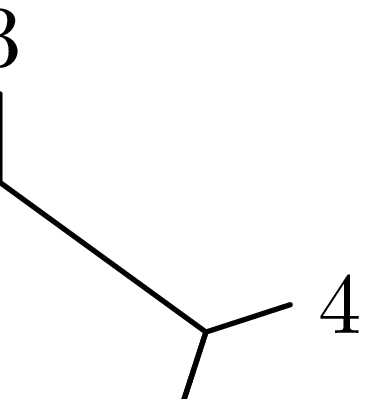}}}%
\hbox to 10pt{}\raise35pt\hbox{$\ - \ $}\hbox to 10pt{}%
\raise0pt\hbox{{\epsfxsize=0.2\hsize\epsfbox{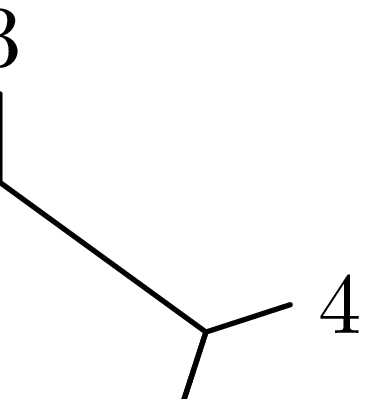}}}
\hbox to 10pt{}\raise35pt\hbox{$\ - \ $}\hbox to 10pt{}%
\raise-5pt\hbox{{\epsfxsize=0.15\hsize\epsfbox{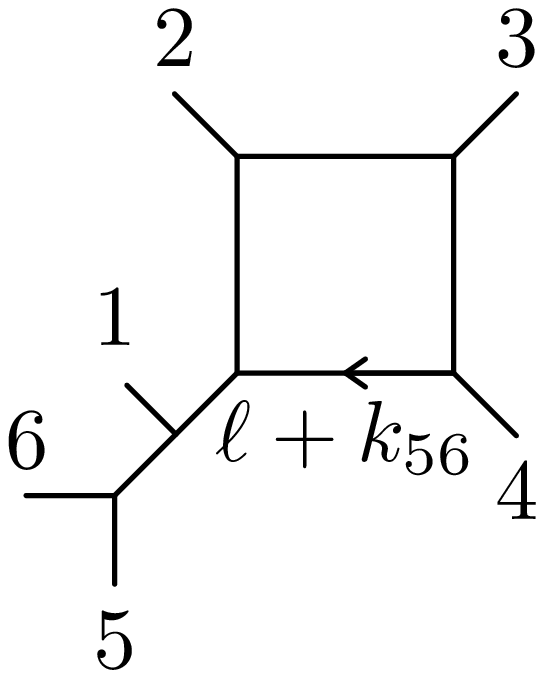}}}
\hbox to 5pt{}\raise35pt\hbox{$\ = \ 0$}\hbox to 5pt{}
}}
\smallskip
\centerline{\hskip-2pt\hbox{
\hbox to 10pt{}\raise40pt\hbox{c)}\hbox to 20pt{}%
\raise0pt\hbox{{\epsfxsize=0.2\hsize\epsfbox{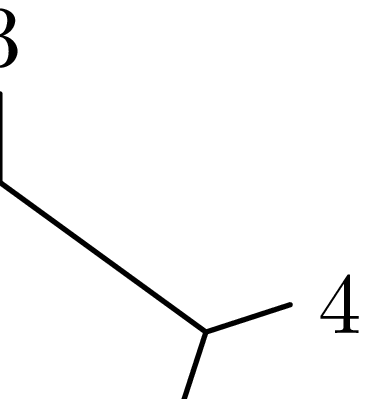}}}%
\hbox to 6pt{}\raise35pt\hbox{$\ - \ $}\hbox to 5pt{}%
\raise5pt\hbox{{\epsfxsize=0.25\hsize\epsfbox{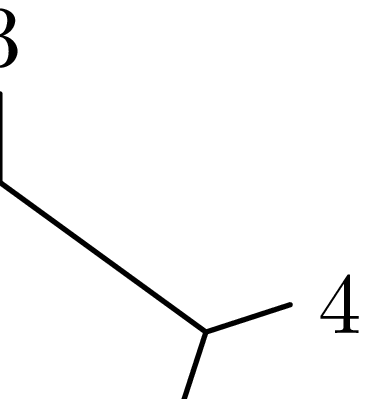}}}
\hbox to 1pt{}\raise35pt\hbox{$\ - \ $}\hbox to 10pt{}%
\raise5pt\hbox{{\epsfxsize=0.16\hsize\epsfbox{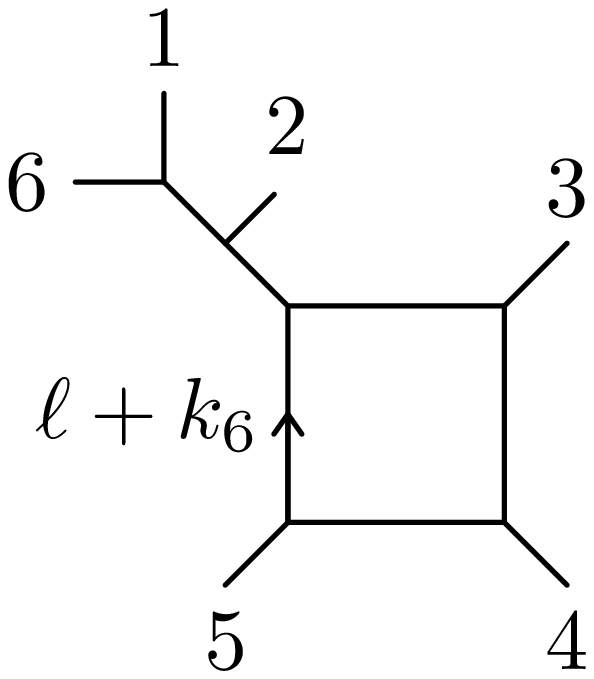}}}
\hbox to 5pt{}\raise35pt\hbox{$\ = \ 0$}\hbox to 5pt{}
}}
\smallskip
\centerline{\hbox{
\hbox to 10pt{}\raise40pt\hbox{d)}\hbox to 20pt{}%
\raise0pt\hbox{{\epsfxsize=0.2\hsize\epsfbox{cubic_graphs-26.eps}}}%
\hbox to 10pt{}\raise35pt\hbox{$\ - \ $}\hbox to 10pt{}%
\raise-20pt\hbox{{\epsfxsize=0.2\hsize\epsfbox{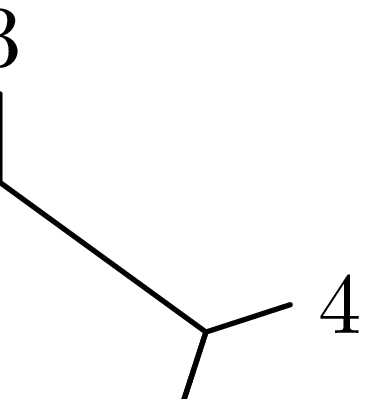}}}
\hbox to 10pt{}\raise35pt\hbox{$\ - \ $}\hbox to 10pt{}%
\raise-5pt\hbox{{\epsfxsize=0.15\hsize\epsfbox{cubic_graphs-25.eps}}}
\hbox to 5pt{}\raise35pt\hbox{$\ = \ 0$}\hbox to 5pt{}
}}

\eqn\PentBoxLimits{
\eqalign{
&A(1,2,6,3,4,5;\ell+k_6),\cr
&A(1,6,5,2,3,4;\ell+k_{56}),\cr
&A(1,3,4,5,2,6;\ell-k_2),\cr
&A(1,5,2,3,4,6;\ell+k_5),
}
\qquad\eqalign{
&a_1=a_2=a_3=a_4=a_5=0,\; a_6=1\cr
&a_1=a_2=a_3=a_4=0,\; a_5=a_6=1\cr
&a_1=a_3=a_4=a_5=a_6=0,\; a_2=-1\cr
&a_1=a_2=a_3=a_4=a_6=0,\; a_5=1
}}
These have been verified to give amplitudes which are BRST invariant and satisfy the
relations a) to d) in the figure above. We will not detail their construction any further, as they can be
obtained by analogous manipulations as discussed above.

%**********************************************************
\newsubsubsec\choices Other parameterization of cubic graphs

Note that the choice of loop momentum to parameterize the cubic graphs plays an important role due
to the inherent asymmetry of the numerators with respect to the label $1$ (which must always be associated with $V_P$).
The cases considered above are the ones which maximize the chances of failure. For instance, if we choose to position $\ell$ in
the edge between $3$ and $4$ in the graphs depicted in a) in the previous figure the resulting triplet of numerators
\medskip
\centerline{\hbox{
%\hbox to 10pt{}\raise40pt\hbox{a)}\hbox to 20pt{}%
\raise0pt\hbox{{\epsfxsize=0.18\hsize\epsfbox{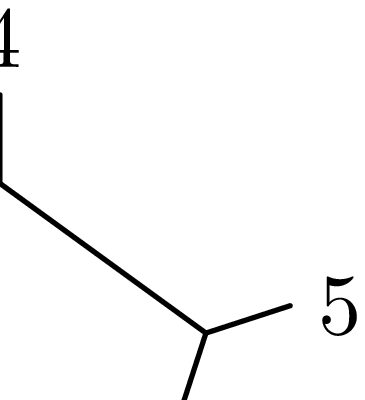}}}%
\hbox to 10pt{}\raise36pt\hbox{$\ - \ $}\hbox to 10pt{}%
\raise0pt\hbox{{\epsfxsize=0.18\hsize\epsfbox{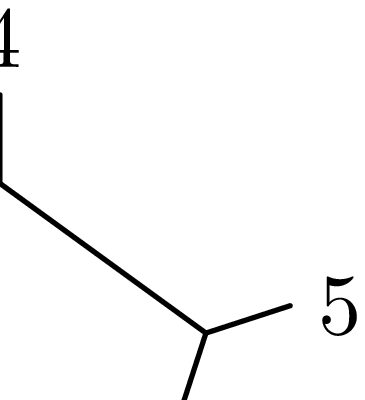}}}
\hbox to 10pt{}\raise36pt\hbox{$\ - \ $}\hbox to 10pt{}%
\raise-5pt\hbox{{\epsfxsize=0.16\hsize\epsfbox{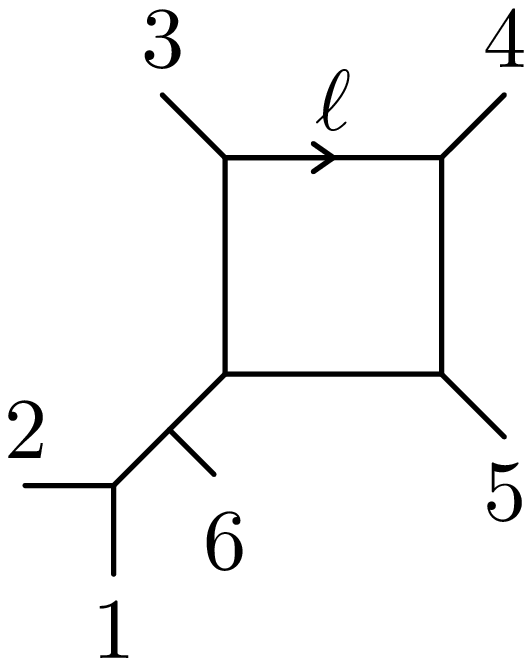}}}
%\hbox to 5pt{}\raise40pt\hbox{$\ = \ 0$}\hbox to 5pt{}
}}
\smallskip\noindent is easily
seen to satisfy the color-kinematics identity. In this case we get,
\eqn\trivsat{
N_{4|5,6,12,3}(\ell) - N_{4|5,12,6,3}(\ell) - N_{4|5,[6,12],3}(\ell) = 0\,.
}
To see this it is enough to use the pentagon \genPent\ to obtain
\eqnn\Nssee
$$\eqalignno{
N_{4|5,6,12,3}(\ell) & =
  V_{12} T^m_{3 , 4 , 5 , 6}\ell_m
 -  {1 \over 2}\,\Big[ V_{12} T_{34 , 5 , 6}
 + V_{12} T_{35 , 4 , 6}
 + V_{12} T_{36 , 4 , 5}&\Nssee\cr
& - V_{12} T_{45 , 3 , 6}
 - V_{12} T_{46 , 3 , 5}
 - V_{12} T_{56 , 3 , 4}
 - V_{123} T_{4 , 5 , 6}
 + V_{124} T_{3 , 5 , 6}\cr
& + V_{125} T_{3 , 4 , 6}
 + V_{126} T_{3 , 4 , 5}\Big]\cr
N_{4|5,12,6,3}(\ell) & =
   V_{12} T^m_{3 , 4 , 5 , 6}\ell_m
 -  {1 \over 2}\, \Big[V_{12} T_{34 , 5 , 6}
 + V_{12} T_{35 , 4 , 6}
 + V_{12} T_{36 , 4 , 5}\cr
& - V_{12} T_{45 , 3 , 6}
 - V_{12} T_{46 , 3 , 5}
 - V_{12} T_{56 , 3 , 4}
 - V_{123} T_{4 , 5 , 6}
 + V_{124} T_{3 , 5 , 6}\cr
& + V_{125} T_{3 , 4 , 6}
 - V_{126} T_{3 , 4 , 5}\Big]\,,
}$$
from which we get $N_{4|5,6,12,3}(\ell) - N_{4|5,12,6,3}(\ell)=-V_{126}T_{3,4,5}$ and \trivsat\ is satisfied since
$N_{4|5,[6,12],3} = V_{[6,12]}T_{3,4,5}=-V_{126}T_{3,4,5}$.

Thus we conclude that the field-theory limit of the genus-one six-point string correlator \SixPointCorr\ for
various color orderings as dictated by the ordering of vertex operators on the boundary of the Riemann surface
satisfies all the color-kinematics identities.

%******************************
\newsubsec\SevenPt Seven points

At seven points, BCJ relations are analogously satisfied. Given their significantly more
complex structure, we will not demonstrate these explicitly here and we will only outline their
construction below.

As alluded to earlier, at seven points there is an extra complication that must be dealt with: the
refined superfields.
To find the field theory limits of the refined terms, we have to use an alternative
method and partially integrate the worldsheet functions against the Koba-Nielsen
factor. This then means that, when we want to verify BCJ relations, we must rearrange these
refined terms.
For relations in which the loop momentum structure is unchanged between
terms (that is, BCJ relations in which there is always momentum $\ell$ going into leg
$1$), this amounts to canceling all $(\ell\cdot k)$ terms. Take for instance the relation
\eqn\SevenPtJeg{
N_{1|2,3,4,5,6,7}(\ell)-N_{1|2,4,3,5,6,7}(\ell)-N_{1|2,34,5,6,7}(\ell)=0\,,
}
and consider the refined terms
$V_1J_{34|2,5,6,7}$
within it. In the standard ordering correlator, these terms are associated with the
worldsheet function
${\cal Z}_{34|1,2,5,6,7}$
and we would therefore expect the heptagon numerator $N_{1|2,3,4,5,6,7}(\ell)$ to contain
the terms
\eqn\JegNumeratorOne{
-\frac{1}{12}V_1J_{34|2,5,6,7}\Big(\ell\cdot k_4 - \frac{1}{2}k^{12}\cdot k^4 + \frac{1}{2}k^4\cdot k^{567}\Big)\,.
}
Likewise, the other numerators we would expect to contain the terms
\eqnn\likew
$$\eqalignno{
N_{1|2,4,3,5,6,7}(\ell)&\leftrightarrow -\frac{1}{12}V_1J_{43|2,5,6,7}\Big(\ell\cdot k_3 - \frac{1}{2}k^{12}\cdot k^3 +
\frac{1}{2}k^3\cdot k^{567}\Big)&\likew\cr
N_{1|2,34,5,6,7}(\ell)&\leftrightarrow0\,.
}$$
The relation \SevenPtJeg\ is clearly not satisfied with these values.

Instead, we cancel the $\ell\cdot k$ terms. For example, we rewrite \JegNumeratorOne\ as
\eqnn\SevenPtEgHeptCancel
$$\eqalignno{
&-\frac{1}{12}V_1J_{34|2,5,6,7}\Big(\frac{1}{2}(\ell-k^{123})^2-\frac{1}{2}(\ell-k^{1234})^2+k^{123}\cdot k^4 -
\frac{1}{2}k^{12}\cdot k^4 + \frac{1}{2}k^4\cdot k^{567}\Big)\cr
=&-\frac{1}{12}V_1J_{34|2,5,6,7}\Big(\frac{1}{2}(\ell-k^{123})^2-\frac{1}{2}(\ell-k^{1234})^2+\frac{1}{2}k^{3}\cdot k^4\Big)\,.&\SevenPtEgHeptCancel
}$$
We then cancel the $(\ell-k)^2$ terms with the denominator of the Feynman loop integrand $I_{1,2,3,4,5,6,7}(\ell)$
associated with this term, and so they contribute to hexagons instead. Hence 
there is only one term of this form associated with the heptagon,
\eqnn\SevenPTBCJNumeratorOne
$$\eqalignno{
&N_{1|2,3,4,5,6,7}(\ell)\leftrightarrow -\frac{1}{24}s_{34}V_1J_{34|2,5,6,7}\,.&\SevenPTBCJNumeratorOne
}$$
Similarly, the other heptagon numerator undergoes this procedure and is associated
with
\eqn\SevenPTBCJNumeratorTwo{
N_{1|2,4,3,5,6,7}(\ell)\leftrightarrow \frac{1}{24}s_{34}V_1J_{34|2,5,6,7}\,.
}
There are then extra terms in the hexagons arising from the canceled portion of the
terms from the heptagons. 
The $34$-hexagon we are interested in inherits a term from the cancellation
\SevenPtEgHeptCancel . Hence we now have
\eqn\SevenPTBCJNumeratorThree{
N_{1|2,34,5,6,7}(\ell)\leftrightarrow -\frac{1}{12}s_{34}V_1J_{34|2,5,6,7}\,.
}
Note this differs from what may be naively expected from \SevenPtEgHeptCancel\ due
to the hexagon containing an extra $2s_{34}$ in its denominator compared with the
heptagon.
Now plugging \SevenPTBCJNumeratorOne , \SevenPTBCJNumeratorTwo , \SevenPTBCJNumeratorThree\ into the relation \SevenPtJeg\  we see it is now satisfied
\eqn\sucbcj{
-\frac{1}{24}s_{34}V_1J_{34|2,5,6,7}
-\frac{1}{24}s_{34}V_1J_{34|2,5,6,7}
-\Big(-\frac{1}{12}s_{34}V_1J_{34|2,5,6,7}\Big)=0\,.
}
Similar manipulations hold for other BCJ relations of this sort. Additional complications arise when
the BCJ relations involve terms of different loop momentum structure, and we have yet to
identify a general algorithm for these situations. However, by explicitly rearranging
amplitudes term by term, we have been able to structure them so that they satisfy all of
the BCJ relations we have tested. Namely, we have been able to simultaneously satisfy
the following more complex relations
\eqnn\sevbcj
$$\eqalignno{
N_{1|2,3,4,5,6,7}(\ell)-N^{a_7=1}_{1|7,2,3,4,5,6}(\ell)-N_{[7,1]|2,3,4,5,6}(\ell)&=0\,,&\sevbcj\cr
N_{1|2,3,4,5,7,6}(\ell)-N^{a_6=1}_{1|6,2,3,4,5,7}(\ell)-N_{[1,6]|2,3,4,5,7}(\ell)&=0\,,\cr
N_{[7,1]|2,3,4,5,6}(\ell)-N^{a_6=1}_{[7,1]|6,2,3,4,5}(\ell)-N_{[6,[7,1]]|2,3,4,5}&=0\,,\cr
N_{[6,[7,1]]|2,3,4,5}(\ell)-N^{a_5=1}_{[6,[7,1]]|5,2,3,4}(\ell)-N_{[5,[6,[7,1]]]|2,3,4}(\ell)&=0\,,\cr
N^{a_6=1}_{[1,6]|2,3,4,5,7}(\ell)+N_{[6,1]|2,3,4,5,7}&=0\,,\cr
N_{[7,1]|2,3,4,5,6}(\ell)+N^{a_7=1}_{[1,7]|2,3,4,5,6}&=0\,.
}$$
Though this is not an exhaustive test, we hope that it is sufficient to serve as a proof
of concept that this method work, and that it should always be possible to rearrange
the refined terms to satisfy the color-kinematics duality.

%**********************************************
\newsubsec\sugrafailsec  Supergravity amplitudes and the double copy

One of the goals in obtaining a parameterization of gauge theory 1-loop integrands
that satisfies the
color-kinematics duality is to construct corresponding supergravity integrands via the double-copy construction \bcjreview.
For five points this construction was carried out explicitly in four dimensions in \CJfive\ while the
ten-dimensional analysis using pure spinor superspace was done in \towardsOne.
In the pure spinor superspace setup, the supergravity
integrand obtained via the double copy must be checked to be BRST invariant, as that guarantees gauge and
supersymmetry invariance of its component expression in terms of polarizations and momenta \psf.

We will now repeat the five-point supergravity construction of \towardsOne\ to highlight that it
is BRST invariant but that it is so only because the numerators satisfy the dihedral symmetries of the
cubic graphs in the cohomology of pure spinor superspace (see \CJfive\ for a discussion of these symmetries).
While at five points our numerators satisfy these symmetries in addition to the color Jacobi identities, the corresponding
symmetries at six points are not satisfied by our
BCJ-satisfying six-point numerators and will prevent the double-copy construction of a BRST-closed
supergravity integrand. Applying the double-copy procedure at six points will be left for a future work.

%*****************************************************************
\newsubsubsec\FivePtSugra The five-point supergravity integrand

Let us construct the five-point supergravity integrand using the double-copy procedure
to highlight the existence of a subtlety:
the consistency of the double-copy construction requires the five-point
numerators not only to satisfy the kinematic Jacobi identities but also the
dihedral symmetries of the cubic graphs. We will see that these symmetries, unlike the
kinematic Jacobi identities, are
satisfied in the cohomology rather than identically.

Starting with the color-dressed integrand \FivePtColDressed\
we replace the color factors by an extra copy of duality-satisfying kinematic numerators. This yields
\eqnn\fiveptsugra
$$\eqalignno{
M_5(\ell)
&= \Big(\half \cN_{1|2,3,45} I_{1,2,3,45} \tilde N_{1|2,3,45}
          + \half \cN_{1|2,34,5} I_{1,2,34,5} \tilde N_{1|2,34,5}&\fiveptsugra\cr
&          + \half \cN_{1|23,4,5} I_{1,23,4,5} \tilde N_{1|23,4,5}
          + \half \cN_{12|3,4,5} I_{12,3,4,5} \tilde N_{12|3,4,5}\cr
&          + \half \cN_{51|2,3,4} I_{51,2,3,4} \tilde N_{51|2,3,4}
          + \cN_{1|2,3,4,5}(\ell) I_{1,2,3,4,5} \tilde N_{1|2,3,4,5}(\ell)
         + \perm(2,3,4,5)\Big)
}$$
Note that the kinematic numerators on the left are written in terms of Berends-Giele numerators $\cN$
of the appendix~\Convapp\ while those on the right are the local numerators $N$.

After setting up the double-copy supergravity integrand \fiveptsugra\ we must check its BRST variation.
Since \fiveptsugra\ is left/right symmetric\foot{The left- or right-moving
terminology refers to the two sides of the double-copy kinematic factors, distinguished by the tildes.}
it is enough to consider the left-moving BRST variation, which we will see
vanishes only if the right-movers are
in the cohomology of the right-moving
pure spinor superspace. To see this surprising fact, consider the variation of the left-moving
pentagon $N_{1|2,3,4,5}(\ell)$ multiplied by the loop-momentum integrand $I_{1,2,3,4,5}$:
\eqnn\QNfive
$$\eqalignno{
QN_{1|2,3,4,5}(\ell)I_{1,2,3,4,5} &=
{1\over 2} V_1 V_2 T_{3,4,5} \big[ (\ell-k_{12})^2 - (\ell-k_1)^2 \big]I_{1,2,3,4,5}&\QNfive\cr
& +{1\over 2} V_1 V_3 T_{2,4,5} \big[ (\ell-k_{123})^2 - (\ell-k_{12})^2 \big]I_{1,2,3,4,5}\cr
& + {1\over 2} V_1 V_4 T_{2,3,5} \big[ (\ell-k_{1234})^2 - (\ell-k_{123})^2 \big]I_{1,2,3,4,5}\cr
& +{1\over 2} V_1 V_5 T_{2,3,4} \big[\ell^2 - (\ell-k_{1234})^2 \big]I_{1,2,3,4,5}\cr
%%%
%%%
&={1\over 2} V_1 V_2 T_{3,4,5} \big[ I_{1,23,4,5} - I_{12,3,4,5}\big]
 +{1\over 2} V_1 V_3 T_{2,4,5} \big[ I_{1,2,34,5} - I_{1,23,4,5}\big]\cr
& + {1\over 2} V_1 V_4 T_{2,3,5} \big[ I_{1,2,3,45} - I_{1,2,34,5}\big]
 +{1\over 2} V_1 V_5 T_{2,3,4} \big[ I_{1,2,3,4} - I_{1,2,3,45}\big]
}$$
where we used identities such as $(\ell-k_1)^2I_{1,2,3,4,5} = I_{12,3,4,5}$ that follow from \DefI. These loop-momentum
identities are trivial but one of them on the last line, namely $\ell^2I_{1,2,3,4,5} = I_{1,2,3,4}$,
has a peculiar behavior:
the right-hand side has no label $5$.
This seemingly innocuous fact will have a surprising implication in the double-copy
construction of the five-point supergravity integrand when \QNfive\ appears multiplied by a right-moving factor $\tilde
N_{1|2,3,4,5}(\ell)$.

The reason is that the right-moving pentagon $\tilde N_{1|2,3,4,5}(\ell)$ depends on the loop momentum and
picks up the shift\foot{In the gauge-theory integrand the term $V_1V_5 T_{2,3,4}I_{1,2,3,4}$ from the last line
of \QNfive\ can be trivially rewritten as $V_1V_5 T_{2,3,4}I_{51,2,3,4}$ since its kinematic factor
is invariant under the shift $\ell\to\ell-k_5$.}
$\ell\to\ell-k_5$ needed when rewriting $I_{1,2,3,4}\to I_{51,2,3,4}$.
More explicitly, one can show that the BRST variation of \fiveptsugra\ contains
\eqnn\shiftedell
$$\eqalignno{
QM_5(\ell) &= \ldots  + \half V_1V_5T_{2,3,4}\big[ I_{1,2,3,4}\tilde N_{1|2,3,4,5}(\ell)
+ I_{51,2,3,4}(\tilde N_{15|2,3,4} - \tilde N_{1|5,2,3,4}(\ell))\big]\cr
&= \ldots  + \half V_1V_5T_{2,3,4}\,I_{51,2,3,4}\big[
\tilde N_{15|2,3,4}
+ \tilde N_{1|2,3,4,5}(\ell-k_5)
- \tilde N_{1|5,2,3,4}(\ell)
\big].\qquad{}&\shiftedell\cr
}$$
On the one hand we know from section~\BCJegs\ that the kinematic Jacobi identity 
\eqn\identic{
\tilde N_{5|1,2,3,4}(\ell) - \tilde N_{1|5,2,3,4}(\ell) + \tilde N_{15|2,3,4} = 0\,.
}
is satisfied\foot{These numerators are readily available to download from \website.}. Therefore
the vanishing of the left-moving BRST variation \shiftedell\ hinges on the dihedral symmetry of the pentagon
$\tilde N_{1|2,3,4,5}(\ell-k_5)=\tilde N_{5|1,2,3,4}(\ell)$. One can show that this symmetry is satisfied
{\it in the cohomology} of the right-moving pure spinor superspace given by the pure spinor
bracket
\eqn\intheco{
\langle \tilde N_{1|2,3,4,5}(\ell-k_5)\rangle =\langle\tilde N_{5|1,2,3,4}(\ell)\rangle\,,
}
where we emphasize that the above would not be true in terms of superfields, i.e. without the
pure spinor brackets.
To see this we use the numerators obtained from the field-theory limits to get
that $\langle \tilde N_{5|1,2,3,4}(\ell) - \tilde N_{1|2,3,4,5}(\ell-k_5)\rangle$ is given by
\eqn\trueid{
\langle \tilde V_1 \tilde T^m_{2,3,4,5} k_5^m
+ \tilde V_{51}\tilde T_{2,3,4} +
\big[\tilde V_1 \tilde T_{52,3,4} + (2\leftrightarrow3,4,5)\big]\rangle = 0\,,
}
as can be seen using the cohomology identity \bcjtrivv.

To summarize, the five-point supergravity integrand is BRST invariant.
But there is a subtlety: the double-copy construction seems to require more than just the kinematic
Jacobi identities, the numerators must also satisfy the dihedral symmetries of
the cubic graphs\foot{At tree level for the double copy construction of supergravity amplitudes
to be BRST invariant it is enough for the numerators to satisfy the kinematic Jacobi identities} (which
are satisfied
in the cohomology of the right-movers).

%***********************************************************************
\newsubsubsec\sixsugrasec Six-point double copy and automorphism symmetries

At six points a naive application of the double-copy procedure with BCJ-satisfying numerators
obtained in the previous sections does not produce a consistent supergravity integrand: it
fails to be BRST invariant in pure spinor superspace. This happens because the numerators, even though they satisfy the
color-kinematics duality they do not satisfy the automorphism symmetries of their associated cubic graphs.

To see this it is enough to use the BCJ-satisfying six-point numerators in
a tentative double-copy construction to obtain, among many others,
the following terms under a left-moving BRST variation $QM_6(\ell)$,
\eqnn\samplew
$$\eqalignno{
& - {1\over 4 s_{23}}V_1 V_{23}T_{4,5,6}\Big(
 I_{123,4,5,6}\tilde N_{1|23,4,5,6}(\ell)
- I_{1,4,5,6}\tilde N_{1|4,5,6,23}(\ell)
- I_{123,4,5,6}\tilde N_{[1,23]|4,5,6}\Big)\cr
&=- {1\over 4 s_{23}}V_1 V_{23}T_{4,5,6}I_{123,4,5,6}\Big(
\tilde N_{1|23,4,5,6}(\ell)
- \tilde N_{1|4,5,6,23}(\ell-k_{23})
+ \tilde N_{231|4,5,6}\Big)\,. &\samplew
}$$
Similarly as described in \QNfive\ at five points,
the missing labels in $I_{1,4,5,5}$ arise from loop-momentum cancellations in $QN_{1|4,5,6,23}(\ell)I_{1,4,5,6,23}$.
This is compensated by the shift $\ell\to\ell-k_{23}$
which is picked up by the right-moving pentagon in the second line.
If the condition  $\tilde N_{1|4,5,6,23}(\ell-k_{23}) = \tilde N_{23|1,4,5,6}(\ell)$ for the automorphism symmetry
of the pentagon was satisfied
then the terms \samplew\ would vanish identically since
\eqn\wehave{
 \tilde N_{1|23,4,5,6}(\ell)
- \tilde N_{23|1,4,5,6}(\ell)
+ \tilde N_{231|4,5,6} = 0\,,
}
as can be verified using
the numerators available to download from \website.
Unfortunately it is not true that $\tilde N_{1|4,5,6,23}(\ell-k_{23}) = \tilde N_{23|1,4,5,6}(\ell)$ and, unlike
the case at five points, this is not true even in the cohomology\foot{Note that the last line of \samplew\ is identical (apart
from the left/right-moving nature of the numerators)
to the BCJ-triplet failure in the representation of \towardsOne, given in equation (6.12) of that reference.
Unlike the representation of \towardsOne,
the six-point integrand of gauge theory found here satisfies all BCJ relations for the left- and right-moving
numerators. However, once terms in the left-moving BRST variation are collected
we see that the BCJ failure of \towardsOne\ in the left-moving sector appears here as
a failure in the right-moving sector due to a
shift of the loop momentum.},
\eqn\sixFAIL{
\langle \tilde N_{1|4,5,6,23}(\ell-k_{23})\rangle \neq \langle\tilde N_{23|1,4,5,6}(\ell)\rangle\,.
}
Therefore the naive application of the double-copy construction at six points is not consistent even though
the numerators satisfy the color-kinematics duality.

It is interesting to observe that the automorphism symmetries of the graphs encoded in the loop momentum shifts
$\ell+a_ik_i$ are satisfied
by the numerators from the integrands with shifted loop momentum $A(\sigma;\ell+a_ik_i)$. In the case of \sixFAIL\  we have
the identity (valid at the superfield level)
\eqn\sixPASS{
N^{a_2=-1,a_3=-1}_{1|4,5,6,23}(\ell) =N_{23|1,4,5,6}(\ell),
}
where the numerator on the left-hand side is the $23$-pentagon from the amplitude with shifted loop
momentum, $A(1,4,5,6,2,3;\ell-k_{23})$. This integrand is
computed with the field-theory limits of section~\FTgs\ with $a_2=a_3=-1$ corresponding
to the shifted loop momentum $\ell-k_2-k_3$.
Unfortunately it
is not clear how to use these numerators directly as functions of $\ell$ rather than as functions of the shift parameters $a_i$.

%**********************************************************************
\newsubsubsec\commentdoublesec Comments on the double-copy construction in pure spinor superspace

The failure of the automorphism symmetry \sixFAIL\ for the $23$-pentagon is a contact term in $s_{23}$ after its
component expansion is evaluated through the pure spinor bracket, that is
$\langle \tilde N_{1|4,5,6,23}(\ell-k_{23}) - \tilde N_{23|1,4,5,6}(\ell)\rangle \sim s_{23}( \ldots)$.
In pure spinor superspace we have
\eqn\same{
N_{23|1,4,5,6}(\ell) - N_{1|4,5,6,23}(\ell-k_{23}) = k^m_{23} V_1T^m_{23,4,5,6}
+ V_{231}T_{4,5,6} + \big[V_1T_{234,5,6} + 4\leftrightarrow5,6\big]
}
which represents the same failure to satisfy the color-kinematics duality as
pointed out in equation (6.12) of \towardsOne.

Since the issue with missing labels in the loop momentum integral as a result of a BRST variation
will always be present
for the BCJ-satisfying numerators obtained in this work, solving this problem seems to require
a different approach to the double-copy construction in the pure spinor superspace context.
Given that the failures are purely contact terms, the generalized double-copy
prescription of \genJac\ may be applicable\foot{We thank Oliver Schlotterer for discussions on this point.}
and it will be interesting to see how BRST invariance is restored.
It is reasonable to speculate that the deformations of the right-moving BCJ triplets by contact terms
as a result of loop momentum shifts due to canceled loop propagators in the left-moving BRST variation
may be a generic feature of the double copy in pure spinor superspace.
If true, the generalized double-copy formalism may be
the norm by which gravity integrands are generated from gauge-theory integrands; a tree-level manifestation
of this behavior was anticipated in \genvanhove.
%This characteristic may
%be especially important at higher loops.
We plan to investigate this problem in future work.

We note that supergravity integrands have been constructed using BCJ numerators
in four dimensions for up to seven points in \MonteiroOx\ and to all multiplicity in \stringinspired\ using
spinor helicity in the MHV sector.
Supergravity amplitudes were also constructed in
\olipartial\ but
using a partial-fraction representation of the loop momentum integrands.

%*******************************
\newnewsec\conculsion Conclusion

In this work we obtained a set of field-theory limit rules for the Kronecker-Eisenstein coefficient functions
present
in the genus-one superstring correlators derived in \refs{\oneloopI,\oneloopII,\oneloopIII}. Using these rules we
found local numerators for ten-dimensional SYM integrands at one loop for five, six and seven points
that satisfy the BCJ color-kinematics duality. These results resolve
the difficulties in an earlier analysis
of the six-point SYM integrands which did not satisfy the color-kinematics duality \towardsOne.

These field-theory limits have an special affinity
with the
pure spinor superspace representation of the superstring correlators. They take into
account arbitrary choices in the parameterization of the loop momentum integrands,
shuffling terms among various numerators preserving BRST invariance
of the SYM one-loop integrands while changing the BRST properties of individual
numerators in a non-trivial way, see the discussion around \notb.
The prescription to find the field-theory limit of the correlator whose parameterization
contains shifts of the loop momentum by arbitrary linear combinations of external particle momenta 
is crucial in
demonstrating all the BCJ color-kinematic identities of our ten-dimensional SYM
representation.

However, in attempting to use the BCJ-satisfying six-point numerators in a double-copy construction
of the supergravity integrand we learned that the numerators must satisfy, in addition to the
kinematic Jacobi identities, also the various graph automorphism symmetries in order for the
supergravity integrand to be BRST invariant. Unfortunately our six-point numerators viewed as functions of
the loop momentum (rather than as the numerators from integrands with general loop momentum as described
at the end of section~\sixsugrasec)
do not
satisfy these symmetries and the double-copy construction initiated here remains incomplete.
However, the contact-term nature of the automorphism symmetry failure indicates that the generalized
double-copy prescription of \genJac\ may resolve this.
We defer the full analysis of this problem to future work.

\bigskip
\noindent{\bf Acknowledgements:} We thank Oliver Schlotterer for discussions and helpful comments on the draft.
EB thanks Kostas Skenderis for useful discussions.
CRM thanks Oliver Schlotterer for collaboration on closely related topics.
CRM is supported by a University Research Fellowship from the Royal Society.

\appendix{A}{Conventions}
\applab\Convapp

\noindent In this appendix we briefly summarize some of the conventions used in the main text.

Sums over deconcatenations are denoted by
$\sum_{A_1...A_n=a_1...a_m}$
They represent the sum over all possible ways of
generating $n$ words from $a_1...a_m$, while maintaining the order.
These words may be empty, but often when they are the terms being summed over will be
zero. So, to give an example, the sum $\sum_{ABC=12}$ denotes the sum over six cases;
three of them are where two of $A$, $B$ and $C$ are empty and the third is $12$, and
the other three are where $A=1, B=2, C=\emptyset$, $A=1,B=\emptyset,C=2$, and $A=\emptyset, B=1,C=2$.

Another notation commonly used is
\eqn\distrib{
(terms)+(a_1,...,a_m\vert N_1,...,N_n),\qquad{}m\leq n.
}
This notation works means a sum over all possible
ways of replacing $a_1,...,a_n$ in the terms with $n$ terms from the ordered
list $N_1,...,N_n$. Further generalizations of this follow naturally,
with $($terms$)+(a_1,...,a_{m_1}\vert b_1,...,b_{m_2}\vert N_1,...,N_n)$ meaning sum
over all ways of generating two ordered lists from $N_1,...,N_n$, one of length $m_1$,
one of length $m_2$, and substituting them in for $a_1,...,a_{m_1}$ and $b_1,...,b_{m_2}$.
For example, in
$V_{[1,23]}T_{[4,56],7,8} + (23\vert 4,56\vert 23,4,56,7,8)$ possible terms are
$V_{[1,4]}T_{[23,56],7,8}$ and $V_{[1,23]}T_{[7,8],4,56}$, but not
$V_{[1,23]}T_{[8,7],4,56}$ as the latter would violate the ordering constraint.

Another summation notation to note is
\eqn\stirnot{
(terms)+[1...n\vert A_1,...,A_m],\qquad{}m\leq n.
}
This denotes the sum over $A_1$, ..., $A_m$ all possible Stirling cycle permutations constructed from $1, ..., n$ \oneloopIII. This means
that you take the set of numbers $1$, ..., $n$, and construct all possible permutation cycles from it, select those involving
$m$ brackets, and canonicalise by having the first term in each cycle be its lowest element, and the cycles ordered by their
lowest elements. Each cycle is then substituted in for an A.
For example, consider the sum $+[1234567|A_1,...,A_4]$. One possible permutation of $1, \ldots, 5$
involving $4$ brackets would be $(12)(64)(3)(57)$, which swaps $1$ with $2$, $6$ with $4$, and $5$ with $7$. We then
begin canonicalising by using that permutation cycles have cyclic symmetry to rewrite this as $(12)(46)(3)(57)$, and then order
the cycles by their lowest values, $(12)(3)(46)(57)$.  Hence, one term in this sum would set $A_1=12$, $A_2=3$, $A_3=46$,
$A_4=57$.
These sums may be thought of as being $A_1=1$ followed by any terms from $2...n$ in any order, then $A_2$ is the next lowest
value left followed by any possible set of values in any order from the numbers left, and so on. So in the above example,
$A_1=15$ would be a possible term, which would mean $A_2$ starting with a $2$ and so it could be $A_2=23$, then $A_3$ starts
with a 4 and so we could have $A_3=4$, and then finally $A_4$ follows the same rules and uses up all remaining letters, so
$A_4=67$.

\newsubsubsec\OtherNotation Lie algebra notation and Berends-Giele currents

We frequently use the notation of words and Lie brackets, especially when indexing SYM multiparticle superfields, see the
discussion on section~3 of \oneloopI.
In any situation where a Lie bracket would be expected but a word $A$ is found instead,
this should be regarded as being the left-to-right Dynkin bracket $\ell(A)$ \Reutenauer,
\eqn\DynkinDef{
\ell(a_1...a_n)\equiv [[...[[a_1,a_2],a_3]...],a_n]\,.
}
For example, $[[[1,23],45],678]$ is interpreted as $[[[1,[2,3]],[4,5]],[[6,7],8]]$ and vice-versa.

A mapping from words to Lie brackets which will be particularly useful is the b-map defined by \PScomb\foot{Note the extra
factor of $\half$ in \bMap\ compared to the definition in \PScomb. This convention leads to local BCJ numerators
which are correctly normalized.}
\eqn\bMap{
b(i)=i,\qquad{}b(P)=\frac{1}{2s_P}\sum_{XY=P}[b(X),b(Y)].
}
For example, $b(12)=\frac{1}{2s_{12}}[1,2]$, and
$b(123)=\frac{1}{4s_{12}s_{123}}[[1,2],3]+\frac{1}{4s_{23}s_{123}}[1,[2,3]]$.

Superfields are described in terms of two broad classes of objects. The first are local and denoted by $V$, $T$, $J$, and $N$.
The composition of the first three of these objects can be found in more detail in \refs{\EOMbbs,\partI}.
The fourth will be used to refer to amplitude numerators and are detailed on a case by case basis.
These objects have a number of slots for indices labelling their
superfield contents, and all such indices will be Lie brackets.
The second class of objects are Berends-Giele (BG) currents. These are related to the
local objects previously described through the use of the b-map on each of their blocks of indices.
The BG current of particular use to us is denoted by ${\cal N}$, defined in terms of local objects $N$ as
\eqn\BGcurrentDefs{
{\cal N}_{A_1\vert A_2,...,A_m}(\ell)\equiv N^{(m)}_{b(A_1)\vert b(A_2),...,b(A_m)}(\ell)
}
For example, a seven-point box Berends-Giele numerator is expanded as
\eqnn\BGcurrentExamples
$$\eqalignno{
{\cal N}_{1|23,456,7}(\ell)&=N_{b(1)|b(23),b(456),b(7)}(\ell)\cr
&=\frac{1}{s_{23}s_{456}}\left(\frac{1}{s_{45}}N_{1|[2,3],[[4,5],6],7}(\ell)+\frac{1}{s_{56}}N_{1|[2,3],[4,[5,6]],7}(\ell)\right).
}$$
It should be noted that generalized Mandelstam invariants are defined with a $\frac{1}{2}$ factor,
\eqn\MandelstamDef{
s_{i_1...i_n}\equiv\frac{1}{2}(k^m_{i_1}+...+k^m_{i_n})^2=\sum_{1\leq a<b\leq n}k_{i_a}\cdot k_{i_b}
}

%************************************************************
\appendix{B}{Cyclic symmetry of the field-theory limit rules}
\applab\SixPointGeneral

\noindent In this appendix we will show that the definitions for the field theory limits we have
given yield the cyclic symmetry relations seen in \CycNotationSym
\eqnn\CycSymRelations
$$\eqalignno{
A(1,2,...,n;\ell+\Sigma_ia_ik_i)&=A(2,3,...,n,1;\ell-k_1+\Sigma_ia_ik_i)&\CycSymRelations\cr
%&=A(n,1,2,...,n-1;\ell-k_{12...(n-1)}+\Sigma_ia_ik_i)\,.
}$$
We refer to terms from $A(1,2,...,n;\ell+\Sigma_ia_ik_i)$ with a (I), and 
$A(2,3,...,n,1;\ell-k_1+\Sigma_ia_ik_i)$ with a (II).  

First, we compare their $b^{(p)}_{ij}$ terms. 
We restrict ourselves to the limit of a single Kronecker-Eisenstein coefficient function, as the limits
of their products are the natural generalization of this and will
follow accordingly.
Referring to \xDef , and using the notation $a_{ji}:=a_j-a_i$, we see that they differ by
\eqnn\bDiffOne
$$\eqalignno{
b^{I\;(p)}_{ij}-&b^{II\;(p)}_{ij}
=\sum_{m=0}^p\Big(\left(\sgn^{12...n}_{ij}\right)^{m}\frac{B_{m}a_{ji}^{p-m}}{m!(p-m)!}
-\left(\sgn^{23...n1}_{ij}\right)^{m}\frac{B_{m}(a_{ji}+\delta_{j1}-\delta_{i1})^{p-m}}{m!(p-m)!}\Big)\cr
&=
\sum_{m=0}^p\frac{B_m}{m!(p-m)!}\Big(\left(\sgn^{12...n}_{ij}\right)^{m}a_{ji}^{p-m}
-\left(\sgn^{23...n1}_{ij}\right)^{m}(a_{ji}-\delta_{j1}+\delta_{i1})^{p-m}\Big)&\bDiffOne
}$$
Clearly in all cases where neither of $i$ or $j$ is $1$ this vanishes. If we suppose
$i=1$, the first $sgn$ function is $1$, and the second is $-1$. Hence this difference
becomes
\eqnn\bDiffTwo
$$\eqalignno{
b^{I\;(p)}_{ij}-b^{II\;(p)}_{ij}&=
\sum_{m=0}^p\frac{B_m}{m!(p-m)!}\Big(a_{j1}^{p-m}
-\left(-1\right)^{m}(a_{j1}+1)^{p-m}\Big)&\bDiffTwo
}$$
This can be shown to vanish.
Taking for instance the $p=3$ case, we have
\eqnn\bDiffThree
$$\eqalignno{
b^{I\;(3)}_{ij}-b^{II\;(3)}_{ij}&=
\frac{B_0}{6}\Big(a_{j1}^{3}-\left(-1\right)^{0}(a_{j1}+1)^{3}\Big)
+\frac{B_1}{2}\Big(a_{j1}^{2}-\left(-1\right)^{1}(a_{j1}+1)^{2}\Big)\cr
&+\frac{B_2}{2}\Big(a_{j1}^{1}-\left(-1\right)^{2}(a_{j1}+1)^{1}\Big)
+\frac{B_3}{6}\Big(a_{j1}^{0}-\left(-1\right)^{3}(a_{j1}+1)^{0}\Big)\cr
&=\frac{1}{6}\Big(a_{j1}^{3}-a_{j1}^3-3a_{j1}^2-3a_{j1}-1\Big)
+\frac{1}{4}\Big(a_{j1}^{2}+a_{j1}^2+2a_{j1}+1\Big)\cr
&+\frac{1}{12}\Big(a_{j1}-a_{j1}-1\Big)
+0 =0\,.&\bDiffThree
}$$
To show that \bDiffTwo\ vanishes in general we expand the bracket $(a_{j1}+1)^{p-m}$,
\eqnn\bDiffExpand
$$\eqalignno{
b^{I\;(p)}_{ij}-b^{II\;(p)}_{ij}&=
\sum_{m=0}^p\frac{B_m}{m!(p-m)!}\Big(a_{j1}^{p-m}
-\left(-1\right)^{m}\sum_{n=0}^{p-m}{p-m \choose n}a_{j1}^n\Big)&\bDiffExpand\cr
&=\sum_{m=0}^p\sum_{n=0}^{p-m-1}\frac{-(-1)^mB_ma_{j1}^n}{m!n!(p-m-n)!}
+\sum_{m=0}^p\frac{B_m(1-(-1)^m)a_{j1}^{p-m}}{m!(p-m)!}\,,
}$$
where we have separated out the terms of order $(p-m)$ in the second line.
In the right hand terms of the above, $(1-(-1)^m)$ vanishes when $m$ is even, and $B_m$ vanishes when $m$ is odd and not $1$.
Hence, this summation reduces to a single term,
\eqn\RHStermsInBdiff{
\frac{2B_1a_{j1}^{p-1}}{1!(p-1)!}=\frac{a_{j1}^{p-1}}{(p-1)!}\,.
}
We then turn to the left hand terms of \bDiffExpand .
Reordering the double summation, these have the form
\eqn\LHStermsInBdiff{
\sum_{n=0}^p\sum_{m=0}^{p-n-1}\frac{-(-1)^mB_m}{m!n!(p-m-n)!}a_{j1}^n
}
We may then use a known identity of Bernoulli numbers \FunctionsBook \foot{Note the definition of the Bernoulli number $B_n$ used in this source differs from that of this paper by a factor $(-1)^n$},
\eqn\BernoulliSum{
\sum_{k=0}^{n-1}{n\choose k}(-1)^kB_k=\delta_{(n-1),0}\,,
} 
to simplify the form of \LHStermsInBdiff\ to
\eqn\LHStermsInBdiffTwo{
\sum_{n=0}^p\frac{-a_{j1}^n\delta_{p-n-1,0}}{n!(p-n-1)!}
=\frac{-a_{j1}^{p-1}}{(p-1)!}\,.
}
The two summations in \bDiffExpand\ therefore reduce to \RHStermsInBdiff\ and \LHStermsInBdiff , which cancel each other and thus this difference vanishes.
Similar will hold if we instead take $j=1$ in \bDiffOne .
 Hence, the $b$ part of the field theory limits matches in both representations.

Then, we move onto the $c$ piece. This difference is given by
\eqnn\bDiffFour
$$\eqalignno{
c^{I\;(p)}_{ij}-c^{II\;(p)}_{ij}=\frac{1}{2(p-1)!}\Big(&
\big(a_{ji}+\sgn^{12...n}_{ij}\dist_4^{12...n}(i,j)\big)^{p-1}&\bDiffFour\cr
-&\big(a_{ji}-\delta_{j1}+\delta_{i1}+\sgn^{23...n1}_{ij}\dist_4^{23...n1}(i,j)\big)^{p-1}\Big)
}$$
Again, we need only consider the cases where one of $i$ and $j$ is $1$. If we take $i=1$ we get
\eqnn\cDiffOne
$$\eqalignno{
c^{I\;(p)}_{ij}-c^{II\;(p)}_{ij}=&\frac{
\big(a_{j1}+\dist_4^{12...n}(1,j)\big)^{p-1}-\big(a_{j1}+1-\dist_4^{23...n1}(1,j)\big)^{p-1}}{2(p-1)!}&\cDiffOne
}$$
We now consider the two pieces of the numerator, and see that these are given by
\eqnn\DistSgnCompare
$$\eqalignno{
(a_{j1}+\dist_4^{12...n}(1,j))^{p-1}
&=\cases{
a_{j1}^{p-1}&$j\leq 4$\cr
(a_{j1}+1)^{p-1}&$j>4$
}\,,&\DistSgnCompare
\cr
(a_{j1}+1-\dist_4^{23...n1}(1,j))^{p-1}
&=\cases{
a_{j1}^{p-1}&$j\leq n-2$\cr
(a_{j1}+1)^{p-1}&$j>n-2$
}
\,.
}$$
When $n=4,5$, the only Kronecker-Eisenstein functions in amplitudes is $g^{(1)}_{ij}$, and we see that setting
$p=1$ in the above gives equivalence. When $n=6$, these coincide in that $n-2=4$.
When $n=7$ and $p>1$, they differ when $j=5$. However, this disagreement will not matter.
At $7$ points a term $g^{(2+)}_{15}$ is multiplied by at most one other $g^{(q)}_{ij}$ function, but we
need at least two Kronecker-Eisenstein coefficient functions in order to make the corresponding $P$ function non-zero.
That is, for example,
\eqnn\DistFailEg
$$\eqalignno{
g^{(2)}_{15}g^{(1)}_{56}\Rightarrow P(15,56)&=\phi_{156|5671}I_{156}=0\,,&\DistFailEg\cr
g^{(2)}_{15}g^{(1)}_{56}g^{(1)}_{67}\Rightarrow P(15,56,67)&=\phi_{1567|5671}I_{5671}\neq 0\,.
}$$
At $8$ points, this will of course become an issue. However, the description of
the $dist$ function was chosen purely for simplicity. 
If we instead think of this function as asking whether the pole being approached crosses
the boundary between particles $n$ and $1$, then consistency should be maintained to
higher points.

%****************************************************
\appendix{C}{The field-theory limit at higher points}
\applab\FThigher

\noindent We anticipate that the field theory limit rules for an arbitrary product of $g^{(n)}_{ij}$ functions
should generalize in the natural way
\eqnn\gGenProd
$$\eqalignno{
\prod_{a=1}^n g^{(p_a)}_{{i_a}{j_a}}&\rightarrow
\sum_{A\in {\cal P}(12...n)}\Bigg(\Big(\prod_{a\in A}b^{(p_{a})}_{{i_a}{j_a}}\Big)\Big(\prod_{b\in A^c}c^{(p_{b})}_{{i_b}{j_b}}\Big)P({{i_{B_1}}{j_{B_1}}},...,{{i_{B_{|B|}}}{j_{B_{|B|}}}})\Big)\Bigg)&\gGenProd
}$$
where ${\cal P}(12...n)$ denotes the power set of $12...n$, $A$ is an element of this,
and $A^c$ its complement.
We stress that the indices of the $c^{(p)}$ and those in the $P$ function are identical.

The general $P$ functions will be as in \Pdef , with $P(i_1j_1,...,i_nj_n)$ chaining
together $i_mj_m$ pairs as much as possible, and then using these as indices for $\phi$ 
and $I$ functions. So, for instance, we would expect
\eqnn\PgenEgs
$$\eqalignno{
P(12,23,34,45,56,67)&\leftrightarrow\hat\phi(\sigma|1234567)I_{1234567}&\PgenEgs\cr
P(15,32,56,24)&\leftrightarrow\hat\phi(\sigma|156)\hat\phi(\sigma|324)I_{156,324}
}$$
As for the limits of $b^{(p)}$ and $c^{(p)}$ at higher points, these we expect will
generalize from \gGenLimits\ in the natural way. As evidence of this, we look to the Fay
identity for $g_{12}^{(n)}g_{23}^{(1)}$
\eqn\FayOneN{
g^{(n)}_{12}g^{(1)}_{23}=-g_{13}^{(n+1)}+g_{13}^{(1)}g_{12}^{(n)}-ng_{12}^{(n+1)}+\sum_{j=0}^n(-1)^jg_{13}^{(n-j)}g_{23}^{(1+j)}\,.
}
We begin by looking at $b^{(n)}$, and restrict ourselves to the case $a_i=0$ $\forall\;i$
initially. In these 
circumstances we know that $b^{(1)}_{ij}=\frac{1}{2}\sgn_{ij}^{12...n}$, and we
would expect the general order $b^{(n)}_{ij}$ to depend only upon the order of
$i$ and $j$ with respect to
the color ordering. Hence, we substitute into \FayOneN\ the values
\eqn\gValuesSubIn{
g^{(1)}_{13},g^{(1)}_{23}\rightarrow\frac{1}{2}\,,\qquad{}
g^{(n)}_{12},g^{(n)}_{13},g^{(n)}_{23}\rightarrow b^{(n)}\,.
}
Upon rearranging this gives us the recursion relation
\eqn\gNplusOne{
b^{(n+1)}=-\frac{1}{n+1-(-1)^n}\sum_{j=1}^n(-1)^jb^{(n-j+1)}b^{(j)}\,.
}
This can be seen to vanish for $n$ even, $n>0$, by virtue of the symmetry in the $gg$ terms
and the antisymmetry of the $(-1)^j$.
For $n$ odd, it simplifies to
\eqn\bEvenReccursion{
b^{(2n)}=-\frac{1}{2n+1}\sum_{j=1}^{2n-1}(-1)^jb^{(2n-j)}b^{(j)}
=-\frac{1}{2n+1}\sum_{j=1}^{n-1}b^{(2n-2j)}b^{(2j)}\,,
}
where the second equality follows from the vanishing of the $b$ with odd indices.
It may then be proved by induction that this is solved by
\eqn\xGenForm{
b^{(n)}= \frac{B_n}{n!}\,,
}
where $B_n$ is the $n^{th}$ Bernoulli number.
Showing this requires an identity due to Euler \eulermaster,
\eqn\BernoulliIdentity{
\sum_{k=1}^{n-1}{2n\choose 2k}B_{2k}B_{2n-2k}=-(2n+1)B_{2n}\,,\quad{}n\geq 2\,.
}
Hence, we speculate that when $a_i=0$ $\forall i$, the field theory limit of a
general term from the Kronecker-Eisenstein series away from poles is given by \xGenForm.
The first few (non-zero) values are
\eqnn\xFirstCases
$$\eqalignno{
&b^{(0)}=1\,,\quad{}
b^{(1)}=\frac{1}{2}\,,\quad{}
b^{(2)}=\frac{1}{12}\,,\quad{}
b^{(4)}=-\frac{1}{720}\,,\quad{}
b^{(6)}= \frac{1}{30240}\,,&\xFirstCases\cr
&b^{(8)}= -\frac{1}{1209600}\,,\quad{}
b^{(10)}= \frac{1}{47900160}\,,\quad{}
b^{(12)}= -\frac{691}{1307674368000}\,.
}$$
We can then extend this to the general $a_i$ case, though with less elegance. If instead
of making the substitution \gValuesSubIn\ into \FayOneN , we instead use the
general $a_i$ values of the $b^{(1)}$ terms, we find the relation
\eqnn\xGenRecursion
$$\eqalignno{
\left(\frac{1}{2}+a_3-a_2\right)b^{(n)}_{12}=-b_{13}^{(n+1)}&+\left(\frac{1}{2}+a_3-a_1\right)b_{12}^{(n)}-nb_{12}^{(n+1)}&\xGenRecursion\cr
&+\left(\frac{1}{2}+a_3-a_2\right)b_{13}^{(n)}+\sum_{j=1}^n(-1)^jb_{13}^{(n-j)}b_{23}^{(1+j)}\,.
}$$
This cannot be as easily rearranged into a recursion relation. However, if we assume
that $b_{ij}^{(n)}$ is a polynomial in $a_j-a_i$ up to order $n$, we may use the above
to identify the polynomial coefficients. Doing this reveals the value of $b^{(4)}_{ij}$
as would be expected from \gGenLimits\ as the unique solution. And then we have verified
that the relation above is satisfied in a number of further cases if we assume this
general form of $b^{(n)}$.

We can perform a similar exercise for the $c^{(n)}_{ij}$ pole terms.
In its current form \FayOneN\ is not the most useful for this,
as we would like the $dist$ functions to be non-zero. 
Instead, we suppose the amplitude we are considering is $A(1,2,...,m)$ for convenience,
and look at an alternative Fay identity,
\eqn\FayOneNBoundary{
g^{(n)}_{1m}g^{(1)}_{m(m-1)}=-g_{1(m-1)}^{(n+1)}+g_{1(m-1)}^{(1)}g_{1m}^{(n)}-ng_{1m}^{(n+1)}+\sum_{j=0}^n(-1)^jg_{1(m-1)}^{(n-j)}g_{m(m-1)}^{(1+j)}\,.
}
We need not restrict ourselves to the $a_i=0$ $\forall i$ case here, as the computation
is simpler. Looking at the $s_{1m}$ single poles leads us to the relation
\eqnn\yGenRecursion
$$\eqalignno{
&c^{(n)}_{1m}\left(-\frac{1}{2}+a_{m-1}-a_m\right)=\left(\frac{1}{2}+a_{m-1}-a_1\right)c^{(n)}_{1m}-nc^{(n+1)}_{1m}\cr
\Rightarrow\quad{}&c^{(n+1)}_{1m}=\frac{1}{n}c^{(n)}_{1m}(1+a_m-a_1)&\yGenRecursion
}$$
Using that we know $c_{1m}^{(1)}=\frac{1}{2}$, this becomes
\eqnn\yGenDef
$$\eqalignno{
c_{1m}^{(n)}=&\frac{1}{2(n-1)!}\left(1+a_m-a_1\right)^{n-1}&\yGenDef
}$$
This agrees with the known values of $c^{(2)}_{17}$ and $c^{(3)}_{17}$ also.
We can also repeat this
calculation for poles of $g_{12}^{(n)}$ to see what would happen if the $dist$ function
were not triggered, and find the similar relation
\eqnn\yGenDefTwo
$$\eqalignno{
c^{(n)}_{12}=&\frac{1}{2(n-1)!}\left(a_2-a_1\right)^{n-1}&\yGenDefTwo
}$$
Hence the form of $c^{(n)}_{ij}$ presented in \gGenLimits\ is the natural generalization,
and we expect \gGenLimits\ to hold to higher points.

We end this discussion though by stressing that this approach is highly speculative, and
we have not tested these values produced in any way beyond the aforementioned
discussion. They are however a strong candidate for what they are attempting to describe.

%*********************************************************
\appendix{D}{The BRST analysis of a seven-point numerator}
\applab\sevAp

\noindent In this appendix we identify the full expression for the $[5,[6,7]]$-pentagon in 
the amplitude $A(1,2,3,4,5,6,7;\ell+4k_4-6k_5)$, and confirm that its variation
has the desired form.
We begin by finding the coefficient of one term contributing to the numerator in
detail, namely $V_1T_{2576,3,4}$.
Within the string correlator this is associated with the worldsheet function
\eqnn\TermFromCorr
$$\eqalignno{
%V_1T_{2576,3,4}
{\cal Z}_{1,2576,3,4}=
%V_1T_{2576,3,4}\Big(
&g^{(1)}_{25}g^{(1)}_{57}g^{(1)}_{76}
+g^{(3)}_{25}
+g^{(3)}_{57}
+g^{(3)}_{76}
-2g^{(3)}_{62}
+g^{(1)}_{25}(g^{(2)}_{57}+g^{(2)}_{76}-g^{(2)}_{62})\cr
+&g^{(1)}_{57}(g^{(2)}_{25}+g^{(2)}_{76}-g^{(2)}_{62})
+g^{(1)}_{76}(g^{(2)}_{25}+g^{(2)}_{57}-g^{(2)}_{62})\,.
%\Big)
&\TermFromCorr
}$$
Only two of these terms contain the $s_{67}s_{567}$ pole structure, 
$g^{(1)}_{25}g^{(1)}_{57}g^{(1)}_{76}$ and $g^{(1)}_{76}g^{(2)}_{57}$.
The contribution of the former was identified in \EndOfEg ,
and the latter follows from \gGenLimitsTwo ,
\eqnn\ExtraTermValues
$$\eqalignno{
&c^{(1)}_{76}c^{(2)}_{57}=\frac{1}{2}\cdot\frac{6}{2}=\frac{3}{2}\,.&\ExtraTermValues\cr
}$$
Summing these together, the $V_{1}T_{2576,3,4}$ contribution to the $[5,[6,7]]$-pentagon is
\eqnn\SixSevenHexSpecificTerm
$$\eqalignno{
\left(-\frac{11}{8}+\frac{3}{2}\right)
V_1T_{2576,3,4}\hat\phi(1234567|576)I_{567}
&=-\frac{1}{8s_{67}s_{567}}V_1T_{2576,3,4}I_{1,2,3,4,567}&\SixSevenHexSpecificTerm
}$$
Similar calculations for all other terms in the correlator yield the numerator expression
\eqnn\PentEgFullDef
$$\eqalignno{
N^{a_4=4,a_5=-6}_{1|2,3,4,[5,[6,7]]}(\ell)&=
6V_1T^{mn}_{2,3,4,5,67}k^m_5k^n_{67}
+V_1T^{m}_{2,3,4,[5,67]}(\ell^m-6k_5^m+6k_{67}^m)\cr
&-6\left((V_1T^m_{25,3,4,67}k^m_{67}+(2\leftrightarrow 3,4))+V_{15}T^m_{2,3,4,67}k^m_{67}+(5\leftrightarrow [6,7])\right)\cr
&+\frac{1}{2}\left(V_{12}T_{3,4,[5,67]}+(2\leftrightarrow 3,4,[5,67])\right)\cr
&+\frac{1}{2}\left(V_{1}T_{23,4,[5,67]}+(2,3|2,3,4,[5,67])\right)&\PentEgFullDef\cr
&+6\left((V_1T_{25,[3,67],4}+(2,3|2,3,4))+(2\leftrightarrow 3)\right)\cr
&+6\left((V_{15}T_{[2,67],3,4}+(2\leftrightarrow 3,4))+(5\leftrightarrow [6,7])\right)\cr
&+6\left((V_{1}T_{2675,3,4}+(2\leftrightarrow 3,4))-(6\leftrightarrow 7)\right)\cr
&+6\left(V_{1675}T_{2,3,4}-(6\leftrightarrow 7)\right)
+4\left(V_1T_{24,3,[5,67]}+(2\leftrightarrow 3)\right)\cr
&
%-\frac{1}{2}V_{1}T_{[25,67],3,4}
+4V_{14}T_{2,3,[5,67]}
-4V_{1}T_{2,3,[4,[5,67]]}
+6V_1J_{5|2,3,4,6,7}^m(k_6^m-k_7^m)\cr
&+6s_{67}\left((V_1J_{5|27,3,4,6}+(2\leftrightarrow 3,4,6))+V_{17}J_{5|2,3,4,5,6}-(6\leftrightarrow 7)\right)
}$$
The $VJ$ terms above are those which arise naively by looking to the $s_{67}s_{567}$ poles in the correlator.
As discussed previously it may be that they require some rearrangement to be in a BCJ representation, but for illustrating the field theory limit methods we give the numerator in the above form. 
A lengthy calculation yields the variation
\eqnn\PentEgExpansion
$$\eqalignno{
&QN^{a_4=4,a_5=-6}_{1|2,3,4,[5,[6,7]]}(\ell)=
\frac{1}{2}V_1V_2T_{3,4,[5,67]}\left((\ell-k_{12}+4k_4-6k_5)^2-(\ell-k_1+4k_4-6k_5)^2\right)\cr
&+\frac{1}{2}V_1V_3T_{2,4,[5,67]}\left((\ell-k_{123}+4k_4-6k_5)^2-(\ell-k_{12}+4k_4-6k_5)^2\right)\cr
&+\frac{1}{2}V_1V_4T_{2,3,[5,67]}\left((\ell-k_{1234}+4k_4-6k_5)^2-(\ell-k_{123}+4k_4-6k_5)^2\right)\cr
&+\frac{1}{2}V_1V_{[5,67]}T_{2,3,4}\left((\ell-k_{1234567}+4k_4-6k_5)^2-(\ell-k_{1234}+4k_4-6k_5)^2\right)\cr
%%%%%%%%%%
% s_67 terms
%%%%%%%%%%
+(k^6\cdot k^7)&\bigg(
\big(6V_1V_{26}T_{3,4,5,7}^mk^m_5+(2\leftrightarrow 3,4,5)\big)
+V_1V_{57}T_{2,3,4,6}^m\big(\ell^m+6k_{57}^m\big)\cr
&+6V_1V_7T_{2,3,4,5,6}^{mn}k^m_5k^n_{67}
+6k^m_5(V_1V_6T_{27,3,4,5}^m+(2\leftrightarrow 3,4,5))&\PentEgExpansion\cr
&+\big(\ell^m+6k^m_{67}\big)V_1V_7T_{2,3,4,56}^m
+6(V_1V_6T_{25,3,4,7}^mk_{67}^m+(2\leftrightarrow 3,4))\cr
&+6V_{15}V_6T_{2,3,4,7}^mk_{67}^m
+6V_{17}V_6T_{2,3,4,5}^mk_{5}^m
\cr
&+6V_1V_5T_{2,3,4,6,7}^{mn}k_5^mk_7^n
+V_{16}V_5T_{2,3,4,7}^mk_5^m\cr
&+(6V_1V_{25}T_{3,4,6,7}^mk_6^m+(2\leftrightarrow 3,4))
+(6V_1V_{5}T_{26,3,4,7}^mk_5^m+(2\leftrightarrow 3,4,7))\cr
%%%%%%%%%%
% End of T^m terms
%%%%%%%%%%
&+ \frac{1}{2}(V_1V_{[2,57]}T_{3,4,6}+(2\leftrightarrow 3,4))\cr
&-\frac{1}{2}(V_1V_{26}T_{3,4,57}+(2\leftrightarrow 3,4))
-\frac{1}{2}(V_1V_{56}T_{23,4,7}+(2,3| 2,3,4,7))\cr
&+\frac{1}{2}(V_1V_7T_{[2,3],4,56}+(2,3|2,3,4,56))
+\frac{1}{2}(V_{12}V_{57}T_{3,4,6}+(2\leftrightarrow 3,4))\cr
&+\frac{1}{2}(V_{12}V_7T_{3,4,56}+(2\leftrightarrow 3,4,56))
-\frac{1}{2}V_{17}V_{56}T_{2,3,4}
+\frac{1}{2}V_{175}V_6T_{2,3,4}\cr
&+6\big((V_1V_{27}T_{[3,5],4,6}+(3\leftrightarrow 4,6))+(2\leftrightarrow 3,4)\big)
+6(V_1V_7T_{[26,5],3,4}+(2\leftrightarrow 3,4))\cr
&+6(V_1V_7T_{25,36,4}+V_1V_7T_{26,35,4}+(2,3|2,3,4))
+6(V_{15}V_{27}T_{3,4,6}+(2\leftrightarrow 3,4))\cr
&+6(V_{15}V_7T_{26,3,4}+(2\leftrightarrow 3,4))
+6(V_{16}V_7T_{25,3,4}+(2\leftrightarrow 3,4))
%+V_{16}V_7T_{25,3,4}-V_{16}V_{57}T_{2,3,4}\big)
\cr
&+6\big((V_1V_{25}T_{37,4,6}+(3\leftrightarrow 4,6))+V_{17}V_{25}T_{3,4,6}+(2\leftrightarrow 3,4)\big)\cr
&+ 6V_{165}V_7T_{2,3,4}
+6(V_1(V_{257}+V_{275})T_{3,4,6}+(2\leftrightarrow 3))
+6V_1V_{576}T_{2,3,4}
\cr
&+4(V_1V_{57}T_{24,3,6}+(2\leftrightarrow 3,6))+4V_{14}V_{57}T_{2,3,6}
+4(V_1V_7T_{24,3,56}+(2\leftrightarrow 3,56))\cr
&+4V_{14}V_7T_{2,3,56}
+4V_1V_{46}T_{2,3,57}
+2V_1V_{457}T_{2,3,6}
+20V_1V_{475}T_{2,3,6}
\cr
&+6V_1Y_{2,3,4,5,6,7}^mk_7^m+6(V_1Y_{26,3,4,5,7}+(2\leftrightarrow 3,4,5,7))
+6V_{16}Y_{2,3,4,5,7}\cr
&-(6\leftrightarrow 7)\bigg)\cr
%%%%%%%%%%
% End of s_67 terms
%%%%%%%%%%
+(k^{5}\cdot k^{67})\bigg(\Big(&
\frac{1}{2}(V_1V_{[2,67]}T_{3,4,5}+V_{12}V_{67}T_{3,4,5}+(2\leftrightarrow 3,4))
+4V_1V_5T_{2,3,67}
+4V_{14}V_{67}T_{2,3,5}\cr
&+\frac{1}{2}(V_{1}V_{67}T_{23,4,5}+(2,3|2,3,4,5))
+4(V_1V_{67}T_{24,3,5}+(2\leftrightarrow 3,5))
-(5\leftrightarrow 67)
\Big)\cr
&-\frac{1}{2}(V_{15}V_{67}T_{2,3,4}+(5\leftrightarrow 6,7))
+6\big((V_1V_25T_{3,4,67}-(25\leftrightarrow 67))+(2\leftrightarrow 3,4)\big)\cr
&-6V_{15}V_{67}T_{2,3,4}
-6V_1Y_{2,3,4,5,67}
\bigg)\cr
+6(k^6\cdot k^7)&(k^5\cdot k^6)V_1V_5(J_{7|2,3,4,6}+J_{6|2,3,4,7})
-6(k^5\cdot k^{67})(k^6\cdot k^7)V_1V_5J_{7|2,33,4,6}
}$$
This has intentionally been expressed with factors $(\ell\cdot k)$ reformulated in terms
of propagators. For an $n$-point amplitude in the canonical ordering with
arbitrary loop momentum structure, this is done with
\eqnn\GeneralEllDotk
$$\eqalignno{
(\ell\cdot k_{i(i+1)...j})&=
-\frac{1}{2}(\ell+\sum_{m=1}^na_mk_m-k_{12....j})^2
+\frac{1}{2}(\ell+\sum_{m=1}^na_mk_m-k_{12....(i-1)})^2\cr
&\qquad{}\qquad{}-k_{i(i+1)...j}\cdot\left(\sum_{m=1}^na_mk_m-\frac{1}{2}k_{i(i+1)...j}\right)\,.&\GeneralEllDotk
}$$
We may then be reassured of the validity of this numerator expression, as those terms
in the variation proportional to propagators cancel terms from other box numerators.
For example, one such set of terms is
\eqnn\VarCancelsBoxes
$$\eqalignno{
V_1V_3T_{2,4,[5,67]}&\left((\ell-k_{123}+4k_4-6k_5)^2-(\ell-k_{12}+4k_4-6k_5)^2\right)I^{a_4=4,a_5=-6}_{1,2,3,4,567}\cr
&=V_1V_3T_{2,4,[5,67]}\left(I^{a_4=4,a_5=-6}_{1,2,34,567}-I^{a_4=4,a_5=-6}_{1,23,4,567}\right)&\VarCancelsBoxes
}$$
This then cancels one term in the variation of the $[3,4],[5,[6,7]]$-box, and one from
the $[2,3],[5,[6,7]]$ box.
Similar holds true for all other terms in the variation, and the remaining terms in
\PentEgExpansion\ are canceled themselves by analogous results in
the variation of hexagons.

%**************************************************************
\appendix{E}{The five-point color-dressed integrand}
\applab\fivecolorapp

\noindent In this appendix the five-point color-dressed integrand will be written
down after the application of the color decomposition techniques of \KKLance.

The five-point color-dressed one-loop integrand can be written as
\eqnn\FivePtColDressed
$$\eqalignno{
M_5(\ell)
&= \Big(\half \cN_{1|2,3,45} I_{1,2,3,45} B_{1,2,3,45}
          + \half \cN_{1|2,34,5} I_{1,2,34,5} B_{1,2,34,5}&\FivePtColDressed\cr
&          + \half \cN_{1|23,4,5} I_{1,23,4,5} B_{1,23,4,5}
          + \half \cN_{12|3,4,5} I_{12,3,4,5} B_{12,3,4,5}\cr
&          + \half \cN_{51|2,3,4} I_{51,2,3,4} B_{51,2,3,4}
          + \cN_{1|2,3,4,5}(\ell) I_{1,2,3,4,5} P_{1,2,3,4,5}
         + \perm(2,3,4,5)\Big)
}$$
where ${\cal N}$ denotes the Berends-Giele counterpart of the $n$-gon numerator as described in the appendix~\Convapp\ while
the color factors of the box and pentagon cubic graphs are
\eqn\bpex{
B_{12,3,4,5} = f^{a12}f^{eab}f^{b3c}f^{c4d}f^{d5e},\qquad P_{1,2,3,4,5} = f^{a1b}f^{b2c}f^{c3d}f^{d4e}f^{e5a}\,.
}
The factor of $\frac{1}{2}$ in \FivePtColDressed\ compensates the overcounting of graphs due to symmetries.
Note that the box numerators do not depend on the loop momentum.

The color-dressed integrand \FivePtColDressed\ is BRST closed. To see this we expand all color factors in terms of their
pentagon constituents using the Jacobi identity as $B_{12,3,4,5} = P_{1,2,3,4,5} - P_{2,1,3,4,5}$ \KKLance\ and
consider the terms proportional to $P_{1,2,3,4,5}$. Using the five-point numerators of section~\GenProcedureFivePts\ these are
\eqnn\StandardOrderTerms
$$\eqalignno{
M_5(\ell)\Big|_{P_{1,2,3,4,5}} &= \cN_{1|2,3,4,5}(\ell)I_{1,2,3,4,5}
+ \half\Big(\cN_{12|3,4,5} I_{12,3,4,5}
-\cN_{21|3,4,5} I_{1,3,4,5} &\StandardOrderTerms\cr
&+\big[\cN_{1|23,4,5}
-\cN_{1|32,4,5} \big]I_{1,23,4,5}
+\big[\cN_{1|2,34,5}
-\cN_{1|2,43,5}\big]I_{1,2,34,5}\cr
&+\big[\cN_{1|2,3,45}
-\cN_{1|2,3,54}\big]I_{1,2,3,45}
+\cN_{51|2,3,4}I_{1,2,3,4}
-\cN_{15|2,3,4}I_{15,2,3,4}\Big)\,.
}$$
After using $N_{ij|k,l,m} = - N_{ji|k,l,m}$ by \BoxDef\ and performing the loop momentum shifts $\ell^\prime =\ell-k_2$
in $I_{1,3,4,5}$ and $\ell^\prime =\ell+k_5$ in $I_{15,2,3,4}$ these terms become the integrand $A(1,2,3,4,5;\ell)$ of \fivexcB,
\eqnn\SymTwo
$$\eqalignno{
M_5(\ell)\Big|_{P_{1,2,3,4,5}} &=
\cN_{1|2,3,4,5}(\ell)I_{1,2,3,4,5}
+\cN_{1|23,4,5}I_{1,23,4,5}
+\cN_{1|2,34,5}I_{1,2,34,5}\cr
&+\cN_{1|2,3,45}I_{1,2,3,45}
+\cN_{12|3,4,5}I_{12,3,4,5}
+\cN_{51|3,4,5}I_{1,2,3,4}\,.&\SymTwo
}$$
Hence, after considering all the permutations the color-dressed integrand \FivePtColDressed\ becomes
\eqn\fivecolor{
M_5(\ell) = A(1,2,3,4,5;\ell)\, P_{1,2,3,4,5} + \perm(2,3,4,5)
}
and it is manifestly BRST closed. The rewriting \fivecolor\ agrees with the general result of \KKLance\ (see e.g. equation (3.4)
of \veronneu).

\listrefs
\bye